\documentclass[mksc,nonblindrev]{informs3}

\makeatletter
\providecommand{\theARTICLETOP}{}
\makeatother

\DoubleSpacedXII

\usepackage{standalone}
\usepackage{hyperref}
\newcounter{numquote}

\newcommand\quoteref[1]{\csname#1\endcsname}
\usepackage{algorithm,algorithmic}
\usepackage{tikz}
\usetikzlibrary{arrows.meta,decorations.pathreplacing,positioning,calc}
\tikzstyle{spinner} = [circle, minimum width = 3cm, draw=black, fill=red!30]
\usepackage{listings}
\usepackage{color}
\definecolor{lightgray}{rgb}{.9,.9,.9}
\definecolor{darkgray}{rgb}{.4,.4,.4}
\definecolor{purple}{rgb}{0.65, 0.12, 0.82}
\usetikzlibrary{calc}
\lstdefinelanguage{JavaScript}{
  keywords={typeof, new, true, false, catch, function, return, null, catch, switch, var, if, in, while, do, else, case, break},
  keywordstyle=\color{blue}\bfseries,
  ndkeywords={class, export, boolean, throw, implements, import, this},
  ndkeywordstyle=\color{darkgray}\bfseries,
  identifierstyle=\color{black},
  sensitive=false,
  comment=[l]{//},
  morecomment=[s]{/*}{*/},
  commentstyle=\color{purple}\ttfamily,
  stringstyle=\color{red}\ttfamily,
  morestring=[b]',
  morestring=[b]"
}
\usetikzlibrary{positioning,arrows.meta,calc,shapes.geometric,decorations.pathreplacing}
\usetikzlibrary{arrows.meta,positioning,calc}
\usetikzlibrary{fit,positioning,calc,arrows.meta,backgrounds}

\tikzset{
  person/.pic={
    \draw[line width=0.8pt, line cap=round, line join=round]
      (0,0) circle (0.08)               
      (0,-0.08) -- (0,-0.40)            
      (-0.22,-0.22) -- (0,-0.18) -- (0.22,-0.22)   
      (-0.18,-0.40) -- (0,-0.30) -- (0.18,-0.40);  
  },
}

\tikzstyle{arrow} = [thick,->,>=stealth]

\tikzset{
  adimg/.style args={#1}{
    inner sep=0pt, outer sep=0pt,
    rounded corners=1.5pt,
    draw=black!50, line width=0.3pt,
    minimum width=2.2cm, minimum height=1.5cm,
    path picture={
      \node[anchor=center, inner sep=0pt] at (path picture bounding box.center)
        {\includegraphics[width=\linewidth,height=\height,keepaspectratio]{#1}};
    }
  }
}

\usepackage{graphicx}
\usepackage{pgfplots}
\usetikzlibrary{positioning}
\usepackage{adjustbox}
\usepackage{booktabs}
\usepackage{pifont}
\usepackage{natbib}
 \bibpunct[, ]{(}{)}{,}{a}{}{,}%
 \def\bibfont{\small}%
\TheoremsNumberedThrough     

\EquationsNumberedThrough    

\pgfplotsset{compat=1.18} 
\begin{document}


 \RUNAUTHOR{}

\RUNTITLE{A Privacy Budgeting Framework for Online Experimentation}

\TITLE{A Privacy Budgeting Framework for Online Experimentation}

\ARTICLEAUTHORS{%
\AUTHOR{Gilian R. Ponte, Alina Ferecatu}\vspace{1em}
\AFF{Rotterdam School of Management, Erasmus University}
}

\ABSTRACT{%
Firms perform online experiments with multi-armed bandits to personalize what consumers are shown while balancing exploration and exploitation. However, third-parties can infer consumers’ underlying segments from observing which banners, ads, or recommendations consumers receive. To control this inference, we propose a privacy arisk budget that firms can set ex ante to bound such third party belief updating using differential privacy.  To spend this privacy risk budget, we propose two strategies: a constant privacy risk strategy and a dynamic privacy risk strategy that spend privacy
risk differently across visitor. We study how privacy risk budgets affect experimentation performance in two applications—website design and a recommendation system—under these strategies. For both strategies, we analytically find privacy risk budgets that optimally balance exploration and exploitation. We then extend the idea of an experiment-level privacy risk budget to a firm-wide privacy risk budget. We apply this firm-wide privacy risk budget in an empirical setting with 78 experiments. We find that the dynamic strategy is particularly valuable in longer and more complex experiments, and that optimizing the allocation of a firm-wide privacy risk budget across experiments substantially improves learning performance.
}%



\maketitle

\vspace{-1.5cm}
\section{Introduction}
Firms routinely experiment with website banners, links, and recommendations to improve engagement and profitability. However, online experimentation also raises privacy concerns, particularly because third-parties can track what consumers are shown and use those exposures to infer their interests and behaviors \citep{goldfarb_tucker_2011,Summers_2016, Kim_2018,shaddy_2026}. These concerns translate into economic costs for firms: consumers increasingly opt out of tracking \citep{Tucker_2014, Johnson_2020, Pengyuan_2024, MILLER2024241} and place a premium on privacy-respecting brands \citep{Jordan_2025}.

To illustrate the privacy risk of experimentation, consider Jamie, a frequent reader of fantasy novels who visits the online bookstore Barnes \& Noble. Third-party software that monitors the website, which we refer to as a tracker, may hold a prior belief that Jamie is a fantasy reader based on public information, such as their Goodreads account. Barnes \& Noble first segments customers by book-genre preferences and then experiments to learn which output, such as a book recommendation, performs best for each segment.\footnote{Barnes \& Noble personalizes content and measures marketing effectiveness with third-party partners. \cite{BarnesNobleCookiePolicy}: ``This is done so that we can personalize and enhance your browsing and shopping experience."} To improve its recommendations, Barnes \& Noble balances learning—randomly displaying different recommendations—with earning—using experimental results to select the currently best-performing recommendation.

This balance between learning and earning shapes Jamie's privacy risk. Under optimization, the recommendation closely reflects Jamie’s segment and underlying preference, therefore introducing a privacy risk. By contrast, experimentation implies randomization, which makes recommendations stochastic and weakens the link to Jamie’s segment. Thus, optimization improves performance but increases privacy risk, whereas randomization strengthens privacy protection at the cost of displaying a suboptimal recommendation.

\begin{figure}[h]
\centering
\resizebox{0.9\linewidth}{!}{%
\begin{tikzpicture}[
    node distance=0.25cm,
    every node/.style={font=\scriptsize},
    box/.style={
        draw,
        line width=0.4pt,
        rounded corners=2pt,
        align=center,
        text width=3cm,
        minimum height=1.75cm,
        inner sep=4pt
    },
    sidebox/.style={
        draw,
        line width=0.4pt,
        rounded corners=2pt,
        align=center,
        text width=3cm,
        minimum height=1.75cm,
        inner sep=4pt
    },
    op/.style={
        font=\large,
        inner sep=2pt
    }
]

\node[sidebox] (prior) 
{Tracker's\\prior belief\\about segment\\membership};

\node[op, right=of prior] (times1) {$\times$};

\node[box, right=of times1] (output) 
{\textbf{Privacy risk}\\
\textbf{Channel 1}\\[0.25em]
Experimental output\\[0.25em]
\textit{What does the output}\\
\textit{reveal about Jamie?}};

\node[op, right=of output] (times3) {$\times$};

\node[box, right=of times3] (interaction) 
{\textbf{Privacy risk}\\
\textbf{Channel 2}\\[0.25em]
Interaction\\[0.25em]
\textit{What do clicks}\\
\textit{reveal about Jamie?}};

\node[op, right=of interaction] (equals) {$=$};

\node[sidebox, right=of equals] (posterior) 
{Tracker's\\posterior belief\\about segment\\membership};

\draw[
    decorate,
    decoration={brace, mirror, amplitude=3pt},
    line width=0.4pt
]
([yshift=-0.05cm]output.south west) --
node[below=2pt, font=\scriptsize, align=center]
{\textbf{The focus of this} \\\textbf{paper’s privacy guarantee.}}
([yshift=-0.05cm]output.south east);

\end{tikzpicture}%
}
\vspace{-1em}
\caption{A decomposition of the tracker's posterior belief update about segment membership into two privacy risk channels, conditional on access.}
\label{fig:privacy_risk_channels}
\end{figure}
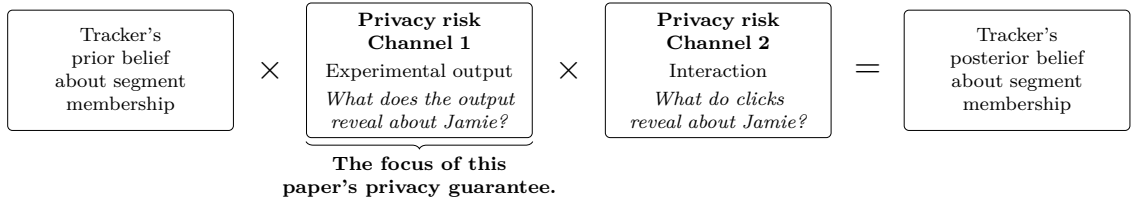

This privacy risk materializes when Jamie’s visit to Barnes \& Noble allows the tracker to update its prior belief about Jamie's book preferences through two information channels displayed in Figure~\ref{fig:privacy_risk_channels}: the experimental output, i.e., the book recommendation, and Jamie’s interaction with the output, i.e., a click on the recommended book (see Web Appendix~\ref{modelprivacyrisk} for a formal decomposition). The information available to the tracker depends on Jamie's GDPR-style consent decision \citep{Johnson_2020, Choi_2023}, the use of an ad blocker \citep{Todri_2022}, and the privacy policies of the firm \citep{Brough_2022}. When the information is not available, the tracker cannot observe either the experimental output or Jamie’s subsequent interactions with the output. When the information is available, the tracker loads with the page and observes information from the first channel, the experimental output displayed to Jamie. The book recommendation Jamie sees reveals information about Jamie’s interests \citep[e.g., see][]{Summers_2016,Kim_2018}. A fantasy book recommendation allows the tracker to increase their belief that Jamie belongs to the fantasy segment. The second channel is Jamie’s interaction with the displayed output. If Jamie clicks on the fantasy book recommendation, the click provides an additional signal that further increases the tracker’s belief that Jamie is a fantasy reader. Together, these channels show that experimentation creates privacy risk through both exposure and interaction.


\vspace{-2em}
 Our privacy guarantee operates at the impression level, bounding how much an experimental output can update a tracker’s belief about a consumer’s segment membership (see Figure~\ref{fig:privacy_risk_channels}, Channel~1).  We protect every displayed experimental output rather than interactions with that output (see Figure~\ref{fig:privacy_risk_channels}, Channel~2). Protecting impressions is particularly important because most displayed outputs do not generate clicks in low-CTR settings \citep{Jason_2026}, yet experimental outputs may still reveal behavioral information that is used to personalize them \citep{Castelluccia_2012, Xin_2023, Chen_2026}.

\begin{figure}[h]
\centering
\begin{tikzpicture}[
    >=Latex,
    font=\small,
    visitor/.style={circle, draw, inner sep=1.2pt, fill=white},
    jamie/.style={circle, draw, very thick, inner sep=1.5pt, fill=gray!20},
    box/.style={
        draw,
        rounded corners,
        align=center,
        minimum width=2.4cm,
        minimum height=10mm,
        inner sep=4pt
    },
    arrow/.style={->}
]


\node[align=center] (step2e) at (3.5,15.35)
{\scriptsize \textbf{Level 1: Customer-level privacy risk $\xi$}:\\[-4mm]
\scriptsize The firm assigns privacy risk to each visitor.};

\coordinate (mainarrow_east) at (9.0,12);
\coordinate (mainarrow_near_west) at (-2.1,12);

\draw[thick, ->] (-2.1,12) -- (mainarrow_east);

\node[visitor, label=below:{1}] at (-1.3,12) {};
\node[visitor] at (-0.3,12) {};
\node[visitor] at (0.8,12) {};
\node[visitor] at (1.9,12) {};
\node[jamie, label=below:{$t$}] (jamie) at (3.8,12) {};
\node[visitor] at (4.8,12) {};
\node[visitor] at (5.9,12) {};
\node[visitor] at (6.9,12) {};
\node[visitor, label=below:{$T$}] at (7.9,12) {};

\node (bestbook) at (0,14)
    {\includegraphics[width=1cm,height=1.5cm]{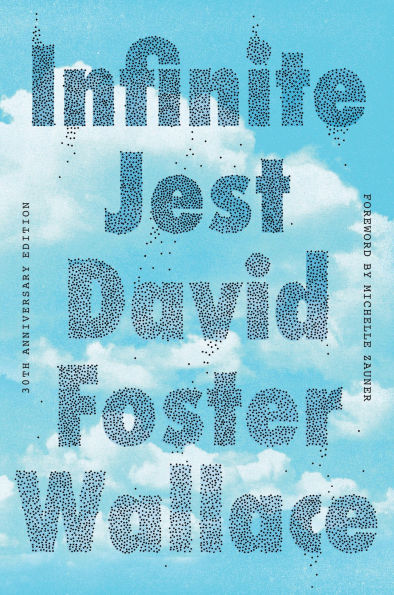}};

\node[align=center] (Jamie) at (3.8,13)
{\scriptsize Jamie.};

\node[align=center] (serve_rule) at (0,12.9)
{\scriptsize Firm serves book \\[-4mm] \scriptsize with highest clicks.};

\draw[thick, arrow]
  ([xshift=2mm,yshift=3mm]serve_rule.east) --
  node[midway, above, xshift=-0.25cm, align=center]
  {\scriptsize exploitation\\[-4mm]\scriptsize (w.p. $1-\varepsilon(\xi)$)}
  (Jamie.west);

\draw[thick, arrow]
  (jamie.north) -- (Jamie.south);

\node (b1) at (6.2,14)
    {\includegraphics[width=1cm,height=1.5cm]{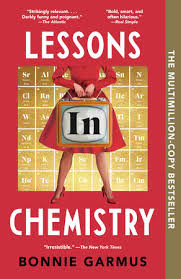}};
\node (b2) at (8.5,14)
    {\includegraphics[width=1cm,height=1.5cm]{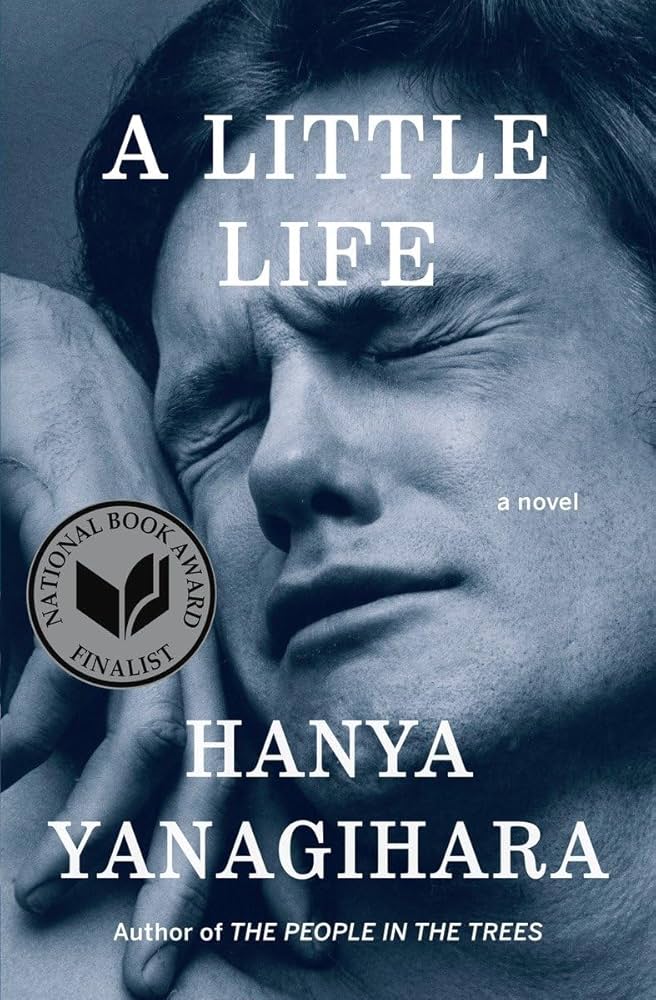}};
\node (b4) at (7.35,14)
    {\includegraphics[width=1cm,height=1.5cm]{figures/infinite.jpg}};

\draw[decorate,decoration={brace,mirror,amplitude=4pt},thick]
    (5.7,13.2) -- (9,13.2)
    node[midway,yshift=-10pt,align=center]
    {\scriptsize Firm serves a random book};

\coordinate (randomserve_west) at (5.2,13.2);

\draw[thick, arrow]
  (randomserve_west) --
  node[midway, above, xshift=-0.25cm, align=center]
  {\scriptsize exploration\\[-4mm]\scriptsize (w.p. $\varepsilon(\xi)$)}
  (Jamie);


\draw[decorate,decoration={brace,mirror,amplitude=4pt},thick]
    (-2.1,11.2) -- (9,11.2)
    node[midway,yshift=-25pt,align=center](step1e){\scriptsize \textbf{Level 2: Experiment-level privacy budget $\gamma$}: \\[-4mm] \scriptsize For each experiment, the largest visitor-level\\[-4mm] \scriptsize privacy risk is bounded by the experiment-level budget.};
     

\node[align=center] (step2f) at (4,5.75)
{\scriptsize \textbf{Level 3: Firm-wide privacy budget $\Gamma$}: \\[-4mm]\scriptsize Across experiments, the firm-wide budget bounds\\[-4mm]\scriptsize the total experiment-level privacy risk.};

\node (exp1title) at (-1,9.2) {\scriptsize Experiment 1:};
\node at (-1,8.8) {\scriptsize coming of age};

\node (e1a) at (-1,7.8)
{\includegraphics[width=1.1cm,height=1.6cm]{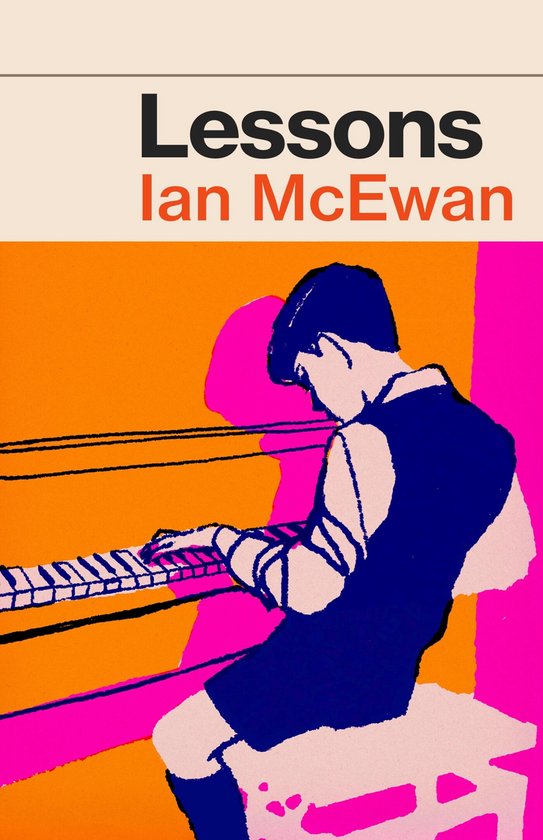}};
\node (e1b) at (-0.6,7.5)
{\includegraphics[width=1.1cm,height=1.6cm]{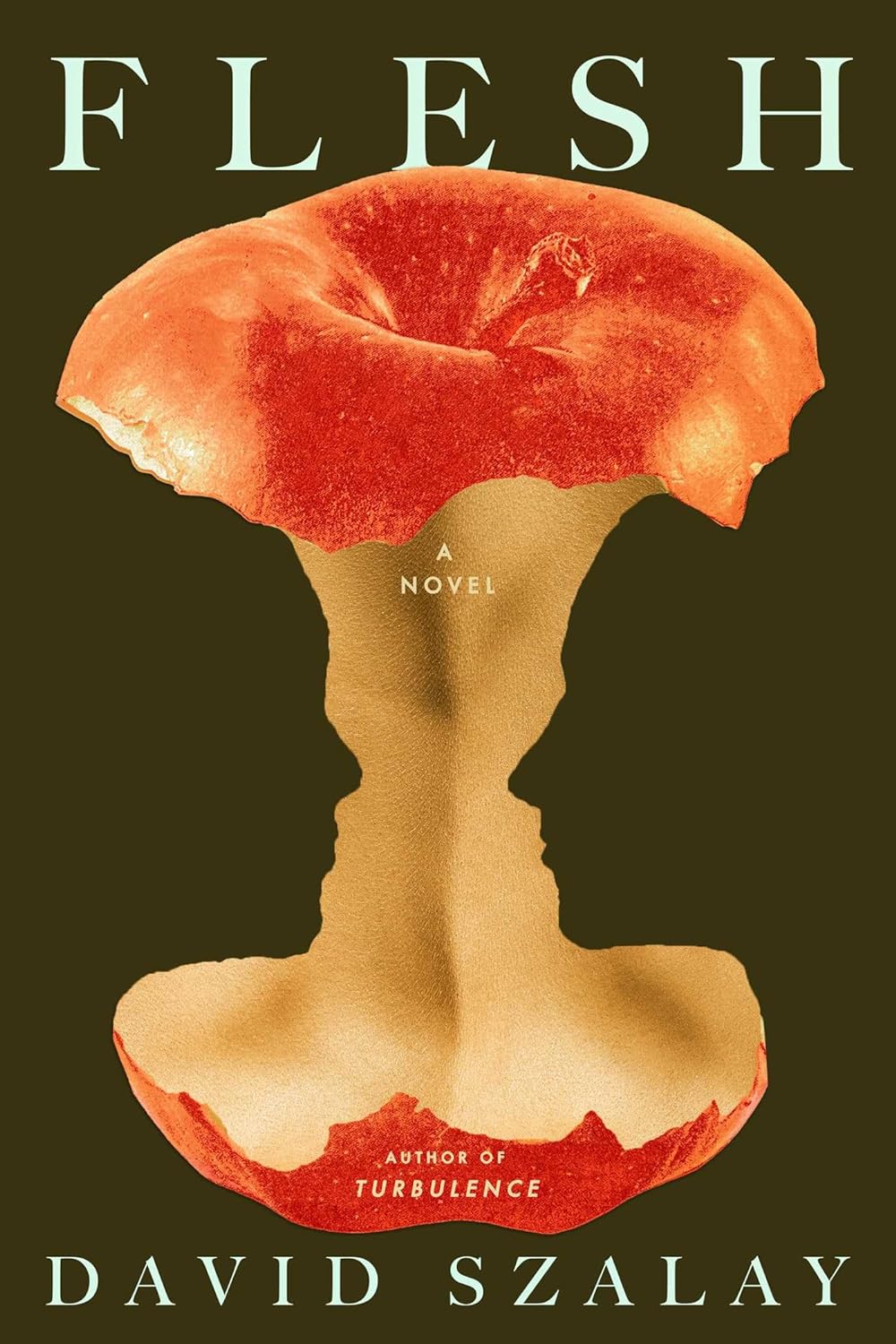}};

\node (exp2title) at (2.5,9.2) {\scriptsize Experiment 2:};
\node at (2.5,8.8) {\scriptsize fantasy};

\node (e2a) at (2.5,7.8)
{\includegraphics[width=1.1cm,height=1.6cm]{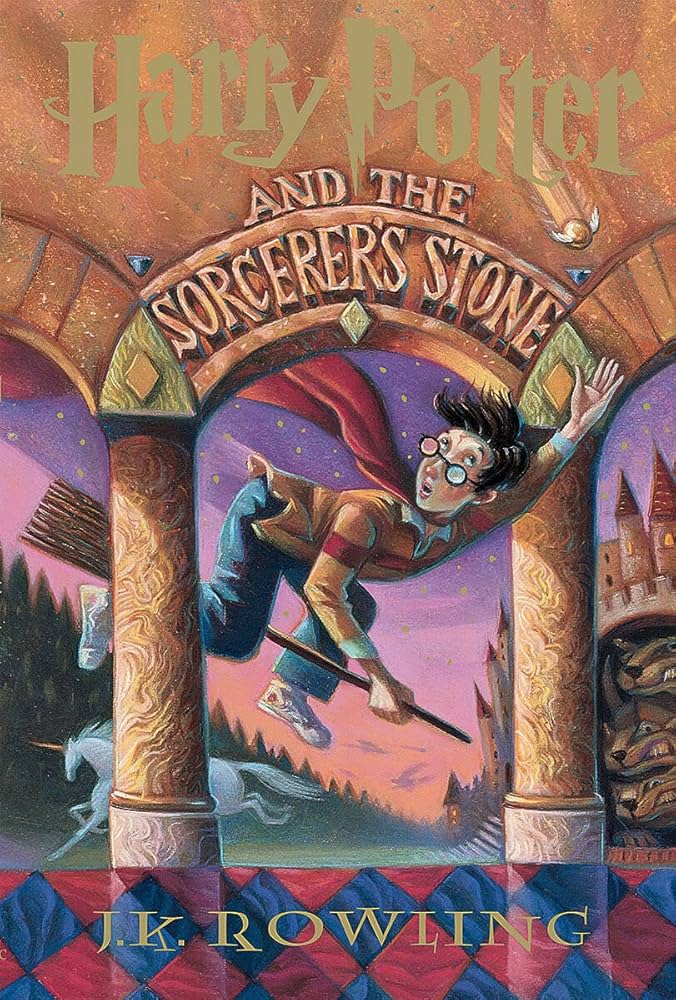}};
\node (e2b) at (2.9,7.4)
{\includegraphics[width=1.1cm,height=1.6cm]{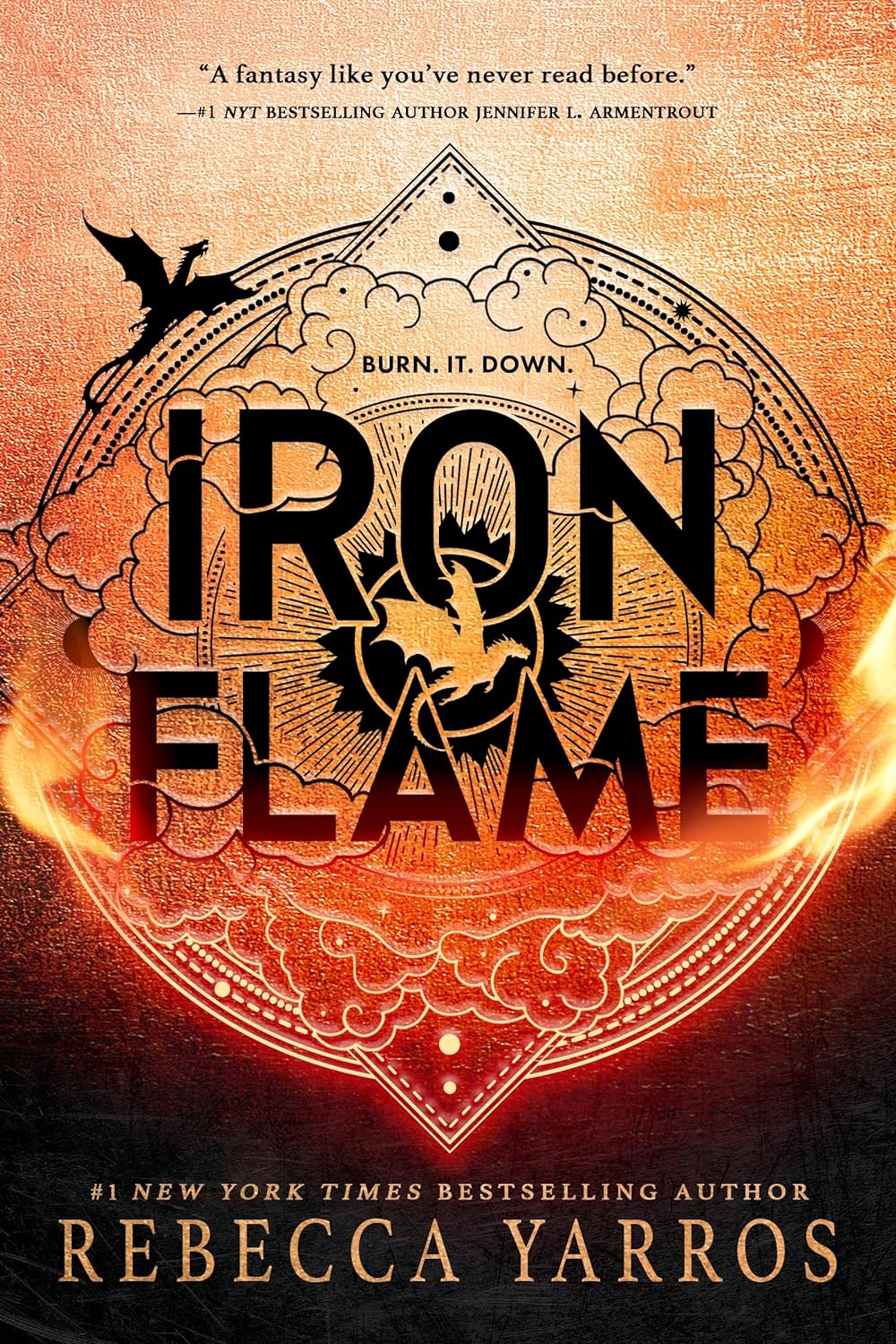}};

\node (exp3title) at (5.5,9.2) {\scriptsize Experiment 3:};
\node at (5.5,8.8) {\scriptsize science fiction};

\node (e3a) at (5.5,7.8)
{\includegraphics[width=1.1cm,height=1.6cm]{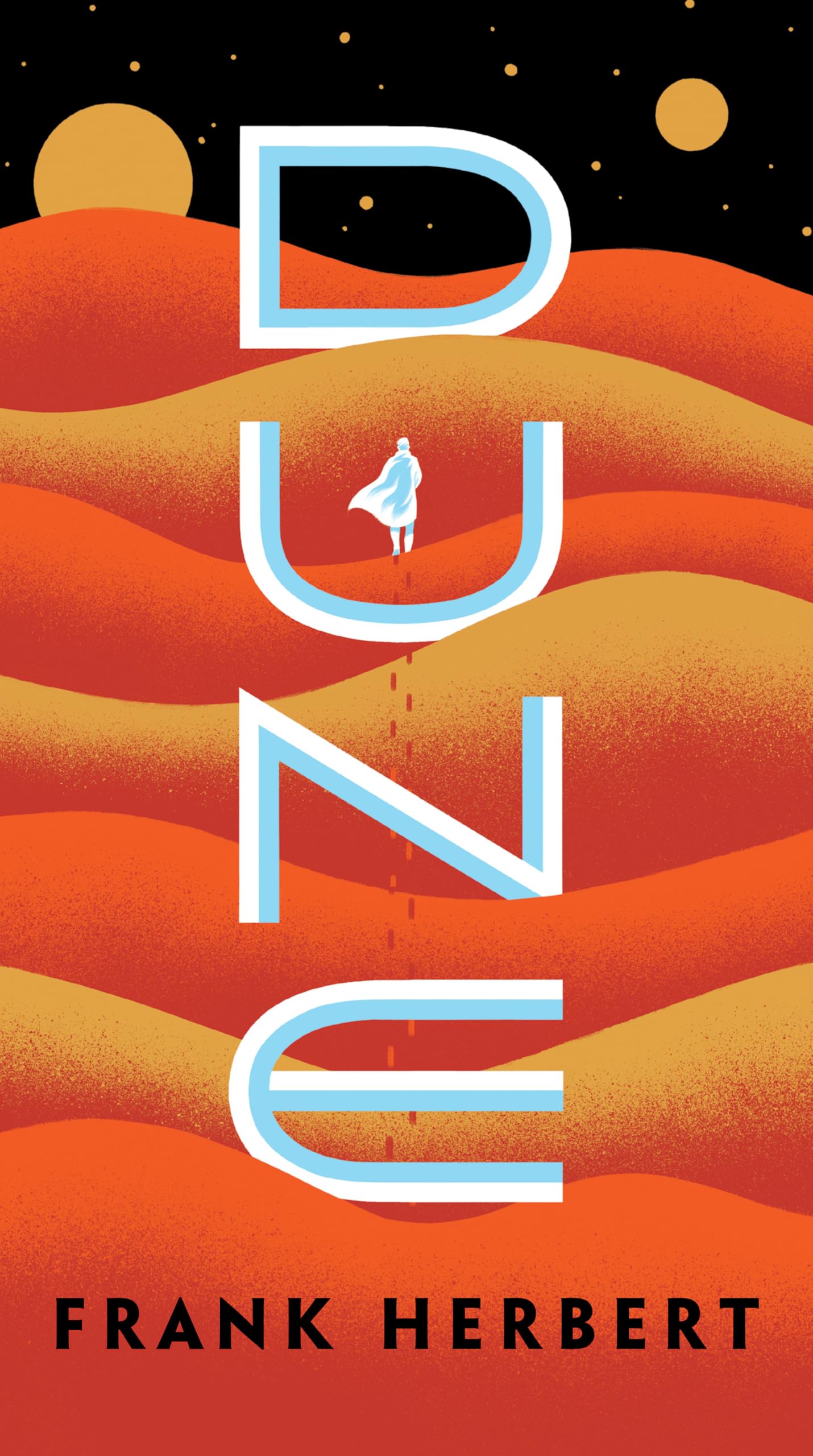}};
\node (e3b) at (5.9,7.5)
{\includegraphics[width=1.1cm,height=1.6cm]{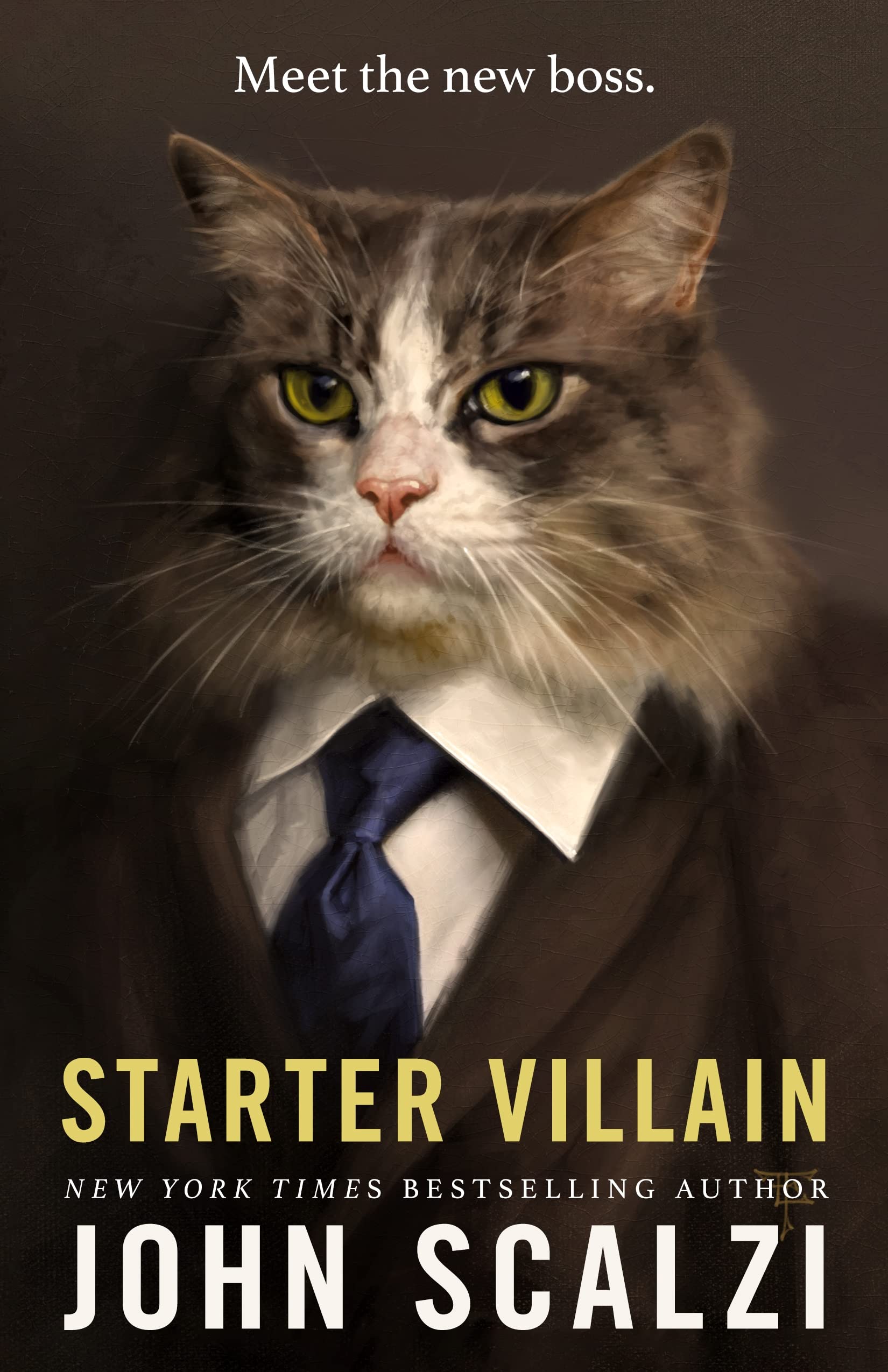}};

\node (exp4title) at (8.0,9.2) {\scriptsize Experiment 4:};
\node at (8.0,8.8) {\scriptsize literary fiction};

\node (e4a) at (8.0,7.8)
{\includegraphics[width=1.1cm,height=1.6cm]{figures/chemistry.jpg}};
\node (e4b) at (8.4,7.5)
{\includegraphics[width=1.1cm,height=1.6cm]{figures/hanya.jpg}};
\node (e4c) at (8.8,7.2)
{\includegraphics[width=1.1cm,height=1.6cm]{figures/infinite.jpg}};

\draw[->, thick] (e1b) --
    node[below] {}
    (step2f.north);

\draw[->, thick] (e2b) --
    node[below, yshift=0.35cm, xshift=0.2cm] {}
    (step2f.north);

\draw[->, thick] (e3b) --
    node[right, yshift=0.3cm, xshift=0cm] {\scriptsize}
    (step2f.north);

\draw[->, thick] (e4c) --
    node[below, yshift=0cm, xshift=0.15cm] {}
    (step2f.north);




\coordinate (customerNW) at (-2.5,16.0);
\coordinate (customerSE) at (9.5,10);

\coordinate (experimentNW) at (-3,10);
\coordinate (experimentSE) at (9.5,9.6);

\coordinate (firmNW) at (-3,9.6);
\coordinate (firmSE) at (9.5,4.9);

\node[fit=(customerNW)(customerSE), inner sep=0pt] (customerpanel) {};
\node[fit=(experimentNW)(experimentSE), inner sep=0pt] (experimentpanel) {};
\node[fit=(firmNW)(firmSE), inner sep=0pt] (firmpanel) {};

\draw[
    black
] (customerNW) rectangle (firmSE);

\draw[
    black
] (-2.5,11.25) -- (9.5,11.25);

\draw[
    black
] (-2.5,9.6) -- (9.5,9.6);





\end{tikzpicture}

\caption{Privacy budgeting in online experimentation operates at three connected levels: (1) the customer, (2) the experiment, and (3) the firm level.}
\label{fig:privacy_hierarchy}

\end{figure}

In this paper, we develop a privacy-budgeting framework that enables firms to quantify and manage the trade-off between the privacy risk of experimentation—at the customer, experiment, and firm levels—and firm performance (see Figure~\ref{fig:privacy_hierarchy}).
We make three contributions. First, we use differential privacy to bound the privacy risk of experimentation: the extent to which third-parties can update their belief about a consumer's segment membership after observing the experimental output \citep{Dwork_2014}. Building on this guarantee, we develop a privacy-budgeting framework for online experimentation. We quantify the privacy risk of experimentation by leveraging the link between differential privacy and multi-armed bandits (MABs) experimentation under the $\varepsilon$-greedy policy \citep{Sutton1998}. Under $\varepsilon$-greedy, the firm serves a random option with probability $\varepsilon$ and the best-performing option with probability $1-\varepsilon$ (see Figure \ref{fig:privacy_hierarchy}, Level 1). The degree of randomization is governed by the most granular privacy risk parameter, $\xi$, set under differential privacy. This parameter $\xi$ bounds how much observing the experimental output can update a tracker's posterior beliefs about a consumer's segment membership. Greater randomization in experimental outputs lowers privacy risk: as the exploration probability $\varepsilon$ increases, the privacy risk parameter $\xi$ decreases.

Second, we propose an experiment-level privacy risk \textit{budget} $\gamma$ that extends customer-level privacy risk across the full experiment (cf. \citealp{Christoph2026PrivacyBudget}; see Figure~\ref{fig:privacy_hierarchy}, Level~2). To allocate privacy budget across visitors, we propose two strategies: a constant strategy that allocates the same privacy risk to every customer and a dynamic strategy that varies privacy risk across customers to optimally balance learning and rewards. Privacy protection is costly; increasing privacy protection limits the firm's rewards from experimentation. We quantify the tradeoff between privacy protection and rewards using a regret analysis \citep[also see][]{Hsu_2014}. We derive a privacy elasticity of regret to measure the percent change in regret induced by a 1\% increase in the privacy budget \citep[cf.][]{NBERw30215}. The privacy elasticity helps managers choose privacy protection levels that minimize regret. We then extend $\gamma$ to a firm-wide privacy budget $\Gamma$ and use privacy elasticities to allocate privacy risk across experiments according to their highest marginal returns, analogous to allocating a marketing budget across campaigns \citep[e.g.,][]{Fischer_2011, Peers_2017, ZiaRao2019} (see Figure~\ref{fig:privacy_hierarchy}, Level~3).

Third, we use empirically-grounded simulations to evaluate how privacy budgets affect learning performance within and across experiments. Within experiments, we compare the constant and dynamic strategies in website-design and recommendation settings. Across experiments, we study how firms should allocate a fixed firm-wide privacy budget across multiple experiments to maximize rewards. We find that portfolio-level optimization using the regret bounds substantially improves rewards relative to benchmark allocations under the same firm-wide budget.

The rest of the paper proceeds as follows. Section~\ref{sec:related_work} reviews related work on privacy protection and experimentation via MABs. Sections~\ref{sec:privacy_risk}--\ref{sec:elasticity} introduce the strategies, evaluate their performance across applications, and derive the elasticities. Section~\ref{sec:multi_exp_portfolio} introduces the firm-wide privacy budget and applies the resulting portfolio model to a set of contemporaneous experiments. Section~\ref{sec:conclusion} concludes.

\section{Privacy protection and experimentation via MABs}\label{sec:related_work}
We embed a privacy protection mechanism, differential privacy, into online experimentation performed via MABs. We experiment using MABs because their link with differential privacy allows us to quantify the privacy risk of experimentation and set different privacy budgets across experiments.\footnote{Unlike MABs, A/B testing does not allow for variation in privacy protection. During the learning phase, visitors face no privacy risk because of the random assignment of different outputs. During the earning phase however, all visitors are assigned to the best-performing option. During this stage, privacy risk is unbounded. An experiment set up as an A/B test cannot deviate from this privacy protection schedule, and visitors have either a zero or an infinite privacy risk depending on whether they visit the website during the learning or the earning stage. Firms cannot set up privacy budgets and ensure different levels of privacy protection to their visitors.} 

We next review the literature related to both building blocks of our privacy budgeting framework and connect the bodies of work.

\subsection{Experimentation via MABs}
In MAB experimentation, the goal is to repeatedly choose between several outputs, labeled as ``arms,'' to maximize overall rewards over a predefined time horizon. To do so, one must balance exploring different outputs to learn their profitability with exploiting the option that maximizes total rewards. The marketing literature has incorporated MABs into display advertising \citep{Schwartz_2017}, website morphing \citep{Hauser_2009, hauser2014website, Liberali_2022}, pricing \citep{Misra_2019}, house-ad recommendations \citep{Aramayo_2023}, and ranked search results \citep{delossantos_2017}. The primary focus of these studies is to empirically validate experimental methods and algorithms. Our focus is instead on embedding privacy protection into experimentation via MABs. We focus on the $\varepsilon$-greedy policy because the inherent stochasticity of this heuristic allows us to link it with differential privacy. In marketing, \cite{Wang_2025} applied MABs and specifically the $epsilon$-greedy policy to optimize recommender systems.

Given our focus on segment specific experimentation, our bandit method is also similar to contextual bandits, which improves assignment by taking into account both user and context features to optimize learning. Our method differs from contextual bandits in two important ways: first, we focus on a limited number of segments, unlike a contextual bandit that typically takes as input a large set of user and context variables \citep{li2010contextual}. Second, and linked to the first, because of the large number of features, the privacy threat is hard to define and contain. We propose a clear measure of privacy risk, directly linked to what third-party can infer about consumers.  

\subsection{The privacy protection literature}
Our work focuses on protecting the privacy of consumers when exposed to experimental output (see Figure \ref{fig:privacy_risk_channels}). We position our contribution relative to privacy-related work focused on the privacy risk channels shown in Figure \ref{fig:privacy_risk_channels}: experimental output exposure, and consumer's interaction with the experimental output. The availability of these channels to a tracker depends on consumers accepting tracking technologies and ad blocking decisions. We therefore first review research on the factors that determine tracking access and then discuss the literature related to each privacy risk channel.

\subsubsection{Tracking access as a precondition for privacy risk.}
A stream of work studies the economic costs of privacy choices and regulations that restrict firms' ability to observe, track, or link consumers across contexts. \citet{Johnson_2020} document that consumers who opt out of behavioral advertising generate 52\% less revenue than impressions from consumers who allow behavioral targeting. \citet{Aridor_Che_Hollenbeck_Kaiser_McCarthy_2024} show that Apple's App Tracking Transparency (ATT) policy led to revenue declines of 8\% to 40\% for affected firms. \cite{Kraft2023Economic} show a decrease of 20\% in advertising revenue when users are allowed to gradually control their privacy settings via the Apple's ATT. \citet{MILLER2024241} show that limiting cookie-based tracking could put 904 million euros in annual revenue at risk. Similarly, \citet{Korganbekova_Zuber_2024} find that Safari’s seven-day cookie policy destroys 13.1\% of the personalization gain. Regarding ad blocking, \citet{Todri_2022} finds that ad blockers reduce online consumer spending by lowering consumers' search activity and shifting purchases away from new brands.

\citet{Korganbekova_Zuber_2024} argue that limiting tracking may not fully protect privacy because firms can still learn about consumers from consumers with related characteristics (also see \cite{Wernerfelt_2025}; \cite{Aguiar_Peukert_Schafer_Ullrich_2026}). This shows that in addition to consent, the experimental output may consequently require protection even when consumers did not consent to its use for inference—or even to being shown the output in the first place. Consumers who are shown the same output may be placed in the same inferred segment, allowing a tracker to conclude that they share similar preferences. 

\subsubsection{Privacy risk channel 1: Exposure to experimental output.}
\citet{Summers_2016} find that consumers react negatively when targeted messages use sensitive or unexpected personal information. \citet{Kim_2018} demonstrate that personalization can increase effectiveness while also raising privacy concerns when consumers infer that firms possess or use personal information. \citet{Lin_2022} documents that consumers value the economic benefits of targeted advertising, while also exhibiting a baseline resistance to being targeted that reflects privacy concerns. \citet{Jerath_Miller_2024} argue that privacy concerns may arise from the perceived inferences embedded in targeted marketing actions. 

In response to these privacy concerns, \citet{Ponte2025ActOfTargeting} use differential privacy to quantify and control privacy risk from targeting consumers with a one-time discount coupon. In contrast, we incorporate differential privacy directly into the experimental design rather than applying privacy protection only after data have been collected. As customers enter the experiment, we optimally balance exploration and exploitation subject to a specified privacy budget. We also expand this privacy-minded way of experimentation to optimize privacy protection to the firm-level portfolio of experiments.

\subsubsection{Privacy risk channel 2: Post-exposure behavior.}
A stream of work studies how firms can share data containing customer behavior while limiting privacy risk. \citet{anand_2023} propose sharing a generative model instead of customer data sets, allowing external parties to learn useful patterns without directly accessing sensitive data. \citet{PONTE2024} propose a differentially private data-sharing framework to quantify and control privacy risk in marketing applications. \citet{Tian2024} show that fusing anonymous customer survey data with CRM records can create reidentification risk; they propose a differentially private data-fusion framework to reduce this risk. We differ from this work as we intervene during the experimentation process and integrate privacy protection in the experimental design.

Prior computer science research uses differential privacy to protect the information that MABs obtain from consumer responses rather than exposure to experimental outputs. Typically, these methods add noise to measures such as total clicks before using them to choose which option to display. For example, \citet{Mishra_2015} develop a differentially private MAB algorithm in which the reward statistics are privatized. \citet{tossou_2015} propose differentially private UCB-style algorithms that compute private empirical mean rewards using noisy cumulative reward sums. Our paper is closest to \citet{ren2020}, who use a randomized response mechanism to protect rewards. We instead apply a randomized response mechanism to the experimental output, limiting what experimental outputs reveal about the consumer’s latent segment. 

\section{Quantifying the privacy risk of experimental output}\label{sec:privacy_risk}
In this section, we develop a method that quantifies and integrates the privacy risk visitors face during online experimentation. For example, a firm, such as Barnes \& Noble, conducts experiments on its website to identify the experimental output that is most effective for each consumer segment, where segments reflect visitors’ interests and preferences. Our goal is to bound what a third-party tracker could infer about visitors' segment membership from their exposure to an experimental output. 



Our method blends the $\varepsilon$-greedy policy used in MAB experimentation with differential privacy to protect experimental output. We rely on these techniques because we can link the exploration probability used in the $\varepsilon$-greedy policy to satisfy $\xi$-differential privacy \citep{Dwork_2014}. This yields a direct mapping between the privacy risk parameter $\xi$ and the exploration parameter $\varepsilon$. Building on this mapping, we introduce an experiment-level privacy budget $\gamma$ that a firm can set ex ante and study how different strategies for spending this budget affect the performance of experimentation.

\subsection{Experimentation via MABs}

Let \(S_t\in\mathcal S\) denote the segment of visitor \(t\), where \(\mathcal S\) is a finite set of consumer segments. We assume that the firm observes \(S_t\) before assigning an experimental output, for example through its first-party customer information, whereas the third-party tracker whose inference we seek to limit does not directly observe \(S_t\).

The experiment contains a common set of \(K\) arms, \(\mathcal A=\{a_1,\ldots,a_K\}\), where an arm is an experimental output that can be displayed during experimentation, such as a book recommendation or an ad. We assume that every arm in \(\mathcal A\) can, in principle, be displayed to every segment in \(\mathcal S\). Thus, the firm does not determine ex ante which arm is optimal for a given segment. Instead, experimentation is used to learn which output performs best for each segment.

Let \(A_t \in \mathcal A\) denote the arm displayed to visitor \(t\). A success is defined as a click on the displayed arm. We allow the expected reward of an arm to vary across consumer segments. Specifically, conditional on displaying arm \(a_k\) to a visitor in segment \(s\), the reward satisfies
\begin{equation}
\label{eq:segment-reward}
R_t \mid \{A_t=a_k, S_t=s\}
\sim \operatorname{Bernoulli}\!\left(\mu_k(s)\right),
\end{equation}
where
\[
\mu_k(s)
=
\mathbb{P}\!\left(
R_t = 1
\mid
A_t=a_k, S_t=s
\right).
\]

The parameters \(\mu_k(s)\) are unknown to the firm and are learned during experimentation. Importantly, we do not assume ex ante that different segments necessarily respond differently to the same arm. The model merely allows \(\mu_k(s)\) to differ across segments, and experimentation reveals whether and how such heterogeneity arises.

For each arm--segment pair \((a_k,s)\), let
\[
N_t(a_k,s)
=
\sum_{u=1}^{t-1}
\mathbf{1}\{S_u=s,\;A_u=a_k\}
\]
denote the number of previous visitors in segment \(s\) who were displayed arm \(a_k\). The firm's estimated reward for that arm--segment pair is
\begin{equation}
\label{eq:segment-Q}
Q_t(a_k,s)
=
\frac{
\sum_{u=1}^{t-1}
\mathbf{1}\{S_u=s,\;A_u=a_k\}R_u
}{
N_t(a_k,s)
},
\end{equation}
whenever \(N_t(a_k,s)>0\).

Accordingly, \(Q_t(a_k,s)\) is the empirical click-through rate of arm \(a_k\) among previously observed visitors in segment \(s\). Because the firm maintains a separate estimate for each arm--segment pair, the same arm may acquire different estimated rewards across segments as experimentation progresses.

We maintain stationarity within each arm--segment pair: for every \(a_k\in\mathcal A\) and \(s\in\mathcal S\), the mean reward \(\mu_k(s)\) remains fixed over the experimental horizon. Conditional on the displayed arm and segment, rewards are independent draws from the corresponding Bernoulli distribution.

The time horizon \(T\) denotes the total number of visitors allocated to the experiment. The objective is to learn which arm maximizes expected reward for each segment while maximizing cumulative rewards over the experimental horizon.
\subsubsection{The $\varepsilon$-greedy policy.}
The $\varepsilon$-greedy policy is a stochastic, near-optimal heuristic for solving the MAB problem and is widely used in both industry and academia \citep[e.g.,][]{Wang_2025, OptimizelyMultiArmedBanditGlossary}.\footnote{Gittins and Jones (1979) proposed an index policy to solve this $K$-dimensional dynamic program that finds the optimal path to maximize expected rewards. The optimal path is deterministic; at every visitor \textit{t}, an arm-specific index $GI_{tk}$ is computed, and the arm with the highest index is chosen. The policy is optimal for infinite horizon problems, and it was shown to be near optimal for finite-horizon problems. Because Gittins index is deterministic, it is not privacy preserving. When in exploitation, visitors belonging to a segment will have the same arm assignment revealing their preferences and interests.}
A myopic, greedy policy selects
\begin{equation}
\label{eq:greedya_t}
a_t
=
\arg\max_{a_k\in\mathcal A}
Q_t(a_k,S_t),
\end{equation}

Thus, the greedy arm depends on the segment of the current visitor through the
segment-specific reward estimates \(Q_t(a_k,S_t)\). Two visitors belonging to
different segments may therefore have different greedy arms even when they
arrive at the same stage of the experiment.

To encourage exploration of alternative arms, a $\varepsilon$-greedy policy explores with probability $\varepsilon$ and exploits with probability $1 - \varepsilon$. The arm served at customer \textit{t}, $a^{*}_t$, is
\begin{equation}
\label{eq:epsilon-greedy}
a_t^*
=
\begin{cases}
a_t
=
\displaystyle
\arg\max_{a_k\in\mathcal A}
Q_t(a_k,S_t),
&
\text{with probability }1-\varepsilon_t,
\\[8pt]
\operatorname{Uniform}(\mathcal A),
&
\text{with probability }\varepsilon_t.
\end{cases}
\end{equation}

Importantly, exploitation is segment-sensitive because it uses \(Q_t(a_k,S_t)\), whereas exploration draws from the same experiment-wide arm space \(\mathcal A\) for every segment. Note that the firm does not specify which output belongs to a particular segment ex ante. Rather, differences in assignment across segments arise endogenously as the experiment learns segment-specific arm effectiveness.

To evaluate how well the $\varepsilon$-greedy policy solves the MAB problem, we next analyze the regret generated by its exploration and exploitation decisions.

\subsubsection{The expected regret of the $\varepsilon$-greedy policy.} \label{sec:epsilon_greedy_regret}
Intuitively, solving the MAB problem with the $\varepsilon$-greedy policy implies minimizing the regret of making mistakes. To minimize regret, \cite{Panageas_Ghosal_Lin_2020} show that we can set the per-round exploration probability at: 
\begin{equation} \label{eq:opt_epsilon}
\begin{aligned}
\varepsilon_t  &= \left(\frac{K \log t}{t}\right)^{1/3},\\
\end{aligned}
\end{equation}

which yields the following bound on the expected cumulative regret
\begin{equation}
\label{eq:cumulative_regret}
\mathbb{E}\!\left[R(T)\right]
=
\mathcal{O}\!\left(
T^{2/3}K^{1/3}(\log T)^{1/3}
\right).
\end{equation}

Equation \eqref{eq:opt_epsilon} implies a higher exploration probability early on, to learn about the effectiveness of the arms, and higher exploitation later on, to maximize rewards by sampling the best-performing arms. Deviating from the above level of exploration to ensure a certain level of privacy increases regret, and decreases the performance of the experiment. 
We next link the $\varepsilon$-greedy policy to differential privacy.

\subsection{Linking the $\varepsilon$-greedy policy to differential privacy}
First, we define the privacy risk under differential privacy. We then connect it to experimentation using the $\varepsilon$-greedy policy. Last, we define the visitor-level privacy risk $\xi$.

\subsubsection{The privacy mechanism: differential privacy.}
As defined in Equation \eqref{eq:greedya_t}, $a_t$ is the greedy arm, the arm that myopically maximizes rewards up to visitor $t$, and $a_t^*$ denotes the arm displayed after randomization through the $\varepsilon$-greedy policy. Following Equation~\eqref{eq:epsilon-greedy}, we define $\mathbb{P}\!\left(a_t^*=k \mid a_t=j\right)=p_{kj}, \text{ where } k,j\in\{1,\dots,K\}$, as the probability that the displayed arm is \(k\) when the greedy arm is \(j\).

A privacy mechanism introduces randomization in arm assignment and masks optimal assignment maximizing rewards, i.e., the greedy arm. In our setup, the privacy mechanism is represented by a \(K\times K\) stochastic matrix
\begin{equation}
\label{eq:matrix1}
\mathbf{P}
=
\begin{pmatrix}
p_{11} & p_{12} & \cdots & p_{1K}\\
p_{21} & p_{22} & \cdots & p_{2K}\\
\vdots & \vdots & \ddots & \vdots\\
p_{K1} & p_{K2} & \cdots & p_{KK}
\end{pmatrix},
\end{equation}
where each column corresponds to a greedy arm \(j\), each row corresponds to a displayed arm \(k\), and
$p_{kj}\ge 0 \text{ and }
\sum_{k=1}^K p_{kj}=1
\quad\text{for every } j\in\{1,\dots,K\}$.
Thus, column \(j\) gives the full distribution of displayed arms conditional on the greedy arm being \(j\). In particular, the diagonal entry \(p_{jj}\) is the probability that arm \(j\) is displayed when arm \(j\) is also the greedy arm, while an off-diagonal entry \(p_{kj}\) with \(k\neq j\) is the probability that the mechanism displays arm \(k\) instead of the greedy arm \(j\).

To guarantee a privacy level \(\xi\) under differential privacy, \cite{wang2016using} define the \(K\)-ary randomized-response mechanism in Equation \eqref{eq:matrix1} and set each diagonal entry of the stochastic matrix to\begin{equation}\label{eq:offdiag}
p_{jj}=\frac{e^{\xi}}{K-1+e^{\xi}},
\qquad j\in\{1,\dots,K\},
\end{equation}

\noindent and each off-diagonal entry is set to
\begin{equation}
p_{kj}=\frac{1}{K-1+e^{\xi}},
\qquad k\neq j.
\end{equation}

This choice of diagonal and off-diagonal entries ensures that, for any displayed arm \(k\) and any two underlying arms \(a\) and \(a'\), the ratio of output probabilities is bounded by \(e^\xi\). Hence, \(\xi\) directly determines how much observing the displayed arm can reveal about the underlying greedy arm (see Web Appendix \ref{sec:dp-bandits}). The privacy mechanism under differential privacy is therefore represented by
\begin{equation}\label{eq:matrix}
\mathbf P
=
\begin{pmatrix}
\frac{e^{\xi}}{K-1+e^{\xi}} & \frac{1}{K-1+e^{\xi}} & \cdots & \frac{1}{K-1+e^{\xi}}\\
\frac{1}{K-1+e^{\xi}} & \frac{e^{\xi}}{K-1+e^{\xi}} & \cdots & \frac{1}{K-1+e^{\xi}}\\
\vdots & \vdots & \ddots & \vdots\\
\frac{1}{K-1+e^{\xi}} & \frac{1}{K-1+e^{\xi}} & \cdots & \frac{e^{\xi}}{K-1+e^{\xi}}
\end{pmatrix}.
\end{equation}
\subsubsection{A visitor-level privacy risk \(\xi\).}
We now connect the exploration parameter \(\varepsilon\) in the \(\varepsilon\)-greedy policy to the differential-privacy parameter \(\xi\) by matching the induced stochastic matrix to the \(K\)-ary randomized-response mechanism. Under Equation~\eqref{eq:epsilon-greedy}, if the greedy arm at round \(t\) is \(a_t=j\), then the probability that the displayed arm equals the greedy arm is
\begin{equation}\label{eq:probabilittttt}
p_{jj}
=
\mathbb{P}(a_t^{*}=k \mid a_t=k)
=
(1-\varepsilon)+\frac{\varepsilon}{K},
\end{equation}
where with probability \(1-\varepsilon\) the mechanism exploits and displays the greedy arm, while with probability \(\varepsilon\) it explores uniformly over all \(K\) arms. 

Under uniform exploration, each arm is selected with probability \(1/K\), so the greedy arm is still displayed with probability \(\varepsilon/K\). This uniform randomization implies an equal exploration probability across arms.\footnote{In Web Appendix~\ref{app:heterogeneity}, we relax this restriction by allowing the exploration probabilities to differ across arms, which induces arm-specific privacy risk parameters. This extension allows firms to assign stronger privacy protection to more sensitive arms. We show that such heterogeneity increases the worst-case regret bound relative to an equal privacy guarantee across arms and quantify the magnitude of this increase.} Using Equation \eqref{eq:offdiag}, we solve Equation~\eqref{eq:probabilittttt} for $\varepsilon$ to yield the exploration probability
\begin{equation}\label{eq:exploration}
\begin{aligned}
\varepsilon
&= \frac{K(1-p_{jj})}{K-1} \\[4pt]
&= \frac{K\left(1-\frac{e^\xi}{K-1+e^\xi}\right)}{K-1} \\[4pt]
&= \frac{K}{K-1+e^\xi}.
\end{aligned}
\end{equation}

Equation~\eqref{eq:exploration} shows that stronger privacy guarantees (smaller \(\xi\)) require more exploration (larger \(\varepsilon\)), whereas weaker privacy guarantees (larger \(\xi\)) permit more exploitation (smaller \(\varepsilon\)). In particular, when \(\xi=0\), then \(\varepsilon=1\) and the displayed arm is fully randomized; as \(\xi\to\infty\), then \(\varepsilon\to 0\) and the policy approaches pure exploitation.

Using the connection between the privacy risk defined under differential privacy and the exploration probability of the $\varepsilon$-greedy policy, we combine Equation \eqref{eq:exploration} and Equation \eqref{eq:epsilon-greedy} to define a $\xi$-differentially private $\varepsilon$-greedy policy as:
\begin{equation}\label{eq:differential-greedy}
    a^{*}_t =
\begin{cases} 
a_{t}  = \arg\max_a Q_t(a) & \text{with probability } 1-\varepsilon  = \frac{e^\xi - 1}{K + e^\xi - 1},\\
\text{Uniform}(\{1,\dots,K\}) & \text{with probability } \varepsilon = \frac{K}{K + e^\xi - 1}.
\end{cases}
\end{equation}

This implies that the bandit explores with probability $\varepsilon = \frac{K}{K + e^\xi - 1}$ and exploits with probability $1-\varepsilon=\frac{e^\xi - 1}{K + e^\xi - 1}$. For example, if we set $\xi = 0.05$ and consider two arms, then the exploration probability equals .975 and the exploitation probability equals $1-.975 = .025$. Intuitively, less privacy risk implies more randomness and more randomness implies more exploration.

\begin{figure}[h]
    \centering
    \includegraphics[width=.7\linewidth]{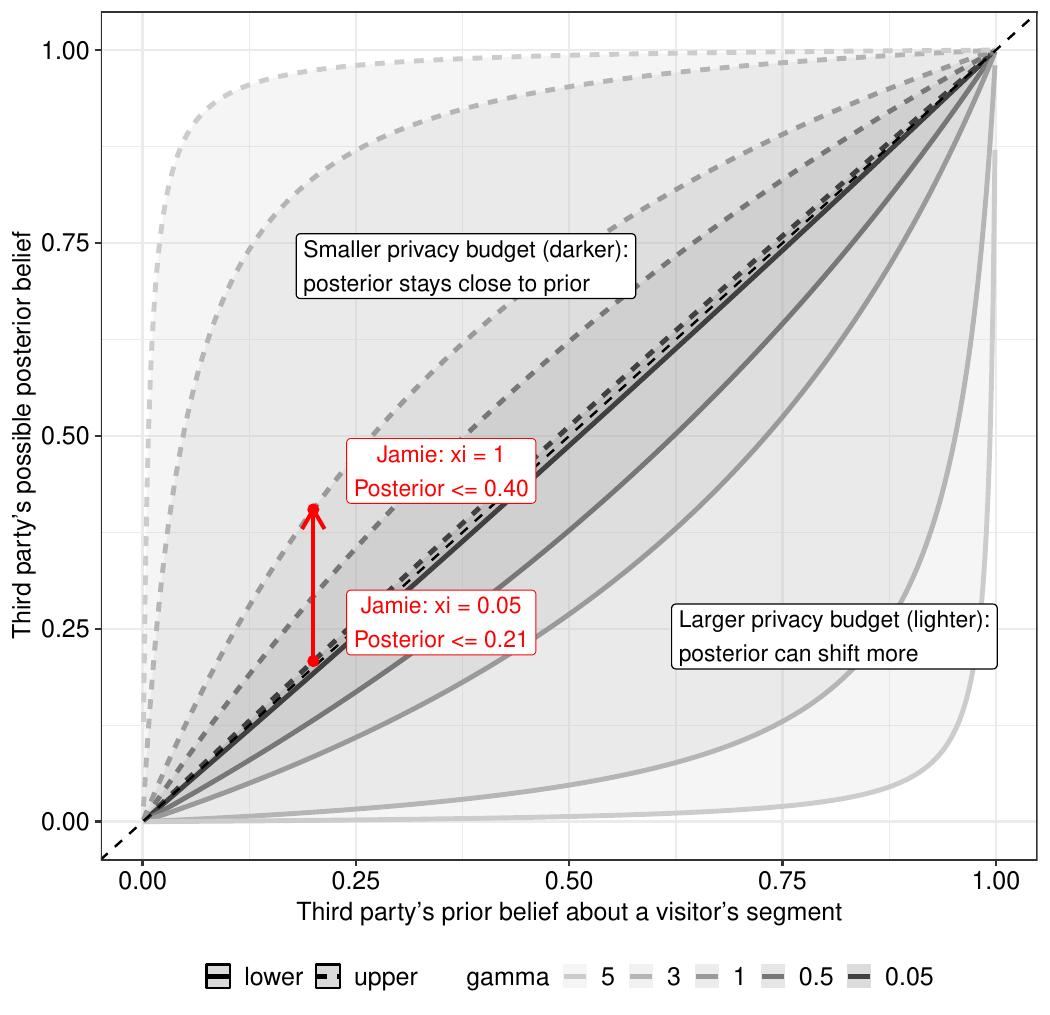}
    \caption{The privacy risk parameter $\xi$ bounds how much a tracker’s belief about a visitor's segment can shift after observing the experimental output. Note: Darker curves (small $\xi$) show less updating, and lighter curves (larger $\xi$) show more updating away from the prior.}
    \label{fig:bound}
\end{figure}

In Figure~\ref{fig:bound}, we provide an interpretation of the privacy risk parameter $\xi$. Under differential privacy, the parameter $\xi$ bounds the extent to which a tracker can update their belief about a visitor' segment membership from exposure to experimental output, without accounting for their interaction with the experimental output. The horizontal axis in Figure \ref{fig:bound} represents a tracker’s prior belief about a visitor's segment, while the vertical axis shows the range of posterior beliefs after observing the output. The diagonal line corresponds to perfect privacy, where the posterior equals the prior—no inference about segment membership is possible.

For small values of $\xi$ (darker curves), the posterior remains close to the prior: returning to our Barnes \& Noble example from the introduction, even after the tracker observes the recommendation shown to Jamie, it cannot tell whether a fantasy-themed recommendation was displayed because Jamie is truly a fantasy reader or simply because of randomization. This is illustrated by the lower red point in Figure~\ref{fig:bound}: starting from a prior belief of $20\%$, a privacy level of $\xi=.05$ bounds the posterior after one observed display at about $21\%$. For larger $\xi$ (lighter curves), the posterior can move further away from the prior, meaning that the observed output is more informative about Jamie’s underlying type; the upper red point shows that with $\xi=1$ the posterior can increase to about $40\%$ after observing the output.

\subsubsection{An experiment-level privacy budget $\gamma$.}\label{privacyriskexperimental}
We have now defined a per-visitor privacy risk level \(\xi\) incurred during online experimentation from experimental output. However, it remains unclear how a firm should allocate privacy risk across visitors. To address this, we introduce an experimental-level privacy budget \(\gamma\) that can be set before experimentation. We assume that each round corresponds to a unique visitor, and the privacy mechanism in round \(t\) is independently applied only to that visitor's assignment (see Web Appendix \ref{app:parallel} for technical details). This implies that the privacy budget is governed by the largest per-visitor privacy risk:
\begin{equation}\label{eq:parallel-composition}
\gamma = \max_t \xi_t.
\end{equation}
Thus, the experiment satisfies the budget $\gamma$ as long as $\xi_t\leq\gamma$ for every visitor $t$. This definition places an upper bound on visitor-level privacy risk but leaves open how privacy risk should be allocated across visitors. We address this question by introducing two privacy budget spending strategies.

\subsection{Privacy budget spending strategies and the trade-off between privacy and performance}
Having established an experimental-level privacy budget, it remains unclear how this budget should be spent across visitors. We develop two strategies that define how to spend the experiment-level privacy budget across visitors: a constant privacy risk strategy, keeping the risk constant across visitors, and a dynamic privacy risk strategy, which allows the risk to change across visitors up to a predefined level.

\subsubsection{Constant privacy risk strategy.}
\begin{figure}[h]
    \centering
    \includegraphics[width=\linewidth]{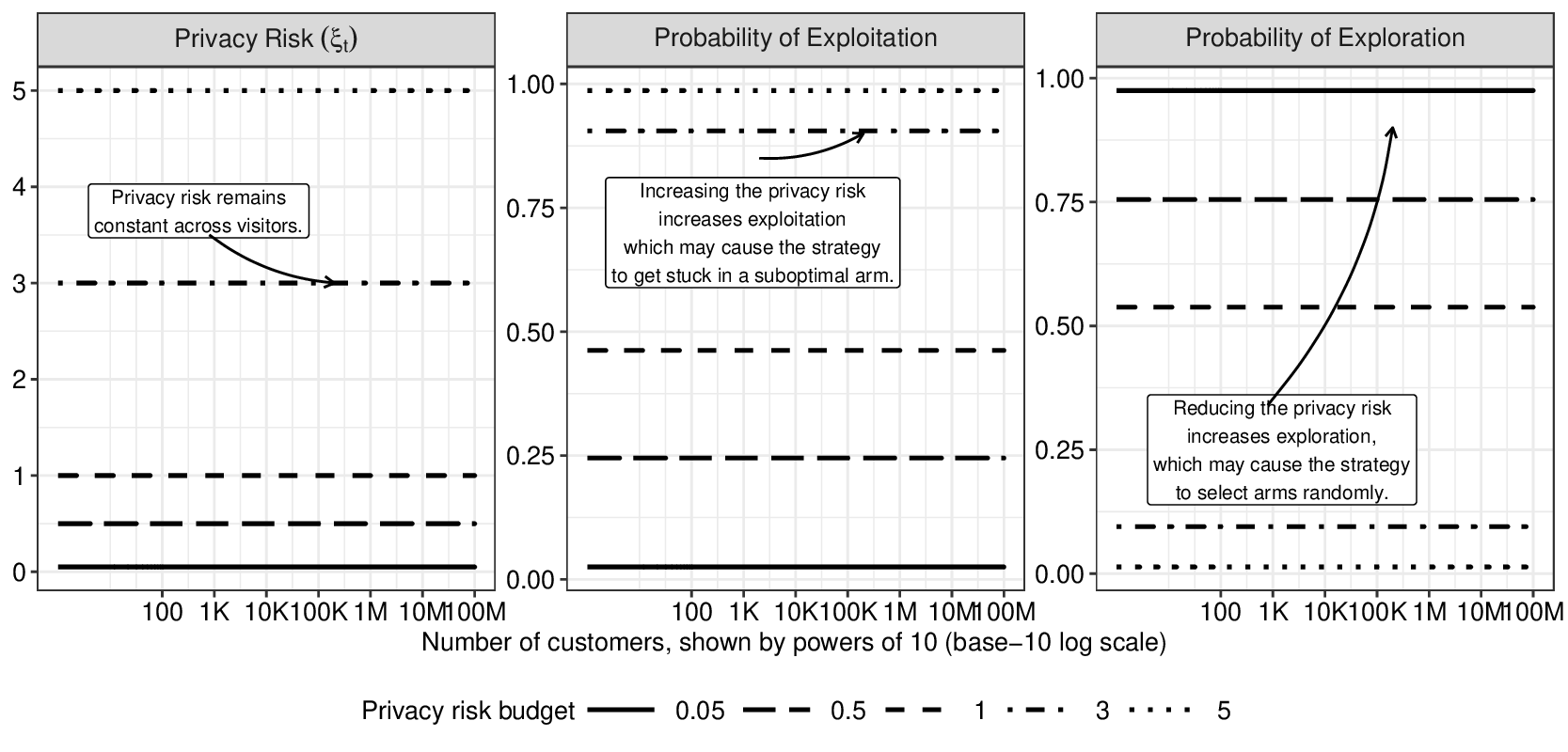}
    \caption{The privacy risk under the constant strategy $\xi_t$ (left panel) and the exploration/exploitation probabilities $\varepsilon_t$ and $1-\varepsilon_t$ (middle/right panel) for different privacy budgets $\gamma \in \{0.05,0.5,1,3,5\}$.}
    \label{fig:2}
\end{figure}

The constant privacy risk strategy assigns the same risk to each visitor, such that $\xi_t=\gamma$ for all $t$. We visualize the strategy in Figure~\ref{fig:2}. The constant strategy has the benefit of simplicity and fairness: all visitors receive the same level of privacy risk (see Figure \ref{fig:2}, left panel). The main drawback is that the strategy limits the potential rewards from experimentation because exploration is especially valuable early in the experiment, when additional information can substantially improve future decisions, whereas exploitation is more valuable later, once the firm has learned which arms perform best. Our regret analysis in \S\ref{sec:epsilon_greedy_regret} and Web Appendix \ref{web:regretanalysis} highlight this adaptive feature of the $\varepsilon$-greedy policy. 

The constant strategy has same privacy risk, therefore the same exploration probability throughout the experiment (middle and right panels), and it cannot adjust to this shift: it may explore too little early on, when learning is most valuable, and too much later on, when the best-performing arms have already become clearer. The expected regret reflects the inflexibility of the constant strategy. By replacing the exploration probability from Equation \eqref{eq:exploration} into Equation \eqref{eq:regret_constant}, the regret bound of the constant strategy becomes (see Web Appendix \ref{app:constant})
\begin{equation}
\begin{aligned}\label{eq:regret_constant}
\mathbb{E}[R_{\text{constant}}(T)]
\;&\le\;
\underbrace{T\,\frac{K}{K + e^{\gamma} - 1}}_{\text{cost of exploration}}
\;+\;
\underbrace{\sqrt{K + e^{\gamma} - 1}\,
\sqrt{T\log T}}_{\text{cost of exploitation}}.
\end{aligned}
\end{equation}

Intuitively, the exploration term $T\,\frac{k}{k+e^{\gamma}-1}$ grows linearly with the number of visitors because, in every round, the bandit explores with constant probability $\frac{k}{k+e^{\gamma}-1}$ and a random pull can incur up to one unit of regret. The second term, $\sqrt{k+e^{\gamma}-1}\,\sqrt{T\log T}$, captures exploitation mistakes due to noisy value estimates. A smaller privacy budget $\gamma$ decreases the exploitation probability, which decreases the cost from uncertainty during learning, but also increases the exploration probability, which raises the exploration cost. This is the core exploration--exploitation trade-off in online experimentation under privacy protection. To incorporate this idea, we next vary privacy risk across visitors so that exploration is concentrated earlier in the experiment and exploitation later.

\subsubsection{Dynamic privacy risk strategy.}
We next derive a dynamic privacy risk schedule that varies risk across visitors to better balance exploration and exploitation than the constant strategy and gets closer to the expected regret bound of the \(\varepsilon\)-greedy policy without privacy protection (see Equation \ref{eq:cumulative_regret}). Under this dynamic strategy, the privacy risk assigned to visitor \(t\) is given by (see Web Appendix \ref{web:regret_difprivacy}):

\begin{equation}\label{privacystrategy}
\xi_t = \log\left(1 - K + \frac{K}{\left(\frac{K \log(t)}{t}\right)^{1/3}}\right).
\end{equation}

\noindent Under this schedule, the firm can protect visitors' privacy without degrading learning performance relative to an \(\varepsilon\)-greedy policy without privacy protection. However, to ensure that privacy risk never exceeds the predetermined budget \(\gamma\) in Equation \eqref{eq:parallel-composition}, we cap the per-visitor privacy risk as follows:
\begin{equation}\label{eq:budgetgamma}
\xi_t \;=\; 
\min\!\left\{
\gamma,\;
\log\left(1 - K + \left(\frac{K}{\left(\frac{K \log(t)}{t}\right)^{1/3}}\right)\right)
\right\}.
\end{equation}

This cap implies that once the budget binds, the exploration and exploitation probabilities are fixed, and the privacy risk \(\xi_t\) remains set at $\gamma$, ensuring that the experiment-wide privacy guarantee is preserved. The cap effectively binds the extent of exploitation the bandit can engage in, and ensures that even after many visitors the bandit still explores and randomizes assignment, thus allowing plausible deniability. 

\begin{figure}[h]
    \centering
    \includegraphics[width=\linewidth]{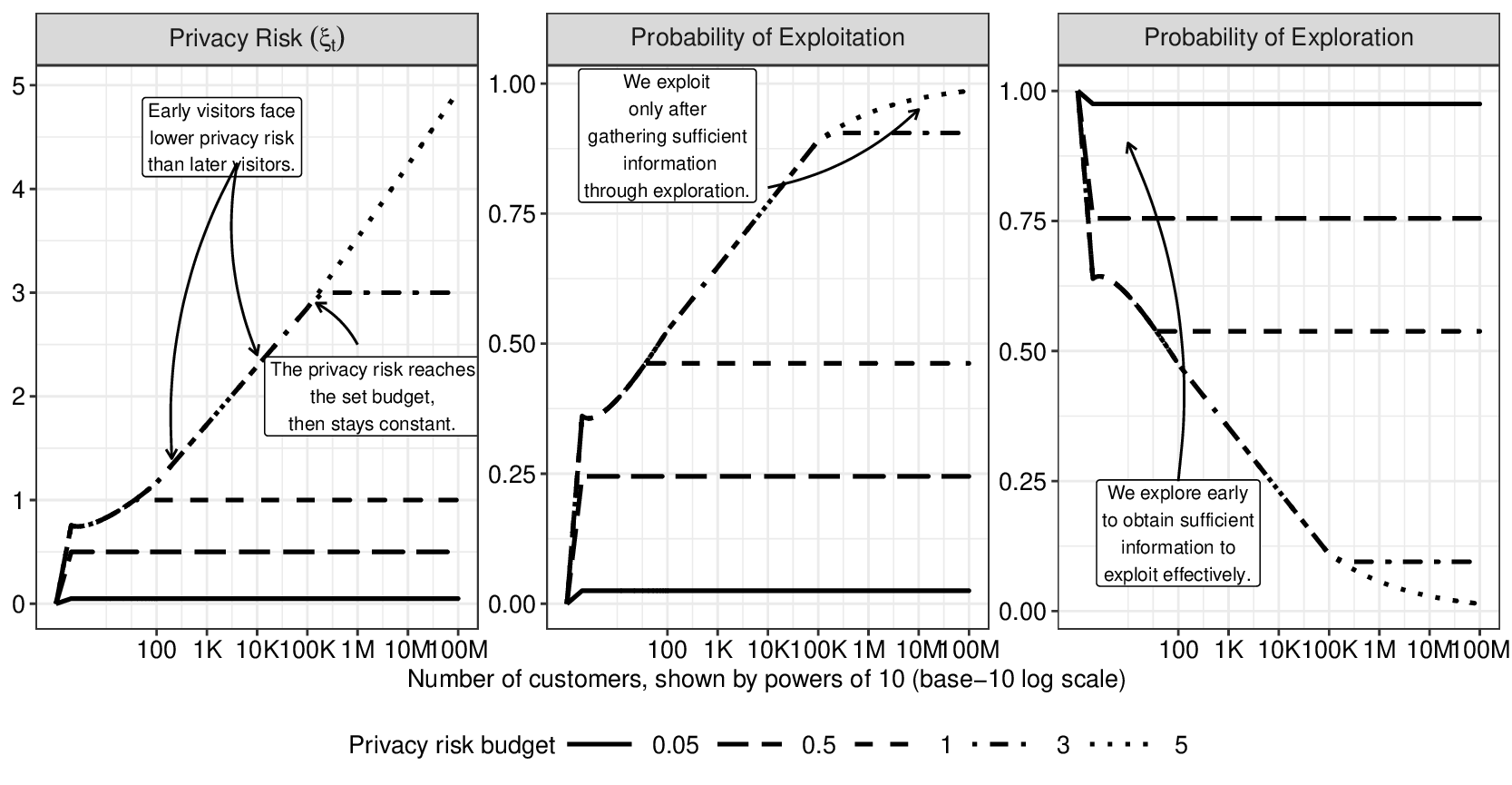}
    \caption{The privacy risk under the dynamic strategy $\xi_t$ (left panel) and the exploitation/exploration probabilities $1-\varepsilon_t$ and $\varepsilon_t$ (middle/right panel). We use different line types to show the varying the privacy budget levels $\gamma \in \{0.05,0.5,1,3,5\}$.}
    \label{fig:3}
\end{figure}

In Figure \ref{fig:3}, we visualize the privacy risk (left panel), the probability of exploitation (middle panel), and the probability of exploration (right panel) under the dynamic privacy risk strategy. Early visitors face lower privacy risk as the bandit explore more, while later visitors face higher privacy risk, as the bandit exploits more frequently. When $\xi_t>\gamma$, the privacy risk and induced exploration and exploitation probabilities are constant.

The introduction of the budget cap in Equation \eqref{eq:budgetgamma} creates an additional regret component relative to the standard $\varepsilon$-greedy bound: once the cap takes effect, the strategy is forced to randomize more than the uncapped schedule would prescribe late in the experiment, which increases regret. This strategy yields the following expected regret bound (see Web Appendix~\ref{web:regret_dynamic} for a derivation):
\begin{equation}\label{eq:regret_dynamic}
\begin{aligned}
\mathbb{E}[R_{\text{budget}}(T)]
\;\lesssim\;&
\underbrace{t_\gamma^{2/3} K^{1/3} (\log t_\gamma)^{1/3}}_{\text{standard $\varepsilon$-greedy regret (see Equation \ref{eq:cumulative_regret})}}
\\
&+
\underbrace{(T - t_\gamma)\frac{K}{K + e^{\gamma} - 1}}_{\text{cost of exploration after budget runs out}}
+
\underbrace{\sqrt{K + e^{\gamma} - 1}\,(\sqrt{T} - \sqrt{t_\gamma})\sqrt{\log T}}_{\text{cost of noisy exploitation after budget runs out}} \, .
\end{aligned}
\end{equation}

where $t_\gamma$ is the visitor where the privacy risk $\xi_t = \gamma$, defined as the first visitor at which the dynamic exploration rate (see Equation \ref{privacystrategy}) falls to the level implied by the privacy budget $\gamma$ (see the left panel in Figure \ref{fig:3}).\footnote{Determining $t_\gamma$ has no closed-form solution but it is straightforward to compute numerically (see Web Appendix~\ref{web:regret_dynamic}). Formally, $t_\gamma := \min\Bigl\{t\in\{1,\dots,T\}:\ \bigl(\tfrac{k\log t}{t}\bigr)^{1/3}\le \tfrac{k}{k+e^\gamma-1}\Bigr\}$.} 

If \(t_\gamma \geq T\), the privacy budget never binds over the experimental horizon, and the regret therefore reduces to the standard \(\varepsilon\)-greedy regret without privacy protection. By contrast, if \(t_\gamma < T\), the budget binds before the experiment ends. From that point onward, the strategy is forced to maintain more randomization than the uncapped dynamic schedule would prescribe, which generates the two additional terms in the regret bound. These terms therefore capture the extra regret incurred after exhaustion, and their magnitude depends on the remaining horizon \(T - t_\gamma\): the earlier the budget binds, the larger the resulting regret penalty.

Having introduced the two strategies, we next apply them to two marketing applications.

\section{Two applications of the privacy risk strategies}\label{sec:applications}
In this section, we apply the strategies to two experiments: an experiment on website design, and one involving a recommendation system for an online fashion retailer. In both applications, CTRs from a randomized control trial (RCT) are available, and we can use these CTRs to perform empirically-grounded simulations that test the effectiveness of our method against several benchmarks.

We study these applications because the experimental output may reveal the visitor's segment. In the recommendation-system application, the displayed fashion items may signal inferred product interests or broader demographic segments \citep{Castelluccia_2012, Xin_2023, Chen_2026}. In the website-design application, assignment different website designs may signal the firm's inference about the visitor's stage in the purchase journey \citep{Hauser_2009, Liberali_2022}.

\subsection{Website design}
\citet{Liberali_2022} report the results of a web design RCT. The experiment was conducted on the MBA website the Rotterdam School of Management (RSM), Erasmus University. The goal was to test two versions of the website, with abstract vs. concrete language, to maximize the total number of clicks on the call-to-action prompting visitors to begin their MBA application.\footnote{The abstract language website used broader, exploratory language (e.g., “Why the RSM MBA? Get the brochure.”) and terms such as Explore, Discover, and Participate in a conversation. In contrast, the concrete language version employed more action-oriented phrasing (e.g., “Ready to start your application? Create an account.”) and concrete terms such as ``Join", ``Apply", and ``Tour (the campus)." The CTRs are .324 for the concrete version and .295 for the abstract version of the website.}

We set $T$=10,000, 100,000, and 1,000,000 visitors, as proxies for small, medium, and large firms. We simulate the constant and dynamic spending strategies under privacy budgets of $\gamma \in \{.05, .5, 1, 3, 5\}$. Assuming that a tracker initially believes that a visitor has a 20\% probability of segment membership, these budgets cap the posterior belief after observing the experimental output at 20.8\%, 29.2\%, 40.5\%, 83.4\%, and 97.4\%, respectively, ranging from strong to weak privacy protection.

We benchmark their performance against four reference strategies: an $\varepsilon$-greedy policy without privacy protection, the Gittins index policy equivalent to optimal learning, a perfect information strategy that assumes the firm knows which option has the highest CTR and always selects the best arm, and a random strategy that chooses an arm at random. To capture uncertainty, we bootstrap each strategy 1,000 times and visualize the resulting uncertainty using 95\% confidence intervals.

\begin{figure}[h]
     \FIGURE
    {\includegraphics[width=\linewidth]{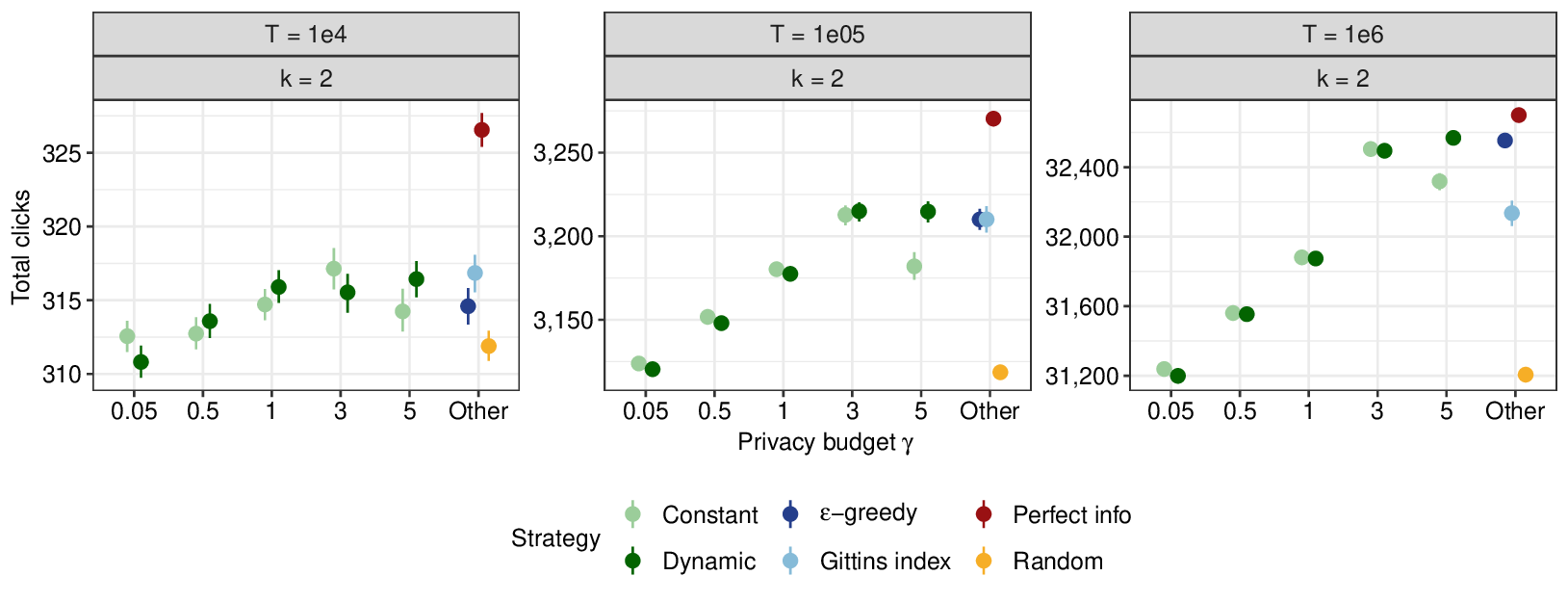}}
    {Website design application --- Total clicks as a function of different strategies.
    \label{fig:app1_total_clicks}}
    {The ``Other'' category on the \textit{x}-axis includes the optimal strategies—the Gittins index and $\varepsilon$-greedy policies—and the theoretical lower and upper bounds represented by random and perfect information, respectively.} 
\end{figure}

Figure \ref{fig:app1_total_clicks} shows the total clicks (rewards) for each strategy across different firm sizes. As expected, the perfect information strategy achieves the highest total clicks, followed by the $\varepsilon$-greedy and Gittins index policies, while the random strategy performs worst. The figure visualizes the empirical implications of our theory (see \S\ref{sec:privacy_risk}). The privacy budget $\gamma$ sets the exploration rate via Equation~\eqref{eq:exploration}, so smaller $\gamma$ implies more randomization and thus more exploration, while larger $\gamma$ allows more exploitation. When $\gamma$ is very small (e.g., $\gamma=0.05$), the exploration rate is high and both the constant and dynamic strategies behave close to random assignment. As $\gamma$ increases, and privacy protection decreases, both strategies improve because the bandit is allowed to exploit more often, shifting probability mass toward the higher-performing arm.

The differences in performance between the two strategies align with the results of our regret analysis (see Equations \ref{eq:regret_constant} and \ref{eq:regret_dynamic}). The dynamic strategy is designed to mimic the ``explore early, exploit later'' pattern of classic $\varepsilon$-greedy, so it benefits most when $\gamma$ is large enough that the cap binds only late (or not at all). In that region, further increases in $\gamma$ relax a constraint that is already slack, so the dynamic strategy exhibits diminishing returns and effectively hits a performance ceiling at high $\gamma$. Consistent with Equation~\eqref{eq:regret_dynamic}, the higher the privacy risk $\gamma$, the longer the bandit can follow the non-private regret bound, and the higher the total rewards. 

By contrast, when implementing the constant strategy, there appears to be an ideal budget level $\gamma$, where performance is maximized given this strategy. When $\gamma$ is very small, the exploration probability is close to one and the strategy behaves nearly at random. As $\gamma$ increases, randomization falls and clicks increase. However, when $\gamma$ becomes too large, the constant strategy explores too little in every round, so early noise can cause the bandit to lock in to a suboptimal arm, reducing total clicks despite the higher budget. This mechanism explains why the constant strategy can peak at an intermediate $\gamma$, whereas the dynamic strategy improves until it reaches its ceiling. 

These differences become more consequential as the experimental horizon increases. With more visitors, the dynamic strategy has more opportunities to benefit from early learning and subsequent exploitation, whereas the constant strategy may continue exploiting an arm selected on the basis of noisy early estimates. Consequently, the performance advantage of the dynamic strategy over the constant strategy becomes more pronounced as $T$ increases. For $T=100{,}000$ and $T=1{,}000{,}000$, the differences in mean performance between the two strategies are statistically significant (see Figure~\ref{fig:app1_total_clicks}).

In the next application, we examine the effectiveness of the constant and dynamic privacy strategies in a more complex setting with multiple arms.

\subsection{Recommendation system}
We examine the robustness of the privacy risk strategies beyond a two-arm setting, and in a different setup involving a recommender system. We use data from a seven-day field experiment conducted by ZOZOTOWN, a large Japanese fashion e-commerce platform \citep{Saito_2020}. For each visitor, ZOZOTOWN displayed three fashion item recommendations selected uniformly at random, then recorded whether the visitor clicked on a recommended item. In total, ZOZOTOWN recorded \(1{,}374{,}327\) visitors across 80 arms corresponding to fashion items.\footnote{CTR values range from $.0012$ to $.0078$. In our simulations, we use the eight highest-performing items, whose CTRs range from $.0012$ to $.0053$. CTRs in this application are approximately an order of magnitude lower than those in the website-design application. Although these low success rates are typical of the setting and provide an informative test of our method, we do not report simulations with $T=10{,}000$ visitors because the bandit does not have enough observations to learn reliably at that sample size.} This setting allows us to vary both the number of arms and the time horizon. We consider \(K \in \{2,4,8\}\) and $T \in \{100{,}000,\;1{,}000{,}000\}$ in our simulations below.

\begin{figure}[h]
     \FIGURE
    {\includegraphics[width=\linewidth]{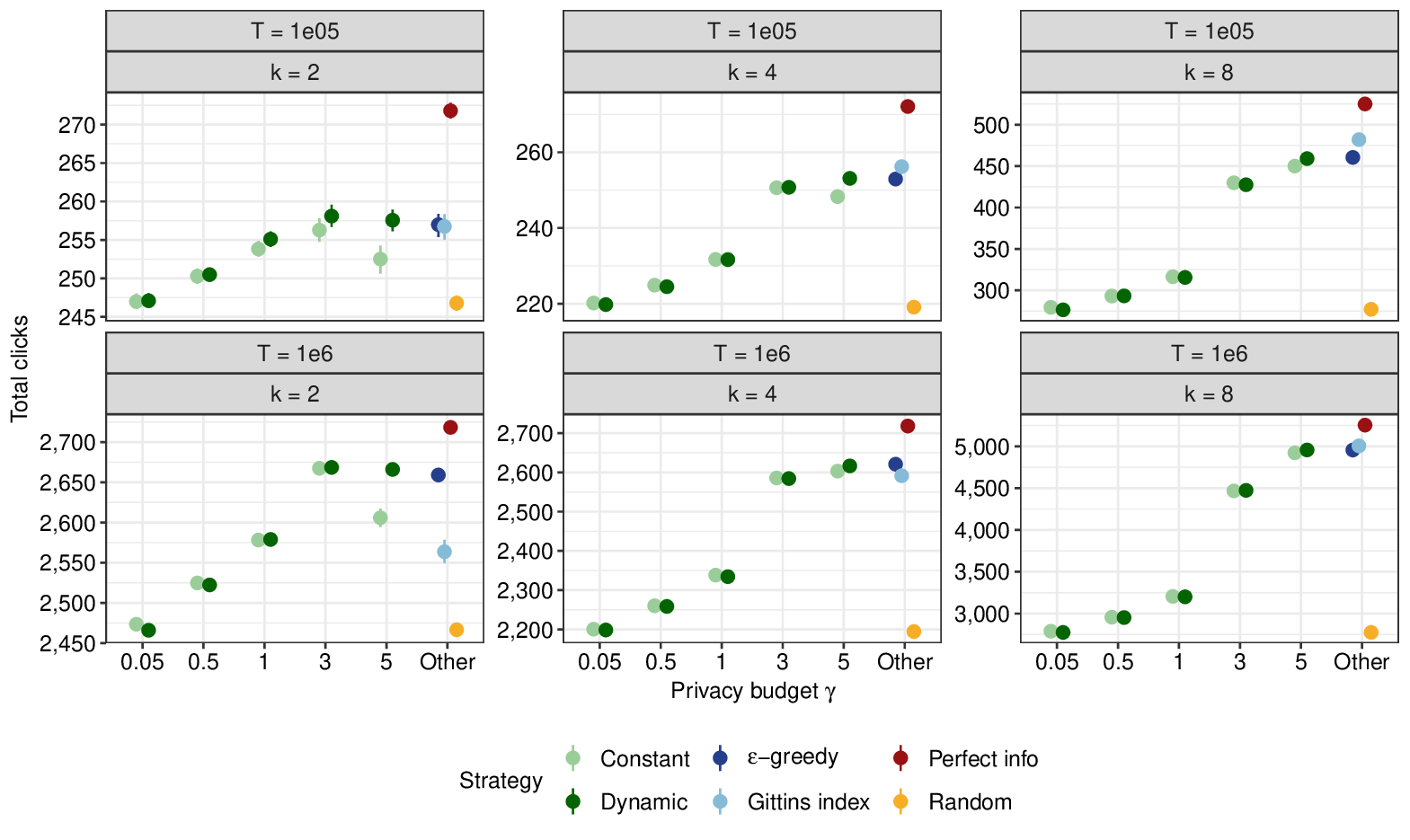}}
    {Recommendations --- Total clicks as a function of different strategies.
    \label{fig:app2_total_clicks}}
    {The ``Other'' category on the \textit{x}-axis includes the benchmark policies---the Gittins-index policy and the non-private $\varepsilon$-greedy policy---as well as the theoretical lower and upper bounds represented by random and perfect information, respectively.} 
\end{figure}

In Figure~\ref{fig:app2_total_clicks}, we plot total clicks on the $y$-axis across different privacy budgets on the $x$-axis. The column panels vary the number of visitors, whereas the row panels vary the number of arms. The results are consistent with those of the previous application. Under the constant strategy, performance initially improves as $\gamma$ increases, then levels off and may decline at higher values of $\gamma$. By contrast, the performance of the dynamic strategy increases with $\gamma$ and eventually plateaus at the optimal-performance benchmark for sufficiently large values of $\gamma$.

The results also highlight a difference from the website-design application. This mainly impacts the constant privacy risk strategy. Figure~\ref{fig:app2_total_clicks} shows that the non-monotonic performance pattern becomes less pronounced when the number of arms exceeds two, relative to the pattern observed in the website design application. This occurs because the exploration probability is determined not only by the privacy budget $\gamma$, but also by the number of arms $K$ (see Equation~\ref{eq:exploration}). For a fixed $\gamma=3$, increasing the number of arms from $K=2$ to $8$ raises the exploration probability from 10\% to 30\%. To achieve a similar number of total clicks as $K$ increases, we must set a larger privacy budget to allow for sufficient exploitation. As a result, for $K \in \{4,8\}$, total clicks under the constant strategy generally increase with $\gamma$: a higher $\gamma$ mainly enables more exploitation while maintaining non-trivial exploration.

In the next section, we derive privacy risk elasticities of regret to capture the marginal regret of stronger privacy protection. Whereas Figures~\ref{fig:app1_total_clicks} and~\ref{fig:app2_total_clicks} compare outcomes of a few discrete values of $\gamma$, elasticities quantify how sensitive regret is to small changes in $\gamma$. This allows us to identify regions of diminishing returns from increasing privacy risk and, when they exist, regret-minimizing privacy budgets that balance the performance gains from additional learning against the costs of greater privacy risk.

\section{Privacy risk elasticities of expected regret}
\label{sec:elasticity}
To quantify the privacy--performance trade-off, we derive the elasticity of expected regret with respect to the privacy budget $\gamma$ for each strategy. The elasticity measures the percentage change in expected regret resulting from a one-percent change in $\gamma$. A negative elasticity indicates that increasing the budget reduces regret, whereas a positive elasticity indicates that it increases regret. A change from negative to positive identifies a regret-minimizing budget. These elasticities extend our empirical findings by helping managers identify regret-minimizing privacy budgets and assess where stronger privacy protection is relatively inexpensive or costly in performance terms.

\subsection{The privacy elasticity of regret for the constant strategy}
\begin{figure}[h]
     \FIGURE
    {\includegraphics[width=\linewidth]{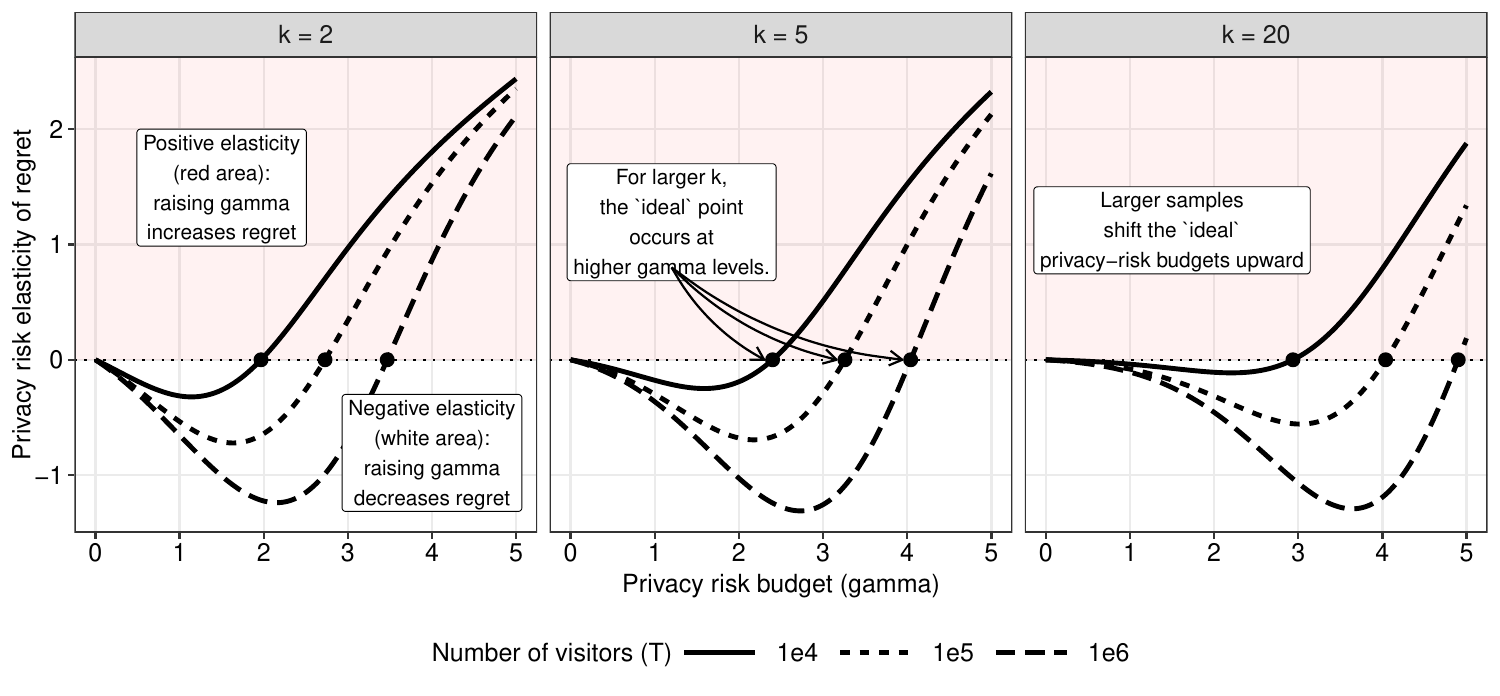}}
    {Privacy risk elasticity of regret as a function of the privacy budget $\gamma$, for different numbers of arms ($K \in {2,5,20}$) and visitors ($T \in {10,000,100,000,1,000,000}$).
    \label{fig:ct_elasticity}}
    {The elasticity is negative where relaxing privacy protection (increasing $\gamma$) reduces regret, and positive where further increases in $\gamma$ raise regret. The zero-crossings mark is the ideal privacy budgets that minimize regret for each $(k,T)$ combination.}
\end{figure}

We derive the privacy elasticity of regret for the constant strategy in Web Appendix~\ref{web:elasticityct}. Figure~\ref{fig:ct_elasticity} plots this elasticity as a function of the privacy budget $\gamma$. The $x$-axis shows the privacy budget $\gamma$ and the $y$-axis shows the percentage change in expected regret resulting from a one-percent increase in $\gamma$. The panels vary the number of arms $K\in\{2,5,20\}$ and the line types represent different numbers of visitors $T$.
 
The sign of the elasticity provides a managerial diagnostic of where the constant strategy is ``too random'' versus ``too greedy.'' When the elasticity is negative, increasing $\gamma$ (i.e., allowing less randomization) reduces regret because the bandit can exploit more effectively. When the elasticity becomes positive, further increases in $\gamma$ increase regret because the constant strategy explores too little and can lock in to a suboptimal arm (see the left panel in Figure \ref{fig:ct_elasticity}). The point where the elasticity crosses zero therefore identifies an ideal privacy budget for the constant strategy. We denote such ideal point by $\gamma_{\text{constant}}^{*}$ for the constant strategy and $\gamma_{\text{dynamic}}^{*}$ for the dynamic strategy.

We observe that the value of $\gamma_{\text{constant}}^{*}$ depends on both the number of visitors and the number of arms (see the middle and right panels in Figure \ref{fig:ct_elasticity}): it shifts upward as either $T$ or $K$ increases. Intuitively, these forces work in opposite directions. As the number of visitors $T$ increase, the bandit has more opportunities to learn, so it can afford to explore less and exploit more as the experiment progresses. By contrast, as $K$ increases, identifying the best arm becomes more difficult, which increases the need for exploration. However, for any fixed $\gamma$, the exploration probability also increases mechanically with $K$ (see Equation \ref{eq:exploration}). The regret-minimizing budget $\gamma_{\text{constant}}^*$ therefore shifts upward to offset this additional randomization. These findings help explain why empirical clicks under the constant strategy initially increase with $\gamma$ but may subsequently plateau or decline in Figure~\ref{fig:app1_total_clicks}, and why this pattern becomes less pronounced as $K$ increases in Figure~\ref{fig:app2_total_clicks}.

These points also clarify where the firm's and the policymaker's objectives align under this strategy. A firm seeking to minimize regret selects the privacy budget $\gamma_{\text{constant}}^*$. Practically, $\gamma_{\text{constant}}^*$ provides managers with a rule of thumb for setting the privacy budget ex ante as a function of the number of arms $K$ and visitors $T$. A policymaker seeking the strongest privacy protection without sacrificing experimental performance also selects $\gamma_{\text{constant}}^*$: to the right of $\gamma_{\text{constant}}^*$, reducing the budget both strengthens privacy and lowers regret, whereas moving to the left of $\gamma_{\text{constant}}^*$ strengthens privacy only at the cost of higher regret. Thus, $\gamma_{\text{constant}}^*$ represents the point at which the firm and the policymaker meet.

\subsection{The privacy elasticity of regret for the dynamic strategy}
\begin{figure}[h]
     \FIGURE
    {\includegraphics[width=\linewidth]{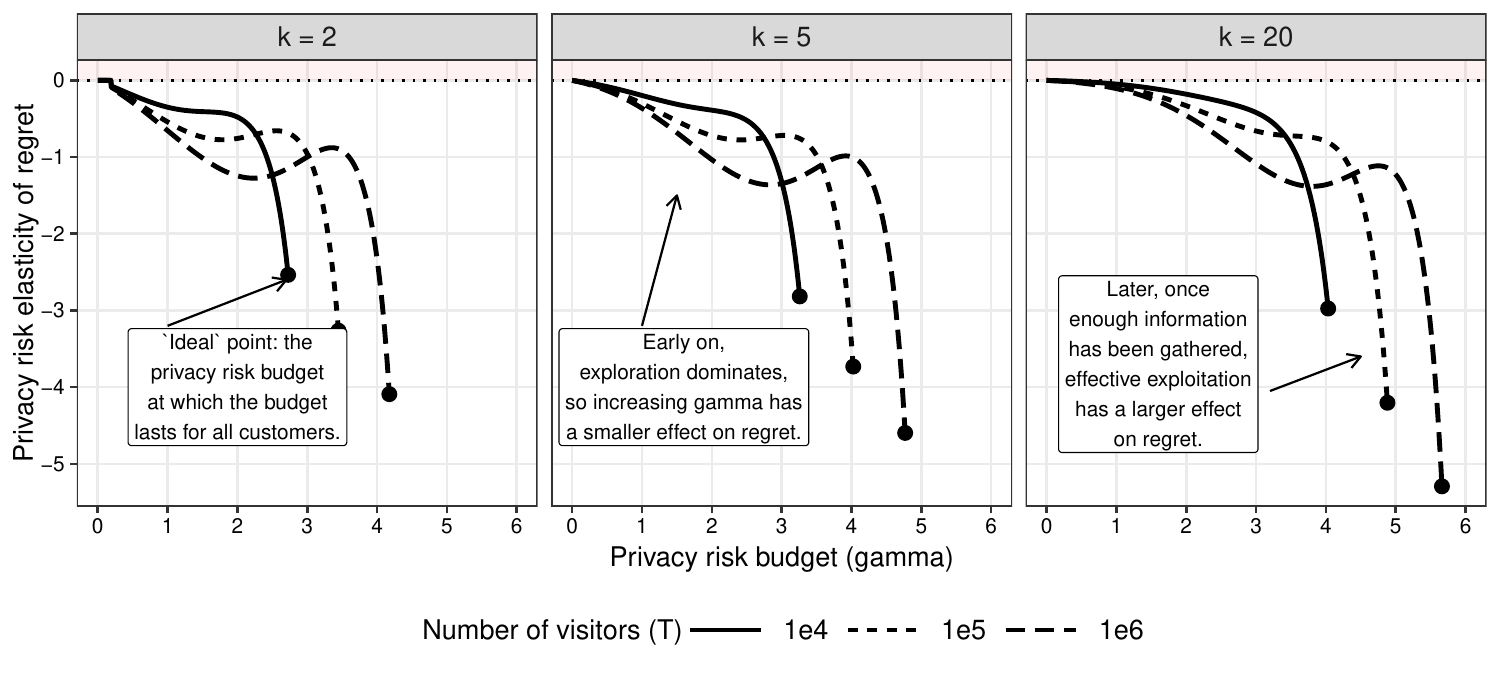}}
    {Privacy risk elasticity of regret as a function of the privacy budget $\gamma$, for different numbers of arms ($K \in {2,5,20}$) and visitors ($T \in {10,000,100,000,1,000,000}$).
    \label{fig:dyn_elasticity}}
    {The elasticity is negative where relaxing privacy protection (increasing $\gamma$) reduces regret, and positive where further increases in $\gamma$ raise regret. The zero-crossings mark is the ideal privacy budgets that minimizes regret for each $(k,T)$ combination.}
\end{figure}

In Figure \ref{fig:dyn_elasticity}, we present the counterpart of Figure \ref{fig:ct_elasticity} for the dynamic strategy (see Web Appendix~\ref{web:elasticitydynamic} for the technical derivation). Unlike the constant strategy, the elasticity here is non-positive throughout: relaxing the privacy budget reduces regret by delaying the point $t_\gamma$ at which the budget cap becomes active (see Equation~\ref{eq:budgetgamma}).

The annotations in Figure~\ref{fig:dyn_elasticity} highlight the three relevant regimes. 
First, as illustrated by the left-panel annotation, the most favorable point $\gamma_{\text{dynamic}}^{*}$ is near the largest value of $\gamma$ for which the privacy budget lasts through all visitors, i.e., $t_\gamma \approx T$ and allows the bandit to follow the optimal exploration/exploitation schedule. 
For small $\gamma$, the curves are relatively flat (see middle panel): early regret is dominated by exploration, so marginally increasing the privacy budget has only limited effect. Exploitation is insufficient to ensure results close to optimal. As $\gamma$ increases, the strategy gathers enough information to exploit effectively (see right panel). This produces a large marginal reduction in regret and therefore a strongly negative elasticity. 

Each curve ends at a finite value of $\gamma$. Beyond that level of $\gamma$, $t_\gamma \geq T$, and the dynamic strategy can follow the exploration/exploitation schedule of the non-private bandit. Further increasing $\gamma$ has no impact on expected regret, and elasticity is zero. This privacy risk cutoff increases as $T$ and $K$ increase, because experimenting with more visitors and more arms require a longer exploration phase before the strategy can exploit effectively.

Taken together, the constant and dynamic strategies yield two distinct experiment-level ideal privacy budgets. Although $\gamma_{\text{constant}}^*$ and $\gamma_{\text{dynamic}}^*$ are values of the same privacy budget, they are not directly comparable because each identifies only the budget at which regret is minimized within its respective strategy. They do not reveal the absolute level of regret attained at those budgets. Comparing the two budgets alone consequently does not establish which strategy delivers better absolute performance. 

This strategy-selection problem becomes especially important when firms conduct multiple experiments over time. Managers must therefore decide not only how to set the privacy budget for an individual experiment, but also how to allocate privacy risk across a portfolio of experiments. We next extend the analysis to the firm level and examine which strategy---constant or dynamic---delivers the highest expected rewards under a given firm-level privacy budget.

\section{Firm-level privacy budget allocation}\label{sec:multi_exp_portfolio}
We next consider a firm with a finite firm privacy budget $\Gamma$ that limits cumulative privacy risk across its portfolio of experiments (see Level~3 in Figure~\ref{fig:privacy_hierarchy}).

\subsection{The optimization problem}
Following the marketing resource-allocation literature
\citep{Fischer_2011, Peers_2017, ZiaRao2019}, we treat the firm-wide privacy budget $\Gamma$ as a scarce resource. The firm must allocate this budget across experiments and select a spending strategy for each experiment to minimize total regret.

\subsubsection{The exogenous environment.}
The firm specifies the firm-wide privacy budget $\Gamma$ and the characteristics of each experiment, including its number of arms and visitors. A visitor may participate in multiple experiments but participates only once in any given experiment. Under sequential composition, the cumulative privacy risk for a visitor is bounded by the sum of the budgets of the experiments in which that visitor participates \citep[see Theorem~3.14 in][]{Dwork_2014}. 

\subsubsection{The objective and feasible solution space.}The firm chooses the allocation that minimizes the regret introduced by privacy protection. The regret bounds for the constant and dynamic strategies in Equations~\eqref{eq:regret_constant} and~\eqref{eq:regret_dynamic} provide the building blocks for this objective. 

Let $e\in\{1,\ldots,E\}$ index experiments and $s\in\{\mathrm{constant},\mathrm{dynamic}\}$ index strategies. Let $z_{e,s}\in\{0,1\}$ indicate whether strategy $s$ is selected for experiment $e$, and let $\gamma_{e,s}\geq0$ denote the privacy budget allocated to that experiment--strategy pair. Finally, let $R_{e,s}(\gamma_{e,s})$ denote the corresponding regret bound. The firm solves
\begin{equation}\label{eq:portfolio_with_explicit_strategy_choice}
\begin{aligned}
\min_{\{z_{e,s},\,\gamma_{e,s}\}}
\quad
& \sum_{e=1}^E\sum_{s\in\mathcal S}
z_{e,s}R_{e,s}(\gamma_{e,s})\\
\text{subject to}\quad
& \sum_{e=1}^E\sum_{s\in\mathcal S}\gamma_{e,s}\leq\Gamma,\\
& \sum_{s\in\mathcal S}z_{e,s}=1
\qquad \forall e,\\
& 0\leq\gamma_{e,s}\leq z_{e,s}\gamma_{e,s}^*
\qquad \forall e,\ s\in\mathcal S,\\
& z_{e,s}\in\{0,1\}
\qquad \forall e,\ s\in\mathcal S.
\end{aligned}
\end{equation}

The first constraint limits total allocated privacy budget to the firm-wide budget $\Gamma$. The second requires the firm to select exactly one strategy for each experiment. The third links the budget allocation to that strategy choice: an unselected strategy receives no budget, whereas a selected strategy may receive any budget up to its strategy-specific performance benchmark $\gamma_{e,s}^*$, derived in \S\ref{sec:elasticity}. The final constraint defines the strategy-selection variables as binary. Restricting allocations to $\gamma_{e,s}^*$ is without loss of optimality because performance does not improve beyond this point and the firm-wide budget need not be fully exhausted.

\subsubsection{The solution space: Return on the privacy budget.} 
The key managerial quantity is the return on the privacy budget. Intuitively, a larger privacy budget gives an experiment more room to learn, which reduces regret. However, the value of additional privacy budget need not be the same across experiments or strategies. For each experiment $e$ and strategy $s$, we capture this return using the marginal learning gain:
\begin{equation}\label{eq:Ge_def_portfolio}
G_{e,s}(\gamma)
\;:=\;
-\,\frac{d}{d\gamma}R_{e,s}(\gamma).
\end{equation}
Because $R_{e,s}(\gamma)$ is a regret bound, a decrease in $R_{e,s}(\gamma)$ is a gain. The minus sign ensures that $G_{e,s}(\gamma)$ is positive when increasing the privacy budget reduces regret. Thus, $G_{e,s}(\gamma)$ measures how much regret is reduced by giving a small additional amount of privacy budget to experiment $e$ under strategy $s$.

This quantity provides a direct way to compare privacy budget allocations. If $G_{e,s}(\gamma)$ is large, then an additional unit of privacy budget produces a large reduction in regret for that experiment--strategy pair. If $G_{e,s}(\gamma)$ is small, then an additional unit of privacy risk has a limited effect on regret. The firm should therefore allocate privacy budget first to the experiment--strategy pairs with the highest marginal learning gains (i.e., highest $G_{e,s}(\gamma)$). At the optimum, all selected pairs whose budgets can still be adjusted must have the same marginal return. That is, there exists a constant 
$\lambda\ge 0$ such that
\begin{equation}\label{eq:equal_marginal_selected_pairs}
G_{e,s}(\gamma_{e,s})=\lambda
\quad \text{for all } (e,s) \text{ such that } z_{e,s}=1
\text{ and } 0<\gamma_{e,s}<\gamma^{*}_{e,s}.
\end{equation}
The constant $\lambda$ is the shadow value of the firm-wide privacy budget: it is the reduction in total regret from marginally relaxing the firm-wide privacy budget.

\begin{figure}[h]
    \centering
    \includegraphics[width=0.9\linewidth]{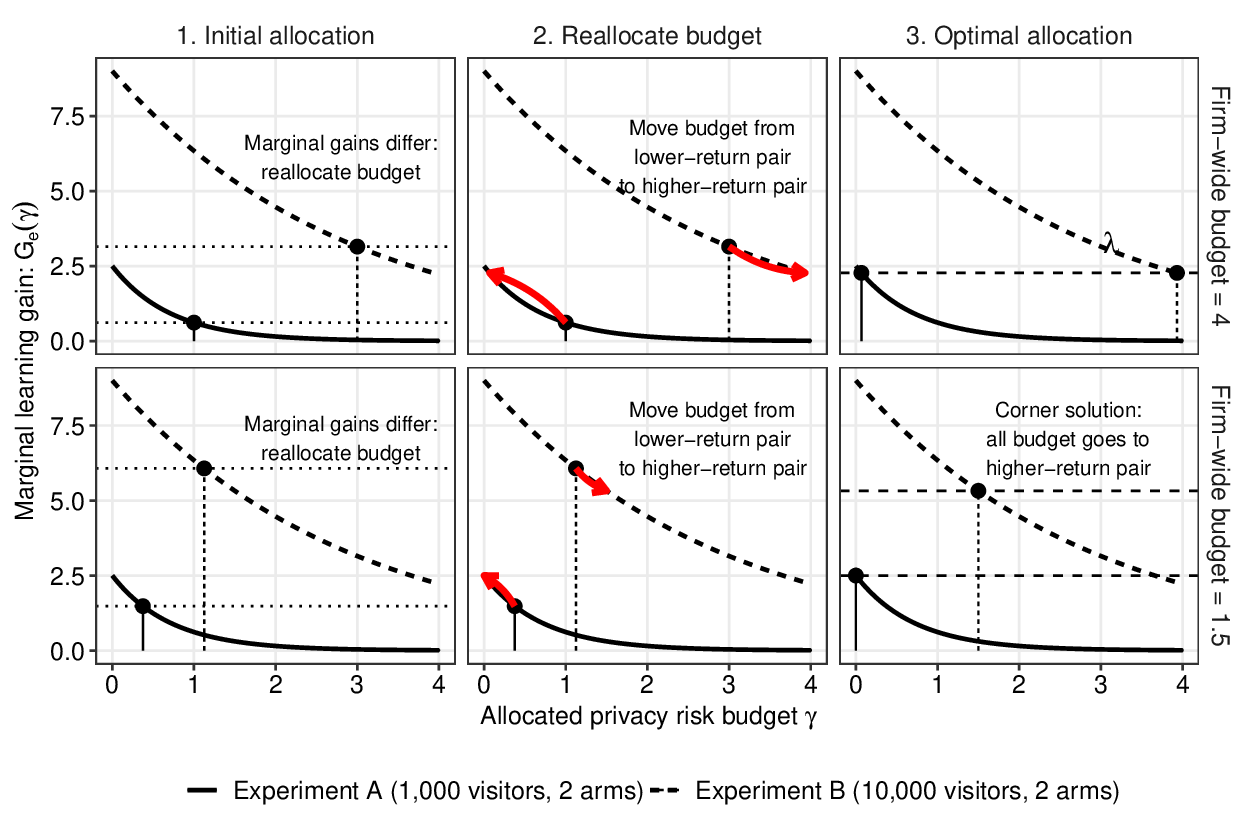}
    \caption{Privacy budget reallocation across two experiments and assuming a single strategy. The $x$-axis shows the allocated privacy budget $\gamma$, and the $y$-axis shows the marginal learning gain $G_{e,s}(\gamma)$. The firm reallocates the budget toward the experiment with the higher marginal learning gain (red arrows).}
    \label{fig:marginal_learning_gain}
\end{figure}

To illustrate our allocation procedure, consider Figure~\ref{fig:marginal_learning_gain} which shows the allocation logic for two hypothetical experiments: Experiment A has 1{,}000 visitors and 2 arms, while Experiment B has 10{,}000 visitors and 2 arms. For simplicity, we assume that only one strategy is available in this illustration. The three columns trace the progression from an initial allocation (see the left panel), through a budget reallocation (see the middle panel), to the optimal allocation (see the right panel). The two rows correspond to two firm-wide privacy budgets: a larger budget ($\Gamma = 4$) in the top row and a smaller budget ($\Gamma = 1.5$) in the bottom row.

In the top row, the initial allocation gives different marginal learning gains across the two experiments. Since Experiment B has the higher marginal gain, the budget is shifted toward Experiment B until the marginal gains are equalized at $\lambda$ (see red arrows). In the bottom row, the smaller firm-wide budget leads to a corner solution: the budget is too limited to equalize marginal gains, so all available budget is allocated to the higher-return experiment.

\subsection{Benchmark policies}\label{sec:benchmarks}
We compare the allocation described above with several benchmark allocations that differ in how they choose strategies and divide the firm-wide privacy budget $\Gamma$ across experiments.

\subsubsection{The upper bound on performance: the $\gamma^*$ benchmark.}
For each experiment, this benchmark assigns the strategy-specific reference level implied by the elasticity analysis (see \S\ref{sec:elasticity}) for the selected strategy:
\[
\gamma_{e,s}=z_{e,s}\gamma^*_{e,s}
\qquad\forall e,\ s\in\{\mathrm{constant},\mathrm{dynamic}\}.
\]

The implied portfolio privacy spending is $\sum_{e=1}^E \sum_s z_{e,s}\gamma^*_{e,s}$, which may exceed $\Gamma$. Hence, this benchmark serves as an unconstrained reference allocation rather than a feasible portfolio solution.

\subsubsection{Managerial heuristic: equal-split benchmark.}
This benchmark allocates the firm-wide privacy budget equally across experiments. With $E$ experiments, each experiment is assigned an equal share $\Gamma/E$, and for each experiment we fix the strategy as the one that minimizes the regret bound at that equal-share cap, i.e., $s_e \in \arg\min_{s\in\{\mathrm{constant},\mathrm{dynamic}\}} R_e^{s}(\Gamma/E).$ The selected cap then satisfies $\sum_{s\in\{\mathrm{constant},\mathrm{dynamic}\}} z_{e,s}\gamma_{e,s}=\Gamma/E$ for all $e$.

\subsubsection{The lower bound on performance: $\Gamma=0$, equivalent to random baseline.}
We include a zero-budget baseline in which no privacy risk is allocated to any experiment (i.e., $\gamma_{e,s}=0$ for all $e,s$). This implies full randomization within each experiment. This benchmark therefore serves as a lower bound reference for learning performance under the strongest privacy protection.

\subsection{Application of the firm-wide privacy budget}\label{sec:asos_application} We next empirically apply the privacy budgeting across multiple experiments under a firm-wide budget. We use a dataset of 78 experiments conducted by ASOS, a British fashion retailer serving customers worldwide \citep{liu2021datasets}. These experiments are online RCTs conducted on ASOS’s e-commerce platform.\footnote{The experiments test product recommendation on the retailer's website. CTRs vary between....,and the number or arms vary between ...... The dataset is publicly available, therefore we can share the data together with our code, to facilitate replication.}
\subsubsection{Setup.}
\begin{figure}[h]
    \centering
    \includegraphics[width=0.7\linewidth]{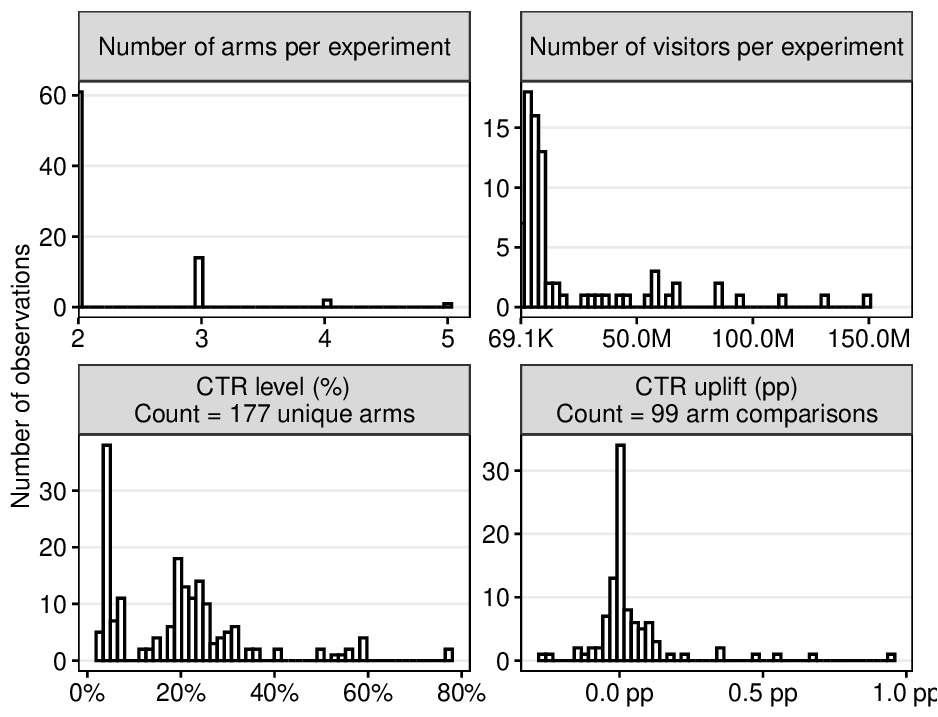}
    \caption{Empirical distributions of key inputs and outputs from the ASOS online experiments.}
    \label{fig:summarystats}
\end{figure}

We visualize summary statistics from these experiments in Figure~\ref{fig:summarystats}. The top-left panel shows the number of arms per experiment, and the top-right panel shows the number of visitors per experiment. The bottom-left panel reports the distribution of arm-level CTRs across all unique arms, while the bottom-right panel shows the distribution of treatment--control CTR uplifts in percentage points for within-experiment arm comparisons.\footnote{Computed as $100\times(\mathrm{CTR}_{\mathrm{arm}}-\mathrm{CTR}_{\mathrm{control}})$, in percentage points.}

Most experiments used two arms, and only a few used four or five arms. The average number of visitors was 21 million, with a minimum of 69{,}000 and a maximum of 149{,}197{,}471. The large visitor counts likely reflect the relatively long runtimes of these experiments. On average, experiments ran for 43.5 days, with a maximum duration of 131 days.

The distribution of arm-level CTRs is broad. The average CTR is 19.55\%, with an interquartile range from 5.69\% to 24.83\%. CTRs range from 2.65\% to 77.24\%, indicating substantial heterogeneity across arms and experiments. The distribution of treatment--control CTR uplifts (in percentage points) is centered near zero, implying that many arms produce modest gains or losses relative to the control, while a smaller number of arms generate large positive uplifts. The distribution is right-skewed, with a long upper tail.

The wide variation in CTRs likely reflects differences in experimental context and click definitions across ASOS’s on-site tests. Although the underlying experimental contexts  are confidential \citep{liu2021datasets}, we can offer plausible interpretations. In some experiments, high CTRs may arise from goal-directed actions, such as checkout steps, whereas low CTRs may reflect more discretionary interactions, such as clicking a secondary fashion item. We use these empirical statistics to construct the experiment-level inputs for the portfolio problem, and then evaluate the benchmark allocations for 10 firm-wide privacy budgets, $\Gamma \in \{.01, .1, 1, 3, 5, 10, 20, 50, 100, 500\}$.

\subsubsection{Results.}
\begin{figure}[h]
    \centering
        \includegraphics[width=0.9\linewidth]{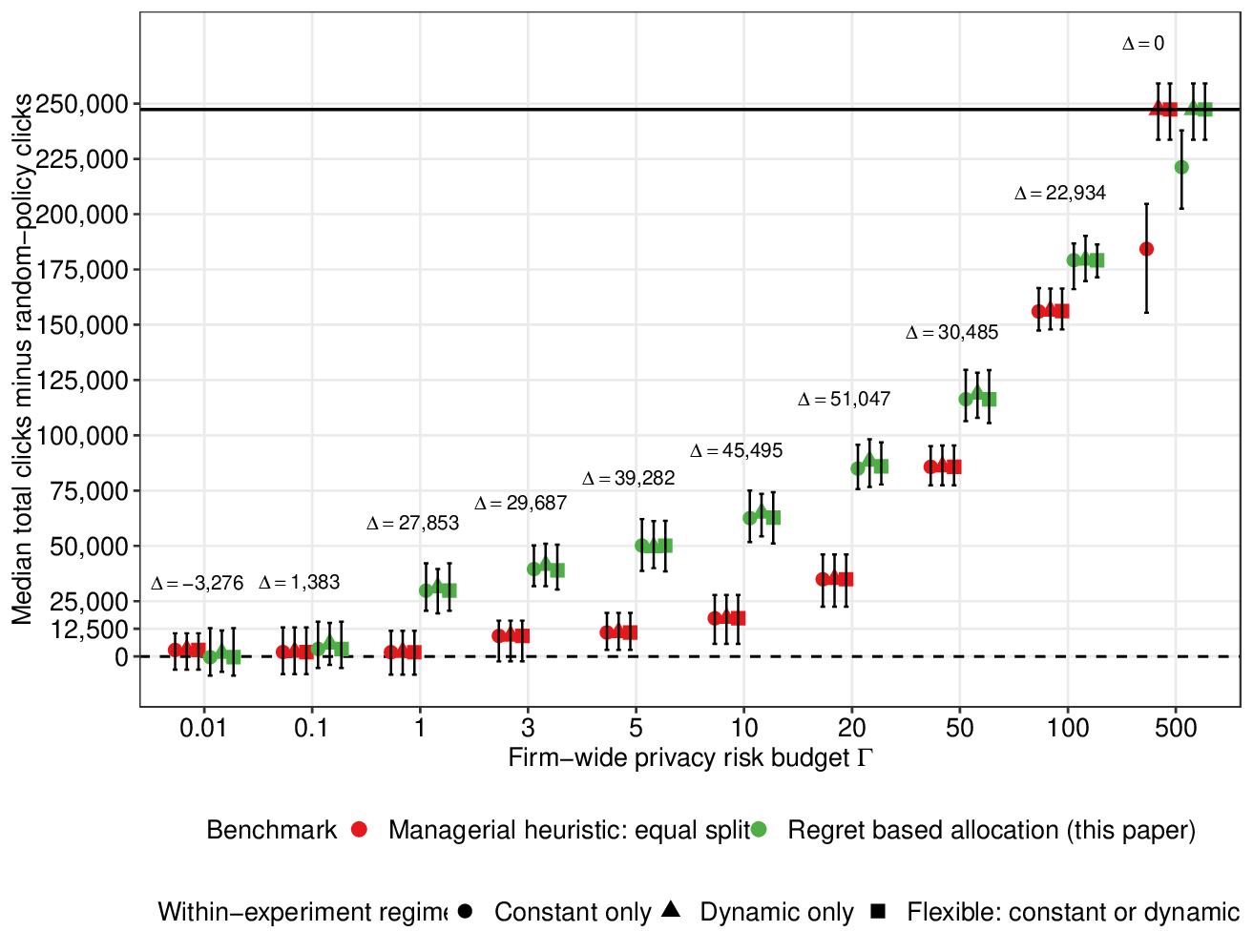}
    \caption{Median improvement in portfolio clicks relative to the random-policy baseline across firm-wide privacy budgets. Points report medians over simulation repetitions, and error bars show the interquartile range. Colors indicate the firm-wide allocation rule, while shapes indicate the within-experiment budgeting regime.}
    \label{fig:firmwide}
\end{figure}

In Figure~\ref{fig:firmwide}, the $y$-axis shows the median improvement in portfolio clicks relative to the random-policy baseline, defined as total clicks under each policy minus random-policy clicks. The $x$-axis shows the privacy budgets $\Gamma$. Points report medians over 100 simulation repetitions, and error bars show the interquartile range. Colors distinguish the benchmark allocation rule (see Section \ref{sec:benchmarks}), while shapes indicate the within-experiment strategy regime. The flexible regime allows the firm to choose, for each experiment, whether to use the constant or dynamic privacy budgeting strategy. By contrast, the constant regime restricts all experiments to use the constant strategy, and the dynamic regime restricts all experiments to use the dynamic strategy. The dashed horizontal line at zero indicates the random (lower bound) benchmark, equivalently the $\Gamma=0$ case. The solid horizontal line indicates the upper bound $\gamma^{*}$ benchmark. We use text annotations to highlight the difference in median clicks between the managerial heuristic and the regret-based allocation benchmark.

Figure~\ref{fig:firmwide} shows that under a fixed firm-wide privacy budget $\Gamma$, the regret-based allocation often delivers substantially more clicks than the managerial equal-split benchmark (see the $\Delta$ annotations for the median difference in clicks under the flexible regime).

The improvement in clicks is largest at intermediate firm-wide privacy budgets (e.g., $\Gamma \in {5,20}$). When the firm-wide privacy budget is very small, both allocation rules have limited ability to learn, so gains in clicks are small (e.g., $\Gamma \in \{0.01, 0.1\}$). When $\Gamma$ is very large, both rules eventually approach the upper-bound benchmark.

This pattern highlights a key managerial choice: firms can improve experimentation outcomes either by increasing the total privacy budget or by allocating the existing budget more effectively. The figure shows that the second option can generate more clicks without increasing firm-wide privacy risk. By directing a fixed budget $\Gamma$ toward experiments where additional information is most valuable, firms can improve performance while keeping total privacy exposure unchanged. The privacy--performance tradeoff therefore depends not only on how much privacy risk the firm permits, but also on how selectively it allocates that risk across experiments.

\section{Discussion}\label{sec:conclusion}
Firms increasingly rely on online experiments to decide what consumers see, such as banners, ads, recommendations, and product rankings. While these decisions are typically studied as learning problems, we highlight the privacy risk that the experimental outputs themselves can create: what a firm displays may reveal information to third-parties before a consumer clicks, purchases, or otherwise responds. Our privacy budgeting framework manages this risk at three levels. At the individual level, privacy risk captures how much a displayed output can reveal about a consumer. At the experiment level, a privacy budget $\gamma$ limits the risk generated by a single experiment and formalizes the trade-off between privacy protection and learning. At the firm level, a firm-wide budget $\Gamma$ limits cumulative risk across potentially linkable experiments, turning privacy protection into a portfolio allocation problem across experiments and budgeting strategies.

A key step in the paper is linking the exploration parameter $\varepsilon$ in the $\varepsilon$-greedy policy to the privacy risk parameter from differential privacy \citep{Dwork_2014}. The parameter $\varepsilon$ determines how often the firm explores rather than exploits the currently best-performing arm. We use this relationship to construct the inputs to the firm-wide portfolio problem. We perform an elasticity analysis that identifies, for each experiment and budgeting strategy, the ideal experiment-level privacy budget $\gamma^*$ at which additional privacy budget has diminishing returns. The firm-wide problem uses these ideal budgets to allocate the available budget $\Gamma$ across these experiment-level opportunities, choosing both which strategy to use and how much budget to assign, in order to improve portfolio clicks subject to the firm-wide privacy budget.

To spend the privacy risk, we propose two strategies a constant and a dynamic strategy. A constant strategy is simple and transparent because it assigns the same privacy budget throughout the experiment. This may be attractive for implementation, communication, and governance. However, it is also rigid: it does not adapt to the changing value of learning over the course of an experiment. A dynamic strategy can allocate privacy risk more flexibly as the experiment evolves, potentially reducing regret for a given privacy budget. The portfolio formulation allows the firm to choose between these strategies experiment by experiment, rather than imposing the same privacy budgeting rule across all experiments.

\subsection{Limitations and future research}
Several choices shape the interpretation of the firm-wide budget. First, the regret bounds are independent of the expected CTRs of the experiments. This is because the regret bounds used in the portfolio problem are worst-case guarantees: they are intended to hold uniformly across possible CTR environments, including those unknown to the firm before experimentation \citep[cf.][]{Joonhwi_2025}. Conditioning the allocation rule on expected CTRs could yield a more tailored bound in a known environment, but would make the result less general and less useful in practice, since these rates are themselves objects of learning. We therefore allocate privacy risk according to the marginal reduction in worst-case regret from an additional unit of privacy budget, rather than according to which experiment has the highest observed or expected CTR.

Second, the firm-wide budget should be interpreted as a conservative portfolio-level constraint. The formulation in Equation \eqref{eq:portfolio_with_explicit_strategy_choice} protects against the worst-case scenario in which the same customer appears in every linkable experiment in the portfolio. In practice, however, most customers will appear in only a subset of experiments. For example, a customer exposed to only 10\% of the experiments will generally face a realized cumulative privacy risk below the portfolio-wide worst-case bound. Thus, $\Gamma$ is an upper bound on cumulative privacy risk across linkable experiments, not a claim that every customer realizes that level of privacy risk.

Third, our framework does not allow privacy sensitivity to vary across arms. Consequently, an arm with a higher CTR does not mechanically receive a higher or lower privacy cost than an arm with a lower CTR. This keeps the privacy accounting focused on the information revealed by the experimental output and on the resulting regret trade-off. A natural extension would introduce arm-specific privacy risk, allowing some displayed items, products, or recommendations—for example, a particularly sensitive book recommendation—to be treated as more privacy-sensitive than others.


Overall, the paper connects privacy protection to the economics of experimentation. Privacy is not treated as a fixed constraint imposed separately on each experiment, but as a scarce resource that must be allocated across a portfolio of learning problems. This perspective is useful for firms that run many concurrent personalization and experimentation systems. It also highlights why privacy governance must operate at the firm level: even when each individual experiment is privacy-aware, cumulative privacy risk can emerge across linkable experiments. A firm-wide privacy budget, combined with regret-based allocation, offers a practical framework for managing the trade-off between learning, performance, and privacy.

\ACKNOWLEDGMENT{%
}
\newpage
\begin{APPENDICES}
\section{A model of privacy risk in experimentation}\label{modelprivacyrisk}
We distinguish between the firm's information and that available to a tracker using Google Analytics (GA) as a running example. GA is used by 51\% of the top one million websites globally \citep{lee2026_google_analytics_statistics}.

Let \(S_i\) denote customer \(i\)'s latent segment. Before the experiment, the firm observes first-party history \(H_i^F\) and forms an internal state $U_i^F=\phi(H_i^F).$ It then generates experimental output \(A_i\) through a mechanism \(M\): $A_i\sim M(U_i^F).$ After exposure, the customer may generate behavior \(Y_i=(Y_i^E,Y_i^O)\), where \(Y_i^E\) is specific to the experimental output and \(Y_i^O\) is other behavior. GA instead observes a potentially different history \(H_i^{GA}\), so generally $H_i^{GA}\neq H_i^F.$ This difference arises because GA receives only information delivered under the measurement architecture \(\Omega\), whereas the firm may retain richer first-party records. 

Let \(G_i\in{0,1}\) indicate whether the GA observation channel is active during the experiment. Its value may depend on consent, technical delivery, browser protections and ad blockers. These factors may themselves reveal information about the customer's segment, but they arise outside the experimental assignment mechanism and are not covered by our privacy guarantee. We therefore condition the analysis on \(G_i=1\).

Conditional on an active channel, \(\Omega\) determines whether GA observes the experimental arm \(A_i\) alone or both \(A_i\) and the experiment-specific response \(Y_i^E\). The arm may be transmitted explicitly through an impression event or inferred from variant-specific URLs, metadata, or page content. Before observing signals from the current experiment, GA's conditional belief about the customer's segment is
\[
\Pr(S_i=s\mid H_i^{GA},\Omega,G_i=1).
\]
This prior incorporates GA's historical information and any information conveyed by the active observation channel. The current experiment can then generate two further updates: one from the displayed output and, when observed, one from the experiment-specific response.

\subsection{Privacy risk from experimental output}
First, GA may observe the displayed experimental output \(A_i\). This corresponds to the case in which the firm's measurement architecture reveals the experimental arm to GA. For architectures in which \(A_i\) is observed by GA, the corresponding output-update factor is
\[
R_A(a)
=
\frac{
\Pr(A_i=a\mid H_i^{GA},\Omega,G_i=1,S_i=s)
}{
\Pr(A_i=a\mid H_i^{GA},\Omega,G_i=1,S_i=s')
}.
\]
This privacy risk is controlled by the privacy-preserving assignment mechanism in Equation \eqref{eq:differential-greedy}. Conditional on GA's historical information, the measurement environment, and the GA channel being active, we bound the posterior-odds multiplier from observing the displayed arm:
\[
R_A(a)\leq e^\gamma.
\]

\subsection{Privacy risk from behavior after experimental output}
Second, if GA also observes behavior directly related to the experimental output, this behavior may create an additional source of posterior updating. For example, after observing that customer \(i\) was shown arm \(A_i=a\), GA may also observe whether the customer clicked on that experimental arm. Conditional on the displayed arm, the experiment-specific behavioral-response factor is
\[
R_{Y^E}(y^E)
=
\frac{
\Pr(Y_i^E=y^E\mid H_i^{GA},\Omega,G_i=1,A_i=a,S_i=s)
}{
\Pr(Y_i^E=y^E\mid H_i^{GA},\Omega,G_i=1,A_i=a,S_i=s')
}.
\]
This factor measures how informative the customer's experiment-specific response is once exposure has already occurred. For instance, if customers in segment \(s\) are more likely than customers in segment \(s'\) to click on arm \(a\), then observing a click on the experimental output increases GA's belief that the customer belongs to segment \(s\).

\subsection{The full posterior update}
The model of privacy risk in experimentation is conditional on the GA observation channel being active. If \(G_i=0\), GA receives no current experimental signal, so the current experiment generates no output- or response-based posterior update. Conditional on \(G_i=1\), the posterior update depends on which experimental signals are observed under the measurement architecture \(\Omega\). If GA observes the displayed arm but not the experiment-specific response, then
\[
\frac{
\Pr(S_i=s\mid H_i^{GA},\Omega,G_i=1,A_i=a)
}{
\Pr(S_i=s'\mid H_i^{GA},\Omega,G_i=1,A_i=a)
}
=
\frac{
\Pr(S_i=s\mid H_i^{GA},\Omega,G_i=1)
}{
\Pr(S_i=s'\mid H_i^{GA},\Omega,G_i=1)
}
R_A(a).
\]

If GA observes both the displayed arm and the raw experiment-specific response, then
\[
\begin{aligned}
&\underbrace{
\frac{
\Pr(S_i=s\mid H_i^{GA},\Omega,G_i=1,A_i=a,Y_i^E=y^E)
}{
\Pr(S_i=s'\mid H_i^{GA},\Omega,G_i=1,A_i=a,Y_i^E=y^E)
}
}_{\text{posterior odds}}
\\[0.75em]
&\qquad =
\underbrace{
\frac{
\Pr(S_i=s\mid H_i^{GA},\Omega,G_i=1)
}{
\Pr(S_i=s'\mid H_i^{GA},\Omega,G_i=1)
}
}_{\text{prior odds conditional on access}}
\underbrace{R_A(a)}_{\text{experimental-output privacy risk}\quad}
\underbrace{R_{Y^E}(y^E)}_{\text{experiment-specific behavior privacy risk}}.
\end{aligned}
\]

\section{Differential privacy in the context of bandits}
\label{sec:dp-bandits}

A randomized mechanism~$\mathcal{M}$ satisfies $\xi$-differential privacy if, for any two neighboring datasets $D$ and $D'$ that differ in at most one individual, and for all measurable events~$S$ in the output space,
\begin{equation}\label{eq:dp-definition}
\mathbb{P}(\mathcal{M}(D)\in S)
\le e^{\xi}\,\mathbb{P}(\mathcal{M}(D')\in S).
\end{equation}

In our setting, let $D=(a_1,\dots,a_T),$ where $a_t\in\{0,1\}$ denotes the non-private arm label associated with visitor $t$. A neighboring dataset is $D'=(a_1,\dots,a_t',\dots,a_T),$ which differs from $D$ only in the $t$-th entry, with $a_t\neq a_t'$.

At round $t$, the $\varepsilon$-greedy rule in Equation~\eqref{eq:epsilon-greedy} induces a random mapping from the non-private arm label $a_t$ to the displayed arm $a_t^*$. Applying this mapping independently across visitors yields the overall mechanism
\[
\mathcal{M}(D)=(a_1^*,\dots,a_T^*).
\]
Under this conditional independence assumption, the joint output distribution factorizes across visitors, so the privacy condition in Equation~\eqref{eq:dp-definition} becomes
\begin{equation}\label{eq:dp-joint}
\mathbb{P}\big(\mathcal{M}(a_1,\dots,a_T)=(a_1^*,\dots,a_T^*)\big)
\le e^{\xi}\,
\mathbb{P}\big(\mathcal{M}(a_1,\dots,a_t',\dots,a_T)=(a_1^*,\dots,a_T^*)\big).
\end{equation}

Because $D$ and $D'$ differ only in the $t$-th entry, all terms except the $t$-th cancel in the likelihood ratio. Hence, the privacy requirement reduces to the one-visitor condition
\begin{equation}\label{eq:difpriv}
\mathbb{P}\big(a_t^*=j \mid a_t=0\big)
\le e^{\xi}\,
\mathbb{P}\big(a_t^*=j \mid a_t=1\big),
\qquad \forall j\in\{0,1\}.
\end{equation}

Equation~\eqref{eq:difpriv} implies that observing a particular displayed output does not allow an outside observer to sharply distinguish between the two possible underlying visitor types. Even if Jamie's underlying arm label were different, the probability of observing the same displayed output could change by at most a factor of $e^{\xi}$.

\section{Parallel composition}\label{app:parallel}
For $T$ visitors, let $A=(a_1,\dots,a_T)$ denote the vector of (possibly adaptive) arm choices, where $a_t \in \{1,\dots,k\}$ is the arm selected by the bandit at round $t$. The displayed arm is generated by a per-visitor randomizer $\mathcal{M}_t:\; a_t \mapsto a_t^*$, implemented via the stochastic matrix $P_t$ (see Equation~\eqref{eq:matrix}). We assume the privacy schedule $(\xi_t)_{t=1}^T$ (equivalently, $(P_t)_{t=1}^T$) is fixed ex ante.  
Let $q_t(\cdot\mid a)$ denote the conditional probability mass function of $\mathcal{M}_t(a)$, i.e.,
$q_t(a_t^*\mid a_t) = (P_t)_{a_t^*,a_t}$.

We make the following assumptions:

\begin{enumerate}
\item[A1.] \textbf{Per-visitor (local) DP.} For all $a,a'$ in the arm domain and all measurable $S_t \subseteq \mathrm{Range}(\mathcal{M}_t)$,
\begin{equation}
\Pr[\mathcal{M}_t(a)\in S_t] \;\le\; e^{\xi_t}\, \Pr[\mathcal{M}_t(a')\in S_t].
\label{eq:dp-t}   
\end{equation}
Equivalently (for discrete outputs), for all possible displayed arms $a_t^*\in\{1,\dots,k\}$,
\[
q_t(a_t^*\mid a) \;\le\; e^{\xi_t}\, q_t(a_t^*\mid a'), 
\qquad \forall a,a'\in\{1,\dots,k\}.
\]

\item[A2.] \textbf{Disjointness and fresh randomness (conditional independence).}
At each round $t$, visitor $t$ contributes a single record: the arm $a_t$ selected by the bandit policy. The policy may be adaptive, so $a_t$ can depend on the past history of privatized displays and rewards. After $a_t$ is chosen, the privacy mechanism $\mathcal{M}_t$ is applied locally to that one record to produce the displayed arm $a_t^*$. In particular, $\mathcal{M}_t$ is memoryless: conditional on the current input $a_t$, the distribution of $a_t^*$ does not depend on past displayed arms or rewards. (Thus, past outputs can influence $a_t$ through the bandit policy, but they do not enter $\mathcal{M}_t$ directly.) 

Moreover, $\mathcal{M}_t$ uses fresh randomization that is independent across visitors and independent of $A$. Consequently, conditional on the realized input vector $A=(a_1,\dots,a_T)$, the privatized displays are independent across visitors:
\[
\Pr(a_1^*,\dots,a_T^* \mid A) \;=\; \prod_{t=1}^T q_t(a_t^*\mid a_t).
\]
This conditional factorization holds even when $A$ is generated adaptively, because conditioning on $A$ fixes the inputs and leaves only the independent privacy randomization in $\{\mathcal{M}_t\}$.
\end{enumerate}

Let $A'$ be a neighboring input vector that differs from $A$ in exactly one coordinate $i$, i.e.,
$A'=(a_1,\dots,a_{i-1},a_i',a_{i+1},\dots,a_T)$ with $a_i'\neq a_i$.
Our goal is to show that for all measurable
$S\subseteq \prod_{t=1}^T \mathrm{Range}(\mathcal{M}_t)$,
\[
\Pr[\mathcal{M}(A)\in S] \;\le\; e^{\gamma}\, \Pr[\mathcal{M}(A')\in S],
\qquad \text{where} \qquad \gamma=\max_t\xi_t.
\]

By Assumption~A2, the probability of any output sequence factorizes:
\[
\Pr[\mathcal{M}(A)\in S]
= \sum_{(a_1^*,\dots,a_T^*)\in S} \prod_{t=1}^T q_t(a_t^*\mid a_t),
\]
and similarly,
\[
\Pr[\mathcal{M}(A')\in S]
= \sum_{(a_1^*,\dots,a_T^*)\in S}
\Big(\prod_{t\neq i} q_t(a_t^*\mid a_t)\Big)\, q_i(a_i^*\mid a_i').
\]
By the single-round DP guarantee in Equation \eqref{eq:dp-t} for $t=i$, we have pointwise for all $a_i^*$:
\[
q_i(a_i^*\mid a_i) \;\le\; e^{\xi_i}\, q_i(a_i^*\mid a_i').
\]
Multiplying by $\prod_{t\neq i} q_t(a_t^*\mid a_t)$ and summing over all $(a_1^*,\dots,a_T^*)\in S$ yields
\[
\Pr[\mathcal{M}(A)\in S] \;\le\; e^{\xi_i}\, \Pr[\mathcal{M}(A')\in S].
\]
Because $A$ and $A'$ may differ at any single coordinate, this implies the uniform guarantee
\[
\Pr[\mathcal{M}(A)\in S] \;\le\; e^{\gamma}\, \Pr[\mathcal{M}(A')\in S],
\qquad \gamma=\max_t \xi_t.
\]
Therefore, the overall mechanism $\mathcal{M}=(\mathcal{M}_1,\dots,\mathcal{M}_T)$ satisfies parallel composition with parameter $\gamma$ over disjoint per-visitor records. The bandit’s adaptivity affects how inputs $A$ are generated but does not affect the privacy accounting, because the privacy guarantee is defined with respect to the per-visitor input record and the released output consists only of the privatized displays $\{a_t^*\}_{t=1}^T$.

\section{Regret analysis}
\label{web:regretanalysis_segment}
We derive the regret bound for the segment-specific $\varepsilon$-greedy policy. Let $\mathcal S$ denote the finite set of consumer segments and let $J=|\mathcal S|$ denote the number of segments. We assume that the firm knows the set of segments but does not know ex ante how many visitors will belong to each segment. Instead, segment membership is observed sequentially as visitors arrive.

For every arm $a_k\in\mathcal A$ and segment $s\in\mathcal S$, the reward is Bernoulli distributed with an unknown stationary mean $\mu_k(s)$. Thus, the firm learns arm effectiveness separately across segments. At visitor $t$, let
\begin{equation}
N_t(s)
=
\sum_{\tau<t}\mathbf{1}\{S_\tau=s\}
\end{equation}
denote the number of previous visitors observed in segment $s$. We define the learning clock for visitor $t$ as
\begin{equation}
\label{eq:segment-clock}
n_t
=
N_t(S_t)+1.
\end{equation}
Hence, $n_t$ is the number of visitors observed so far in the current visitor's segment, including visitor $t$. Importantly, $n_t$ is observed by the firm when visitor $t$ arrives and does not require knowledge of the number of future visitors in that segment.

\subsection{The regret bound of the $\varepsilon$-greedy policy}
To facilitate our analysis, consider a visitor $t$ belonging to segment $S_t=s$. Let $\bar{\mu}_{k,s}$ denote the empirical reward estimate for arm $a_k$ in segment $s$, and let $\mu_k(s)$ denote its true expected reward. We define the clean event $C_t$ as the event that all arm-specific empirical means for the current segment are sufficiently close to their true means:
\begin{equation}
\label{eq:clean-event}
C_t
=
\left\{
\left|
\bar{\mu}_{k,S_t}
-
\mu_k(S_t)
\right|
\leq r_t,
\quad
\forall k\in\{1,\ldots,K\}
\right\},
\end{equation}
where $r_t$ denotes the confidence radius. The bad event $\bar C_t$ is the complement of the clean event.

Under uniform exploration, each arm is selected with probability $\varepsilon_t/K$. Because learning occurs separately within each segment, the relevant number of opportunities for learning about the current segment is $n_t$, rather than the total experiment-wide visitor count $t$. Therefore, each arm receives exploratory observations at a rate of order
\begin{equation}
\frac{n_t\varepsilon_t}{K}.
\end{equation}

More precisely, because $\varepsilon_t$ decreases with the number of observations from a segment, the expected cumulative number of exploratory draws of a given arm by local round $n_t$ is at least of order $n_t\varepsilon_t/K$. Standard concentration inequalities for the exploration counts, combined with Hoeffding's inequality for Bernoulli rewards, therefore give a confidence radius of order
\begin{equation}
\label{eq:confidence-radius}
r_t
=
\sqrt{
\frac{2K\log n_t}
{n_t\varepsilon_t}
},
\end{equation}
up to universal constants. The finite initialization period and the corresponding concentration failure probabilities do not affect the asymptotic regret rate.

Let
\begin{equation}
a^\star(S_t)
=
\arg\max_{a_k\in\mathcal A}
\mu_k(S_t)
\end{equation}
denote the truly optimal arm for the current visitor's segment, and let
\begin{equation}
a_t
=
\arg\max_{a_k\in\mathcal A}
\bar{\mu}_{k,S_t}
\end{equation}
denote the greedy arm selected using the current empirical estimates. On the clean event, because $a_t$ maximizes the empirical reward,
\[
\bar{\mu}_{a_t,S_t}
\geq
\bar{\mu}_{a^\star(S_t),S_t}.
\]
It follows that
\begin{align}
\mu_{a^\star(S_t)}(S_t)-\mu_{a_t}(S_t)
&=
\mu_{a^\star(S_t)}(S_t)
-
\bar{\mu}_{a^\star(S_t),S_t}
\nonumber\\
&\quad+
\bar{\mu}_{a^\star(S_t),S_t}
-
\bar{\mu}_{a_t,S_t}
\nonumber\\
&\quad+
\bar{\mu}_{a_t,S_t}
-
\mu_{a_t}(S_t)
\nonumber\\
&\leq
2r_t
\nonumber\\
&=
2
\sqrt{
\frac{2K\log n_t}
{n_t\varepsilon_t}
}.
\label{eq:segment-exploitation-bound}
\end{align}

Thus, conditional on exploitation, the regret from selecting the empirically best arm rather than the truly optimal arm for the current segment is bounded by the estimation error. Because rewards are bounded between zero and one, exploration generates at most one unit of regret. The expected instantaneous regret at visitor $t$ is therefore bounded by
\begin{align}
\mathbb E[\widetilde R(t)]
&\leq
\Pr(\text{explore})\cdot 1
+
\Pr(\text{exploit})
\cdot
2
\sqrt{
\frac{2K\log n_t}
{n_t\varepsilon_t}
}
\nonumber\\
&=
\varepsilon_t
+
(1-\varepsilon_t)
2
\sqrt{
\frac{2K\log n_t}
{n_t\varepsilon_t}
}
\nonumber\\
&\leq
\varepsilon_t
+
2
\sqrt{
\frac{2K\log n_t}
{n_t\varepsilon_t}
}.
\label{eq:reg}
\end{align}

The final inequality uses only $1-\varepsilon_t\leq1$. Thus, Equation~\eqref{eq:reg} does not require the asymptotic approximation $1-\varepsilon_t\approx1$. To minimize the order of the expected regret, we balance the exploration and exploitation terms:
\begin{equation}
\varepsilon_t
\asymp
\sqrt{
\frac{K\log n_t}
{n_t\varepsilon_t}
}.
\end{equation}
This gives
\begin{equation}
\varepsilon_t^3
\asymp
\frac{K\log n_t}{n_t},
\end{equation}
and therefore
\begin{equation}
\label{eq:optimal-epsilon}
\varepsilon_t
=
\left(
\frac{K\log n_t}{n_t}
\right)^{1/3},
\end{equation}
where we suppress universal multiplicative constants that do not affect the regret rate. Substituting Equation~\eqref{eq:optimal-epsilon} into Equation~\eqref{eq:reg} gives
\begin{equation}
\label{eq:per-round-regret}
\mathbb E[\widetilde R(t)]
=
\mathcal O\left(
\frac{
K^{1/3}(\log n_t)^{1/3}
}{
n_t^{1/3}
}
\right).
\end{equation}

We next obtain an experiment-wide bound that does not require the firm to know how many visitors belong to each segment. Let $T_s$ denote, only for purposes of the ex-post regret analysis, the realized number of visitors in segment $s$. Summing Equation~\eqref{eq:per-round-regret} over the local learning rounds of all segments gives
\begin{align}
\mathbb E[R(T)]
&\leq
\mathcal O\left(
K^{1/3}
\sum_{s\in\mathcal S}
\sum_{n=1}^{T_s}
\frac{(\log n)^{1/3}}{n^{1/3}}
\right)
\nonumber\\
&\leq
\mathcal O\left(
K^{1/3}
(\log T)^{1/3}
\sum_{s\in\mathcal S}
T_s^{2/3}
\right).
\end{align}

Because $\sum_{s\in\mathcal S}T_s=T$ and $x^{2/3}$ is concave,
\begin{equation}
\sum_{s\in\mathcal S}T_s^{2/3}
\leq
J^{1/3}T^{2/3}.
\end{equation}
Hence, the cumulative regret is bounded by
\begin{equation}
\label{eq:cumulative-regret}
\mathbb E[R(T)]
=
\mathcal O\left(
K^{1/3}
J^{1/3}
T^{2/3}
(\log T)^{1/3}
\right).
\end{equation}

Importantly, Equation~\eqref{eq:cumulative-regret} does not require the firm to know the number or proportion of visitors in each segment in advance. The firm only needs to observe the current visitor's segment and the corresponding learning clock $n_t$. The factor $J^{1/3}$ reflects the statistical cost of segment-specific learning: because arm effectiveness may differ across segments, observations cannot generally be pooled across segments.

\subsection{The regret bound under differential privacy}
\label{web:regret_difprivacy}
We next incorporate differential privacy into the regret analysis. The privacy guarantee itself is established in Web Appendix~\ref{sec:dp-bandits}. Here, we use the mapping between the exploration probability $\varepsilon_t$ and the visitor-level privacy-risk parameter $\xi_t$. Under the $K$-ary randomized-response mechanism,
\begin{equation}
\label{eq:epsilon-xi-regret}
\varepsilon_t
=
\frac{K}
{K+e^{\xi_t}-1}.
\end{equation}

To make the randomization required by differential privacy coincide with the
exploration probability that balances learning regret, we equate
Equation~\eqref{eq:epsilon-xi-regret} with
Equation~\eqref{eq:optimal-epsilon}:
\begin{equation}
\frac{K}
{K+e^{\xi_t}-1}
=
\left(
\frac{K\log n_t}{n_t}
\right)^{1/3}.
\end{equation}
Solving for $\xi_t$ yields
\begin{equation}
\label{eq:xi-optimal}
\xi_t
=
\log\left[
1-K
+
\frac{K}{
\left(
\frac{K\log n_t}{n_t}
\right)^{1/3}
}
\right].
\end{equation}

Thus, the privacy-risk parameter depends on the amount of information
accumulated about the current visitor's segment, measured by $n_t$, rather
than on the total number of visitors in the experiment.

Substituting Equation~\eqref{eq:epsilon-xi-regret} into the per-round regret
bound in Equation~\eqref{eq:reg} gives
\begin{align}
\mathbb E[\widetilde R(t)]
&\leq
\frac{K}
{K+e^{\xi_t}-1}
+
2
\sqrt{
\frac{
2K\log n_t
}{
n_t\cdot
\frac{K}{K+e^{\xi_t}-1}
}
}
\nonumber\\
&=
\frac{K}
{K+e^{\xi_t}-1}
+
2
\sqrt{
\frac{
2(K+e^{\xi_t}-1)\log n_t
}{
n_t
}
}.
\label{eq:regret_constant_difP}
\end{align}

When $\xi_t$ is set according to Equation~\eqref{eq:xi-optimal}, the induced
exploration probability is
\[
\varepsilon_t
=
\left(
\frac{K\log n_t}{n_t}
\right)^{1/3}.
\]
Substituting this value into Equation~\eqref{eq:regret_constant_difP} yields
\begin{align}
\mathbb E[\widetilde R(t)]
&\leq
\frac{
K^{1/3}(\log n_t)^{1/3}
}{
n_t^{1/3}
}
+
2\sqrt{2}
\frac{
K^{1/3}(\log n_t)^{1/3}
}{
n_t^{1/3}
}
\nonumber\\
&=
(1+2\sqrt{2})
\frac{
K^{1/3}(\log n_t)^{1/3}
}{
n_t^{1/3}
}
\nonumber\\
&=
\mathcal O\left(
\frac{
K^{1/3}(\log n_t)^{1/3}
}{
n_t^{1/3}
}
\right).
\end{align}

Summing over all visitors and applying the same argument used in
Equation~\eqref{eq:cumulative-regret} gives
\begin{equation}
\label{eq:cumulative-private-regret}
\mathbb E[R(T)]
=
\mathcal O\left(
K^{1/3}
J^{1/3}
T^{2/3}
(\log T)^{1/3}
\right).
\end{equation}

Thus, when $\xi_t$ is chosen such that the randomization required for privacy coincides with the exploration already required by the segment-specific $\varepsilon$-greedy policy, differential privacy does not change the order of the regret bound. The resulting bound differs from the global-bandit case through the factor $J^{1/3}$, which arises from segment-specific learning rather than from differential privacy.

This equivalence holds for the privacy schedule in Equation~\eqref{eq:xi-optimal}. A stricter privacy requirement can force an exploration probability above the regret-balancing level and therefore
increase regret.

\subsection{The regret bound without asymptotic approximation}

Equation~\eqref{eq:reg} does not require the approximation
$1-\varepsilon_t\approx1$. Retaining the exploitation probability explicitly,
the finite-round regret bound is
\begin{equation}
\label{eq:finite-regret}
\mathbb E[\widetilde R(t)]
\leq
\varepsilon_t
+
2(1-\varepsilon_t)
\sqrt{
\frac{2K\log n_t}
{n_t\varepsilon_t}
}.
\end{equation}

Using
\[
\varepsilon_t
=
\frac{K}{K+e^{\xi_t}-1}
\]
and
\[
1-\varepsilon_t
=
\frac{e^{\xi_t}-1}
{K+e^{\xi_t}-1},
\]
Equation~\eqref{eq:finite-regret} becomes
\begin{align}
\mathbb E[\widetilde R(t)]
&\leq
\frac{K}
{K+e^{\xi_t}-1}
\nonumber\\
&\quad+
2
\frac{e^{\xi_t}-1}
{K+e^{\xi_t}-1}
\sqrt{
\frac{
2(K+e^{\xi_t}-1)\log n_t
}{
n_t
}
}.
\label{eq:finite-private-regret}
\end{align}

Thus, the finite-round regret can be evaluated without assuming that
$1-\varepsilon_t$ converges to one. The simpler bound used above follows
directly from $1-\varepsilon_t\leq1$.

\section{Regret of the dynamic strategy}
\label{web:regret_dynamic}
We next derive the regret of the dynamic privacy strategy under segment-specific learning. Recall that
\[
n_t
=
N_t(S_t)+1
\]
denotes the number of visitors observed so far in the current visitor's segment, including visitor \(t\). Thus, \(n_t\), rather than the global experiment round \(t\), determines how much the firm has learned about the current segment. The firm observes \(n_t\) when visitor \(t\) arrives and does not need to know how many visitors will ultimately belong to that segment. We cap the per-visitor privacy-risk parameter at a target level \(\gamma\):
\begin{equation}
\label{eq:dynamic-xi-cap}
\xi_t
=
\min\left\{
\gamma,\,
\log\left[
1-K+
\frac{K}{
\left(
\frac{K\log n_t}{n_t}
\right)^{1/3}
}
\right]
\right\}.
\end{equation}

Let \(\varepsilon_\gamma\) denote the exploration probability implied by
\(\gamma\):
\begin{equation}
\label{eq:epsilon-gamma}
\varepsilon_\gamma
:=
\frac{K}{K+e^\gamma-1}.
\end{equation}

Under the uncapped dynamic schedule, the exploration probability for visitor
\(t\) is
\begin{equation}
\label{eq:dynamic-exploration}
\varepsilon_t^{\mathrm{dyn}}
=
\left(
\frac{K\log n_t}{n_t}
\right)^{1/3}.
\end{equation}
As elsewhere, this expression applies after the finite initialization period and is truncated at one whenever necessary.

Because the mapping from \(\xi\) to \(\varepsilon\) is decreasing, imposing \(\xi_t\leq\gamma\) is equivalent to imposing \(\varepsilon_t\geq\varepsilon_\gamma\). Hence, the capped dynamic strategy can equivalently be written as
\begin{equation}
\label{eq:capped-dynamic-exploration}
\varepsilon_t
=
\frac{K}{K+e^{\xi_t}-1}
=
\max\left\{
\varepsilon_t^{\mathrm{dyn}},
\varepsilon_\gamma
\right\}.
\end{equation}

\subsection{The local exhaustion threshold}
Under segment-specific learning, there is no single experiment-wide exhaustion time at which the privacy cap begins to bind for all visitors. Instead, define the local exhaustion threshold
\begin{equation}
\label{eq:n-gamma-def}
n_\gamma
:=
\min\left\{
n:
\left(
\frac{K\log n}{n}
\right)^{1/3}
\leq
\varepsilon_\gamma
\right\}.
\end{equation}
Thus, \(n_\gamma\) is the number of observations within a segment at which the unconstrained learning schedule would first require less exploration than the privacy cap permits.

Equivalently, on the decreasing region \(n>e\), \(n_\gamma\) solves
\begin{equation}
\label{eq:n-gamma-equation}
\left(
\frac{K\log n_\gamma}{n_\gamma}
\right)^{1/3}
=
\varepsilon_\gamma
\qquad\Longleftrightarrow\qquad
\frac{\log n_\gamma}{n_\gamma}
=
\frac{\varepsilon_\gamma^3}{K}.
\end{equation}
Because \(\log n/n\) is strictly decreasing for \(n>e\), the relevant post-initialization solution is unique and can be obtained numerically.

When all segments share the same \(K\) arms and the same privacy cap \(\gamma\), the local threshold \(n_\gamma\) is common across segments. However, different segments reach this threshold at different experiment-wide times. For segment \(s\), define
\begin{equation}
\label{eq:segment-global-exhaustion}
\tau_{s,\gamma}
=
\inf\left\{
t:
S_t=s
\text{ and }
N_t(s)+1\geq n_\gamma
\right\},
\end{equation}
with \(\tau_{s,\gamma}=\infty\) if segment \(s\) does not reach the threshold within the experimental horizon. Frequently observed segments therefore tend to reach the privacy cap earlier in experiment-wide time, whereas rare segments remain on the unconstrained learning schedule for longer.

Importantly, the firm does not need to know \(\tau_{s,\gamma}\) in advance. At each visitor, it determines whether the privacy cap binds simply by comparing the observed local learning clock
\(n_t\) with \(n_\gamma\).

\subsection{Pre-cap regret}

Consider visitors for whom the current segment has not yet moved beyond the local exhaustion threshold. For these visitors,
\[
\varepsilon_t
=
\left(
\frac{K\log n_t}{n_t}
\right)^{1/3}.
\]
Substituting this expression into the per-round regret bound in Equation~\eqref{eq:reg} gives
\begin{equation}
\label{eq:precap-perround}
\mathbb E[\widetilde R(t)]
\leq
(1+2\sqrt{2})
\left(
\frac{K\log n_t}{n_t}
\right)^{1/3}.
\end{equation}

Let \(J=|\mathcal S|\) denote the number of consumer segments. Each segment can contribute at most \(n_\gamma\) observations before moving into the capped region. Hence, the total number of pre-cap observations is at most
\[
\min\{T,Jn_\gamma\}.
\]
Summing the segment-specific regret terms and applying the same concavity argument used in the general regret analysis yields
\begin{equation}
\label{eq:precap-regret}
\mathbb E[R_{\mathrm{pre}}(T)]
=
\mathcal O\left(
K^{1/3}
J^{1/3}
\min\{T,Jn_\gamma\}^{2/3}
(\log T)^{1/3}
\right).
\end{equation}

This bound requires no knowledge of the eventual number or proportion of visitors in each segment.

\subsection{Post-cap regret}

Once the current segment has moved beyond \(n_\gamma\), the privacy cap binds and the exploration probability is fixed at
\[
\varepsilon_t=\varepsilon_\gamma.
\]
The per-round regret bound is therefore
\begin{equation}
\label{eq:postcap-perround}
\mathbb E[\widetilde R(t)]
\leq
\underbrace{\varepsilon_\gamma}_{\text{exploration regret}}
+
\underbrace{
2\sqrt{
\frac{2K\log n_t}
{n_t\varepsilon_\gamma}
}
}_{\text{estimation error}}.
\end{equation}

Let \(M_\gamma\) denote the total number of visitors whose segment-specific learning clock exceeds \(n_\gamma\) during the experiment. The cumulative exploration cost after the cap is therefore at most
\[
M_\gamma\varepsilon_\gamma.
\]

For the estimation-error component, grouping visitors by segment and summing over their local learning clocks gives
\begin{align}
&
2\sqrt{\frac{2K}{\varepsilon_\gamma}}
\sum_{\substack{t:\,n_t>n_\gamma}}
\sqrt{\frac{\log n_t}{n_t}}
\nonumber\\
&\qquad\leq
4\sqrt{2}
\sqrt{
\frac{KJ M_\gamma\log T}
{\varepsilon_\gamma}
}.
\label{eq:postcap-estimation}
\end{align}
The inequality follows from an integral comparison within each segment and the Cauchy--Schwarz inequality across the \(J\) segments.

Consequently,
\begin{equation}
\label{eq:postcap-regret}
\mathbb E[R_{\mathrm{post}}(T)]
\leq
M_\gamma\varepsilon_\gamma
+
4\sqrt{2}
\sqrt{
\frac{KJ M_\gamma\log T}
{\varepsilon_\gamma}
}.
\end{equation}

To obtain a bound that does not depend on the realized segment sequence, note that if any visitor enters the post-cap region, at least \(n_\gamma\) observations must first have been accumulated within some segment. Hence,
\[
M_\gamma
\leq
(T-n_\gamma)_+,
\qquad
(x)_+:=\max\{x,0\}.
\]
Therefore,
\begin{equation}
\label{eq:postcap-regret-simple}
\mathbb E[R_{\mathrm{post}}(T)]
\leq
(T-n_\gamma)_+\varepsilon_\gamma
+
\mathcal O\left(
\sqrt{
\frac{
KJ(T-n_\gamma)_+\log T
}{
\varepsilon_\gamma
}
}
\right).
\end{equation}

Combining the pre-cap and post-cap components gives the following simple bound for the dynamic strategy:
\begin{equation}
\label{eq:dynamic-total-regret}
\boxed{
\begin{aligned}
\mathbb E[R_{\mathrm{dyn}}(T)]
\leq\;&
\mathcal O\left(
K^{1/3}
J^{1/3}
\min\{T,Jn_\gamma\}^{2/3}
(\log T)^{1/3}
\right)
\\
&+
(T-n_\gamma)_+\varepsilon_\gamma
\\
&+
\mathcal O\left(
\sqrt{
\frac{
KJ(T-n_\gamma)_+\log T
}{
\varepsilon_\gamma
}
}
\right).
\end{aligned}
}
\end{equation}

The first term captures regret while segments remain on the unconstrained learning schedule. The second term captures the cumulative cost of the random exploration required after the privacy constraint becomes binding. The third term captures estimation error during exploitation in the capped region.

A larger value of \(\gamma\), corresponding to a weaker privacy guarantee, implies a smaller \(\varepsilon_\gamma\). This reduces the amount of forced random exploration but increases the estimation-error component because fewer observations are generated through exploration. Conversely, a smaller \(\gamma\) provides stronger privacy protection and forces more exploration. The dynamic strategy balances these effects by following the regret-balancing exploration schedule until the privacy cap becomes binding within each segment.

\section{Regret of the constant strategy}
\label{app:constant}

Under the constant strategy, the per-visitor privacy-risk parameter is fixed at
\(\gamma\):
\begin{equation}
\xi_t
\equiv
\gamma,
\end{equation}
which implies the constant exploration probability
\begin{equation}
\label{eq:constant-epsilon}
\varepsilon_t
\equiv
\varepsilon_\gamma
:=
\frac{K}{K+e^\gamma-1}.
\end{equation}

Although the randomization probability is constant across visitors, learning remains segment-specific. For visitor \(t\), recall that
\[
n_t=N_t(S_t)+1
\]
denotes the number of observations accumulated for the current visitor's segment. Define the clean event
\begin{equation}
C_t
=
\left\{
\left|
\bar{\mu}_{k,S_t}
-
\mu_k(S_t)
\right|
\leq r_t,
\quad
\forall k\in\{1,\ldots,K\}
\right\},
\end{equation}
with confidence radius
\begin{equation}
r_t
=
\sqrt{
\frac{2K\log n_t}
{n_t\varepsilon_\gamma}
}.
\end{equation}

Using the same concentration argument as in Section~\ref{web:regretanalysis}, expected instantaneous regret satisfies
\begin{equation}
\label{eq:constant-perround}
\mathbb E[\widetilde R(t)]
\leq
\varepsilon_\gamma
+
2
\sqrt{
\frac{2K\log n_t}
{n_t\varepsilon_\gamma}
}.
\end{equation}

Summing the exploration component over all \(T\) visitors gives
\[
T\varepsilon_\gamma.
\]
For the estimation-error component, summing over the segment-specific learning clocks and applying an integral comparison gives
\begin{align}
&
2\sqrt{\frac{2K}{\varepsilon_\gamma}}
\sum_{t=1}^{T}
\sqrt{\frac{\log n_t}{n_t}}
\nonumber\\
&\qquad\leq
4\sqrt{2}
\sqrt{
\frac{KJT\log T}
{\varepsilon_\gamma}
},
\end{align}
where the final inequality follows from the Cauchy--Schwarz inequality across the \(J\) consumer segments.

Consequently, cumulative regret under the constant strategy satisfies
\begin{equation}
\label{eq:constant-regret}
\boxed{
\mathbb E[R_{\mathrm{const}}(T)]
\leq
T\varepsilon_\gamma
+
4\sqrt{2}
\sqrt{
\frac{KJT\log T}
{\varepsilon_\gamma}
}.
}
\end{equation}

The first term,
\[
T\varepsilon_\gamma,
\]
is the cumulative cost of random exploration. Because the constant strategy explores with probability \(\varepsilon_\gamma\) at every visitor, this term grows linearly with the experimental horizon for any fixed \(\varepsilon_\gamma>0\).

The second term captures regret from estimation error when exploiting. Importantly, this cost is not constant. Even though the privacy parameter and exploration probability are fixed, estimation error decreases as the firm accumulates more observations within each segment. The factor \(J^{1/2}\) reflects the statistical cost of estimating separate arm-specific reward distributions across the \(J\) consumer segments.

Unlike the dynamic strategy, the constant strategy does not adapt the amount of randomization to the amount of information accumulated about the current segment. Consequently, it does not generally achieve the same regret rate as the segment-specific regret-balancing $\varepsilon$-greedy policy. For a fixed privacy cap, the linear exploration term \(T\varepsilon_\gamma\) eventually dominates the cumulative regret.

\section{Regret of the dynamic strategy}\label{web:regret_dynamic}

We cap the per-visitor privacy controller at a target budget $\gamma$, i.e.,
\[
\xi_t=\min\!\left\{\gamma,\ \log\!\left(1-k+\frac{k}{\bigl(\frac{k\log t}{t}\bigr)^{1/3}}\right)\right\}.
\]
Let $\varepsilon_\gamma$ denote the exploration probability implied by the cap $\gamma$ via Equation~\eqref{eq:exploration},
\[
\varepsilon_\gamma \;:=\; \frac{k}{k+e^{\gamma}-1}.
\]
Under the uncapped dynamic schedule in Equation~\eqref{privacystrategy}, the implied exploration rate is
\[
\varepsilon_t^{\mathrm{dyn}} \;:=\; \Bigl(\tfrac{k\log t}{t}\Bigr)^{1/3}.
\]
Because $\xi\mapsto \varepsilon(\xi)$ is decreasing, imposing $\xi_t\le \gamma$ is equivalent to imposing
$\varepsilon_t \ge \varepsilon_\gamma$. Hence the capped dynamic strategy can be written as
\[
\varepsilon_t
=\frac{k}{k+e^{\xi_t}-1}
=\max\!\left\{\varepsilon_t^{\mathrm{dyn}},\ \varepsilon_\gamma\right\}.
\]

Let $t_\gamma$ be the (implicit) exhaustion time, defined as the first round at which the uncapped schedule would call for less exploration than the cap permits, i.e.,
\[
t_\gamma \;:=\; \min\Bigl\{t\in\{1,\dots,T\}:\ \varepsilon_t^{\mathrm{dyn}}\le \varepsilon_\gamma\Bigr\},
\]
with the convention that $t_\gamma=T+1$ if the set is empty (i.e., if the cap never takes effect within the horizon).
Equivalently, $t_\gamma$ solves
\begin{equation}\label{eq:t-gamma-def}
\Bigl(\tfrac{k\log t_\gamma}{t_\gamma}\Bigr)^{1/3}=\varepsilon_\gamma
\quad\Longleftrightarrow\quad
\frac{\log t_\gamma}{t_\gamma}=\frac{\varepsilon_\gamma^3}{k}.
\end{equation}
For $t>e$, the function $\log t/t$ is strictly decreasing, so when $t_\gamma\le T$ the solution is unique and can be obtained numerically (e.g., by one-dimensional root finding).

\subsection{Pre-cap regret ($t\le t_\gamma$).}

Before we hit the cap, the regret bound follows the optimal $\varepsilon$-greedy regret.
For $t\le t_\gamma$, we have $\varepsilon_t=(k\log t/t)^{1/3}$.
Plugging into Equation \eqref{eq:reg}, the per-round regret up to the cap $t_\gamma$ becomes: 
\[
\mathbb{E}[\tilde R(t)]
\;\le\;
\Bigl(\tfrac{k\log t}{t}\Bigr)^{1/3}
\;+\;2\sqrt{2}\,\Bigl(\tfrac{k\log t}{t}\Bigr)^{1/3}
\;=\;
(1+2\sqrt{2})\Bigl(\tfrac{k\log t}{t}\Bigr)^{1/3}.
\]
Summing and using a standard integral comparison, the cumulative regret up to the cap $t_\gamma$ becomes: 
\begin{equation}\label{eq:precap-sum}
\sum_{t=1}^{t_\gamma}\mathbb{E}[\tilde R(t)]
\;\lesssim\;
k^{1/3}\,t_\gamma^{2/3}(\log t_\gamma)^{1/3}.
\end{equation}

\subsection{Post-cap regret ($t>t_\gamma$).}
For $t > t_\gamma$, the exploration probability $\varepsilon_t = \varepsilon_\gamma$ is constant. 
From the per-round regret bound in Equation~\eqref{eq:reg}, we have
\[
\mathbb{E}[\tilde R(t)]
\;\le\;
\underbrace{\varepsilon_\gamma}_{\text{exploration regret}}
\;+\;
\underbrace{2\sqrt{\frac{2k\log t}{t\,\varepsilon_\gamma}}}_{\text{estimation (exploitation) regret}}.
\]
Summing over $t = t_\gamma + 1, \dots, T$ gives
\begin{equation}\label{eq:sum-start}
\sum_{t=t_\gamma+1}^{T}\mathbb{E}[\tilde R(t)]
\;\le\;
(T-t_\gamma)\,\varepsilon_\gamma
\;+\;
2\sqrt{\frac{2k}{\varepsilon_\gamma}}
\sum_{t=t_\gamma+1}^{T}\sqrt{\frac{\log t}{t}}.
\end{equation}

\noindent We now upper bound the summation term which bounds the exploitation regret to:
\[
S := \sum_{t=t_\gamma+1}^{T}\sqrt{\frac{\log t}{t}}.
\]

\noindent Since $\log t$ is monotone increasing, for all $t \le T$ we have
\[
\sqrt{\frac{\log t}{t}}
\;\le\;
\sqrt{\log T}\cdot \frac{1}{\sqrt{t}}.
\]
Hence,
\[
S
\;\le\;
\sqrt{\log T}\sum_{t=t_\gamma+1}^{T}\frac{1}{\sqrt{t}}
\;\le\;
\sqrt{\log T}\int_{t_\gamma}^{T}\frac{dx}{\sqrt{x}}
\;=\;
2\sqrt{\log T}\,\big(\sqrt{T}-\sqrt{t_\gamma}\big).
\]

\noindent Therefore,
\begin{equation}\label{eq:sum-bound-A}
S
\;\le\;
2\sqrt{\log T}\,\big(\sqrt{T}-\sqrt{t_\gamma}\big).
\end{equation}

\noindent Substituting the bound from \eqref{eq:sum-bound-A} into
Equation~\eqref{eq:sum-start} yields
\[
\sum_{t=t_\gamma+1}^{T}\mathbb{E}[\tilde R(t)]
\;\le\;
(T-t_\gamma)\,\varepsilon_\gamma
\;+\;
2\sqrt{\frac{2k}{\varepsilon_\gamma}}\;
\Bigl(2\sqrt{\log T}\,\big(\sqrt{T}-\sqrt{t_\gamma}\big)\Bigr).
\]
Simplifying constants gives the final result:
\begin{equation}\label{eq:postcap-proof}
\sum_{t=t_\gamma+1}^{T}\mathbb{E}[\tilde R(t)]
\;\le\;
\underbrace{(T-t_\gamma)\,\varepsilon_\gamma}_{\text{exploration cost (linear)}}
\;+\;
\underbrace{\mathcal{O}\left(\sqrt{\frac{k}{\varepsilon_\gamma}}\,
\Bigl((\sqrt{T}-\sqrt{t_\gamma})\,\sqrt{\log T}\Bigr)\right)}_{\text{estimation error (sublinear)}}.
\end{equation}

The first (linear) term, $(T-t_\gamma)\varepsilon_\gamma$ represents the cumulative cost of random exploration: each round explores with probability $\varepsilon_\gamma$, and each random arm pull incurs at most unit regret. The second (sublinear) term arises from estimation error when exploiting. Even when the bandit chooses greedily producing expected instantaneous regret of order $\mathcal{O}\big(\sqrt{\frac{k\log t}{t\varepsilon_\gamma}}\big)$. Summing these decaying errors yields a sublinear term proportional to $\sqrt{(T-t_\gamma)\log T}$. Thus, larger privacy budgets $\gamma$ (smaller $\varepsilon_\gamma$) 
reduce the exploration cost but inflate the estimation error term, 
illustrating the privacy–exploration–regret trade-off.

\section{Regret of the constant strategy}\label{app:constant}
Let the privacy controller be capped constantly,
\[
\xi_t \equiv \gamma
\qquad\Longrightarrow\qquad
\varepsilon_t \equiv \varepsilon_\gamma := \frac{k}{\,k+e^{\gamma}-1\,}.
\]
Define the clean event at round $t$ by
\[
C_t := \bigl\{\,|\bar\mu_a-\mu_a|\le r_t,\ \forall a\in[k]\,\bigr\},
\qquad
r_t := \sqrt{\frac{2k\log t}{t\,\varepsilon_\gamma}},
\]
so that by Hoeffding and a union bound
\begin{equation}\label{unionboundd}
\mathbb{E}[\tilde R(t)]
\;\le\;
\varepsilon_\gamma
\;+\;
2\sqrt{\frac{2k\log t}{t\,\varepsilon_\gamma}}    
\end{equation}
Consequently, the cumulative regret satisfies
\[\;
\mathbb{E}[R(T)]
\;\le\
\underbrace{T\,\varepsilon_\gamma}_{\text{cost of exploration}}
\;+\;
\underbrace{4\sqrt{2}\,\sqrt{\frac{k}{\varepsilon_\gamma}}\;\sqrt{T\log T}}_{\text{cost of exploitation}}
\;
\]

Here we do not match the regret bound that ``standard" $\varepsilon$-greedy introduces, because the exploration and exploitation do not adapt over $t$.

In terms of cost of exploration, with a constant privacy cap $\xi_t\equiv\gamma$, you explore each round with probability $\varepsilon_\gamma$. Every exploratory pull can be suboptimal and costs at most 1 unit of regret, so you pay on average $\varepsilon_\gamma$ per round, adding up linearly to $T\,\varepsilon_\gamma$ over $T$ rounds. The cost of exploitation is also constant.

\section{Privacy risk budget elasticities}\label{web:elasticities}
We first consider the constant strategy followed by the dynamic strategy.
\subsection{Constant privacy risk strategy}\label{web:elasticityct}


Let
\[
\varepsilon(\gamma)=\frac{k}{k+e^\gamma-1},\qquad
R(T,\gamma)=T\,\varepsilon(\gamma)+4\sqrt{2}\,\sqrt{\frac{k}{\varepsilon(\gamma)}}\,\sqrt{T\log T}.
\]
Define \(D:=4\sqrt{2}\,\sqrt{kT\log T}\), so that
\[
R \;=\; T\,\varepsilon + D\,\varepsilon^{-1/2}.
\]
We seek the point elasticity of \(R\) with respect to \(\gamma\):
\[
\mathcal{E}_{R,\gamma}\;=\;\frac{d\ln R}{d\ln\gamma}\;=\;\frac{\gamma}{R}\,\frac{dR}{d\gamma}.
\]

By the chain rule,
\[
\frac{dR}{d\gamma}
= \frac{dR}{d\varepsilon} \cdot \frac{d\varepsilon}{d\gamma}
\]

\noindent We therefore obtain the following 
\[
\frac{d\varepsilon}{d\gamma}
=\frac{d}{d\gamma}\Bigl(\frac{k}{k+e^\gamma-1}\Bigr)
=-\frac{k\,e^\gamma}{(k+e^\gamma-1)^2},\qquad
\frac{dR}{d\varepsilon}
=T-\frac{D}{2}\,\varepsilon^{-3/2}.
\]

\noindent Plugging these results back in

\[
\frac{dR}{d\gamma}
= \frac{dR}{d\varepsilon} \cdot \frac{d\varepsilon}{d\gamma} =\Bigl(T-\frac{D}{2}\varepsilon^{-3/2}\Bigr)\Bigl(-\frac{k\,e^\gamma}{(k+e^\gamma-1)^2}\Bigr),
\]
Using the definition of an elasticity
\[\mathcal{E}_{R,\gamma}\;=\;\frac{\gamma}{R}\,\frac{dR}{d\gamma} =-\frac{\gamma}{R}\Bigl(T-\frac{D}{2}\varepsilon^{-3/2}\Bigr)\frac{k\,e^\gamma}{(k+e^\gamma-1)^2}.\]

Using
\(
\varepsilon=\frac{k}{k+e^\gamma-1},\;
\varepsilon^{-1/2}=\sqrt{\frac{k+e^\gamma-1}{k}},\;
\varepsilon^{-3/2}=\Bigl(\frac{k+e^\gamma-1}{k}\Bigr)^{3/2},
\)
we obtain
\[
R=\frac{Tk}{k+e^\gamma-1}
+4\sqrt{2}\,\sqrt{T\log T}\,\sqrt{k+e^\gamma-1},
\]
\[
T-\frac{D}{2}\varepsilon^{-3/2}
=T-\frac{2\sqrt{2}\,\sqrt{T\log T}}{k}\,\bigl(k+e^\gamma-1\bigr)^{3/2}.
\]
Finally, we obtain the elasticity
\[
\mathcal{E}_{R,\gamma}
=
-\frac{\gamma\,e^\gamma\Bigl(Tk-2\sqrt{2}\,\sqrt{T\log T}\,\bigl(k+e^\gamma-1\bigr)^{3/2}\Bigr)}
{\left(\dfrac{Tk}{k+e^\gamma-1}
+4\sqrt{2}\,\sqrt{T\log T}\,\sqrt{k+e^\gamma-1}\right)\,\bigl(k+e^\gamma-1\bigr)^{2}}
\,.
\]

\subsection{Elasticity of the dynamic privacy risk strategy}
\label{web:elasticitydynamic}

We next derive the elasticity of the regret bound under the capped dynamic privacy risk strategy. Recall that the exploration probability implied by the privacy risk cap is
\[
\varepsilon_\gamma
=
\frac{k}{k+e^\gamma-1},
\]
and that the exhaustion time \(t_\gamma\) is implicitly defined by
\begin{equation}
\label{eq:elasticity-tgamma}
\frac{\log t_\gamma}{t_\gamma}
=
\frac{\varepsilon_\gamma^3}{k}.
\end{equation}

Let
\[
A:=1+2\sqrt{2},
\qquad
B:=4\sqrt{2}.
\]
Using the pre-cap regret bound and the post-cap summation bound, the cumulative regret can be approximated by
\begin{equation}
\label{eq:dynamic-regret-elasticity}
\begin{aligned}
R_{\mathrm{dyn}}(T,\gamma)
={}&
\underbrace{
A k^{1/3}t_\gamma^{2/3}
(\log t_\gamma)^{1/3}
}_{\text{pre-cap regret}}
\\
&+
\underbrace{
(T-t_\gamma)\varepsilon_\gamma
}_{\text{post-cap exploration cost}}
\\
&+
\underbrace{
B\sqrt{\frac{k}{\varepsilon_\gamma}}
\left(\sqrt{T}-\sqrt{t_\gamma}\right)
\sqrt{\log T}
}_{\text{post-cap estimation error}}.
\end{aligned}
\end{equation}

Both \(\varepsilon_\gamma\) and \(t_\gamma\) depend on the privacy risk budget \(\gamma\). Therefore, the total derivative of regret is
\begin{equation}
\label{eq:dynamic-total-derivative}
\frac{dR_{\mathrm{dyn}}}{d\gamma}
=
\frac{\partial R_{\mathrm{dyn}}}{\partial t_\gamma}
\frac{dt_\gamma}{d\gamma}
+
\frac{\partial R_{\mathrm{dyn}}}{\partial\varepsilon_\gamma}
\frac{d\varepsilon_\gamma}{d\gamma}.
\end{equation}

We first differentiate the exploration probability:
\begin{equation}
\label{eq:dynamic-deps-dgamma}
\frac{d\varepsilon_\gamma}{d\gamma}
=
-\frac{k e^\gamma}
{\left(k+e^\gamma-1\right)^2}.
\end{equation}

To obtain the derivative of the exhaustion time, differentiate Equation~\eqref{eq:elasticity-tgamma} with respect to \(\gamma\). Since
\[
\frac{d}{dt}
\left(\frac{\log t}{t}\right)
=
\frac{1-\log t}{t^2},
\]
we obtain
\[
\frac{1-\log t_\gamma}{t_\gamma^2}
\frac{dt_\gamma}{d\gamma}
=
\frac{3\varepsilon_\gamma^2}{k}
\frac{d\varepsilon_\gamma}{d\gamma}.
\]
Thus,
\begin{equation}
\label{eq:dynamic-dtgamma-dgamma}
\frac{dt_\gamma}{d\gamma}
=
\frac{3\varepsilon_\gamma^2}{k}
\frac{d\varepsilon_\gamma}{d\gamma}
\frac{t_\gamma^2}{1-\log t_\gamma}.
\end{equation}

We next calculate the partial derivative of regret with respect to \(t_\gamma\). Differentiating Equation~\eqref{eq:dynamic-regret-elasticity}
gives
\begin{equation}
\label{eq:dynamic-partial-t}
\begin{aligned}
\frac{\partial R_{\mathrm{dyn}}}{\partial t_\gamma}
={}&
A k^{1/3}t_\gamma^{-1/3}
\left[
\frac{2}{3}(\log t_\gamma)^{1/3}
+
\frac{1}{3}(\log t_\gamma)^{-2/3}
\right]
\\
&-
\varepsilon_\gamma
-
\frac{B}{2}
\sqrt{\frac{k}{\varepsilon_\gamma}}
\frac{\sqrt{\log T}}{\sqrt{t_\gamma}}.
\end{aligned}
\end{equation}

The partial derivative with respect to the exploration probability is
\begin{equation}
\label{eq:dynamic-partial-epsilon}
\begin{aligned}
\frac{\partial R_{\mathrm{dyn}}}
{\partial\varepsilon_\gamma}
={}&
(T-t_\gamma)
\\
&-
\frac{B}{2}
\sqrt{k}\,
\varepsilon_\gamma^{-3/2}
\left(\sqrt{T}-\sqrt{t_\gamma}\right)
\sqrt{\log T}.
\end{aligned}
\end{equation}

Substituting Equations~\eqref{eq:dynamic-deps-dgamma},
\eqref{eq:dynamic-dtgamma-dgamma},
\eqref{eq:dynamic-partial-t}, and
\eqref{eq:dynamic-partial-epsilon} into
Equation~\eqref{eq:dynamic-total-derivative} gives
\begin{equation}
\label{eq:dynamic-dR-dgamma}
\begin{aligned}
\frac{dR_{\mathrm{dyn}}}{d\gamma}
={}&
\Bigg\{
A k^{1/3}t_\gamma^{-1/3}
\left[
\frac{2}{3}(\log t_\gamma)^{1/3}
+
\frac{1}{3}(\log t_\gamma)^{-2/3}
\right]
\\
&\qquad
-
\varepsilon_\gamma
-
\frac{B}{2}
\sqrt{\frac{k}{\varepsilon_\gamma}}
\frac{\sqrt{\log T}}{\sqrt{t_\gamma}}
\Bigg\}
\frac{dt_\gamma}{d\gamma}
\\
&+
\Bigg\{
(T-t_\gamma)
-
\frac{B}{2}
\sqrt{k}\,
\varepsilon_\gamma^{-3/2}
\left(\sqrt{T}-\sqrt{t_\gamma}\right)
\sqrt{\log T}
\Bigg\}
\frac{d\varepsilon_\gamma}{d\gamma}.
\end{aligned}
\end{equation}

The point elasticity of dynamic-strategy regret with respect to the privacy risk budget is therefore
\begin{equation}
\label{eq:dynamic-elasticity-final}
\mathcal{E}^{\mathrm{dyn}}_{R,\gamma}
=
\frac{\gamma}{R_{\mathrm{dyn}}(T,\gamma)}
\left[
\frac{\partial R_{\mathrm{dyn}}}{\partial t_\gamma}
\frac{dt_\gamma}{d\gamma}
+
\frac{\partial R_{\mathrm{dyn}}}
{\partial\varepsilon_\gamma}
\frac{d\varepsilon_\gamma}{d\gamma}
\right].
\end{equation}

The elasticity captures two channels through which an increase in the privacy risk budget affects regret. First, increasing \(\gamma\) lowers \(\varepsilon_\gamma\), reducing the amount of random exploration after the cap binds. Second, increasing \(\gamma\) delays the exhaustion time \(t_\gamma\), extending the period during which the uncapped dynamic schedule applies. These effects also influence the estimation-error component because the number of post-cap observations and the amount of post-cap exploration change simultaneously.

A negative elasticity implies that increasing the privacy risk budget reduces regret, whereas a positive elasticity implies that increasing the budget raises regret. The sign therefore depends on the balance between the reduction in exploration cost, the change in the exhaustion time, and the resulting change in estimation error.

If the cap does not become binding within the horizon, then
\[
\varepsilon_\gamma
\leq
\left(\frac{k\log T}{T}\right)^{1/3},
\]
and the dynamic exploration schedule is independent of \(\gamma\) over all rounds \(t\leq T\). In this case,
\begin{equation}
\mathcal{E}^{\mathrm{dyn}}_{R,\gamma}=0.
\end{equation}

\section{Heterogeneity in privacy risk among arms}\label{app:heterogeneity}
So far, we apply the same privacy guarantee to every arm. This corresponds to the standard $\varepsilon$-greedy policy, where the mechanism exploits the greedy arm with probability $1-\varepsilon$ and explores uniformly over all arms with probability $\varepsilon$. Because exploration is typically spread equally across arms, each arm receives the same privacy guarantee.

An extension is to allow privacy guarantees to differ across arms. This is useful when some arms are viewed as more privacy-sensitive than others. Let $\xi_j>0$ denote the privacy parameter associated with displayed arm $j$. A smaller value of $\xi_j$ gives arm $j$ stronger privacy protection, while a larger value allows arm $j$ to be more informative.

To connect arm-specific privacy guarantees to $\varepsilon$-greedy, we keep the same exploitation-exploration structure but replace uniform exploration with weighted exploration.  Specifically, at visitor $t$, the mechanism selects
\begin{equation}
\label{eq:weighted-epsilon-greedy}
a_t^*
=
\begin{cases}
a_t=\arg\max_a Q_t(a), & \text{with probability } 1-\varepsilon,\\[4pt]
\text{Categorical}(w_1,\dots,w_k), & \text{with probability } \varepsilon,
\end{cases}
\end{equation}
where $w_j\geq 0$ and $\sum_{j=1}^k w_j=1$. Under the weighted $\varepsilon$-greedy policy in Equation~\eqref{eq:weighted-epsilon-greedy}, the probability of displaying arm $j$ when the greedy arm is $l$ is
\begin{equation}
\label{eq:weighted-display-prob}
p_{jl}
=
\Pr(a_t^*=j\mid a_t=l)
=
\begin{cases}
1-\varepsilon+\varepsilon w_j, & \text{if } j=l,\\[4pt]
\varepsilon w_j, & \text{if } j\neq l.
\end{cases}
\end{equation}

The diagonal entry in Equation~\eqref{eq:weighted-display-prob} corresponds to the case in which the displayed arm equals the greedy arm. In this case, arm $j$ is displayed either because the policy exploits, with probability $1-\varepsilon$, or because the policy explores and draws arm $j$, with probability $\varepsilon w_j$. The off-diagonal entries correspond to cases in which the displayed arm differs from the greedy arm. In those cases, arm $j$ can be displayed only through exploration, with probability $\varepsilon w_j$.

To give displayed arm $j$ its own privacy upper bound $\xi_j$, we require
\begin{equation}
\label{eq:arm-specific-ratio}
\frac{p_{jj}}{p_{jl}}
=
e^{\xi_j}
\qquad \text{for all } l\neq j.
\end{equation}
Substituting Equation~\eqref{eq:weighted-display-prob} into
Equation~\eqref{eq:arm-specific-ratio} gives
\begin{equation}
\label{eq:ratio-substitution}
\frac{1-\varepsilon+\varepsilon w_j}{\varepsilon w_j}
=
e^{\xi_j}.
\end{equation}
Rearranging Equation~\eqref{eq:ratio-substitution} yields
\begin{equation}
\label{eq:privacy-exploration-link}
1-\varepsilon
=
\varepsilon w_j(e^{\xi_j}-1),
\end{equation}
and therefore
\begin{equation}
\label{eq:epsilon-wj}
\varepsilon w_j
=
\frac{1-\varepsilon}{e^{\xi_j}-1}.
\end{equation}

Equation~\eqref{eq:epsilon-wj} shows how privacy and exploration are linked. If $\xi_j$ is small, then $e^{\xi_j}-1$ is small, so $\varepsilon w_j$ must be large. Thus, more privacy-sensitive arms are sampled more often during exploration, which makes observing those arms less informative to an external observer. Summing Equation~\eqref{eq:epsilon-wj} over all arms gives
\begin{equation}
\label{eq:sum-epsilon-wj}
\sum_{j=1}^k \varepsilon w_j
=
\sum_{j=1}^k
\frac{1-\varepsilon}{e^{\xi_j}-1}.
\end{equation}
Since $\sum_{j=1}^k w_j=1$, the left-hand side of Equation~\eqref{eq:sum-epsilon-wj} equals $\varepsilon$. Hence,
\begin{equation}
\label{eq:epsilon-implicit}
\varepsilon
=
(1-\varepsilon)
\sum_{j=1}^k
\frac{1}{e^{\xi_j}-1}.
\end{equation}
Solving Equation~\eqref{eq:epsilon-implicit} for $\varepsilon$ yields
\begin{equation}
\label{eq:epsilon-heterogeneous}
\varepsilon
=
\frac{
\sum_{j=1}^k \frac{1}{e^{\xi_j}-1}
}{
1+
\sum_{j=1}^k \frac{1}{e^{\xi_j}-1}
}.
\end{equation}
The corresponding exploration weights are
\begin{equation}
\label{eq:weights-heterogeneous}
w_j
=
\frac{
\frac{1}{e^{\xi_j}-1}
}{
\sum_{m=1}^k \frac{1}{e^{\xi_m}-1}
}.
\end{equation}

Therefore, the weighted $\varepsilon$-greedy policy that implements arm-specific privacy guarantees is
\begin{equation}
\label{eq:weighted-policy-final}
a_t^*
=
\begin{cases}
a_t=\arg\max_a Q_t(a), & \text{with probability }
\displaystyle
1-\varepsilon
=
\frac{1}
{1+\sum_{m=1}^k \frac{1}{e^{\xi_m}-1}},\\[16pt]

\text{Categorical}(w_1,\dots,w_k), & \text{with probability }
\displaystyle
\varepsilon
=
\frac{
\sum_{m=1}^k \frac{1}{e^{\xi_m}-1}
}{
1+\sum_{m=1}^k \frac{1}{e^{\xi_m}-1}
}.
\end{cases}
\end{equation}

Equivalently, substituting Equations~\eqref{eq:epsilon-heterogeneous} and \eqref{eq:weights-heterogeneous} into Equation~\eqref{eq:weighted-display-prob} gives
\begin{equation}
\label{eq:weighted-matrix-expanded}
p_{jl}
=
\begin{cases}
\displaystyle
\frac{e^{\xi_j}}
{(e^{\xi_j}-1)\left(1+\sum_{m=1}^k \frac{1}{e^{\xi_m}-1}\right)}
& \text{if } j=l, \\[16pt]
\displaystyle
\frac{1}
{(e^{\xi_j}-1)\left(1+\sum_{m=1}^k \frac{1}{e^{\xi_m}-1}\right)}
& \text{if } j\neq l.
\end{cases}
\end{equation}

For each displayed arm $j$, Equation~\eqref{eq:weighted-display-prob} implies
\begin{equation}
\label{eq:privacy-ratio-final}
\frac{p_{jj}}{p_{jl}}
=
\frac{1-\varepsilon+\varepsilon w_j}{\varepsilon w_j}
=
e^{\xi_j}
\qquad \text{for all } l\neq j.
\end{equation}
Thus,
\begin{equation}
\label{eq:arm-specific-privacy-loss}
\max_{l,l'} \log \frac{p_{jl}}{p_{jl'}}
=
\xi_j.
\end{equation}
Hence, observing displayed arm $j$ has privacy loss bounded by $\xi_j$. The overall differential-privacy guarantee of the mechanism is governed by the largest arm-specific privacy parameter:
\begin{equation}
\label{eq:global-privacy}
\xi^{global}
=
\max_j \xi_j.
\end{equation}
Thus, the mechanism is $\max_j \xi_j$-differentially private.

\subsection{Numerical example}
Consider an experiment with $K=3$ arms and arm-specific privacy parameters
\[
\xi_1=0.5,\qquad
\xi_2=1,\qquad
\xi_3=2.
\]
Because smaller values of $\xi_j$ correspond to stronger privacy protection, arm 1 requires the most privacy protection and arm 3 the least. The term
\[
\frac{1}{e^{\xi_j}-1}
\]
captures the relative amount of exploration needed for arm $j$ to achieve its privacy guarantee. A smaller privacy parameter $\xi_j$ makes $e^{\xi_j}-1$ smaller, and therefore increases this term. Intuitively, a more privacy-sensitive arm must be shown more often through random exploration, so that observing this arm is less informative about whether it was actually the greedy arm.

For the three arms, these relative exploration requirements are
\[
\frac{1}{e^{\xi_1}-1}
=
\frac{1}{e^{0.5}-1}
\approx 1.542,
\]
\[
\frac{1}{e^{\xi_2}-1}
=
\frac{1}{e^1-1}
\approx 0.582,
\]
and
\[
\frac{1}{e^{\xi_3}-1}
=
\frac{1}{e^2-1}
\approx 0.157.
\]
Thus, arm 1 receives the largest relative exploration requirement, because it has the strongest privacy guarantee, while arm 3 receives the smallest. The sum of these values,
\[
\sum_{j=1}^3 \frac{1}{e^{\xi_j}-1}
\approx
1.542+0.582+0.157
=
2.280,
\]
serves as a normalizing constant. It converts the relative exploration requirements into valid exploration weights that sum to one, and also determines the overall exploration probability. Using Equation~\eqref{eq:epsilon-heterogeneous}, the exploration probability is
\[
\varepsilon
=
\frac{2.280}{1+2.280}
\approx
0.695,
\]
and the exploitation probability is
\[
1-\varepsilon
\approx
0.305.
\]

Using Equation~\eqref{eq:weights-heterogeneous}, the exploration weights are
\[
w_1=\frac{1.542}{2.280}\approx0.676,\qquad
w_2=\frac{0.582}{2.280}\approx0.255,\qquad
w_3=\frac{0.157}{2.280}\approx0.069.
\]
Thus, the weighted $\varepsilon$-greedy policy in Equation~\eqref{eq:weighted-policy-final} becomes
\[
a_t^*
=
\begin{cases}
a_t=\arg\max_a Q_t(a), & \text{with probability } 0.305,\\
\text{Categorical}(0.676,0.255,0.069), & \text{with probability } 0.695.
\end{cases}
\]

The off-diagonal probabilities are
\[
\varepsilon w_1\approx0.695\times0.676=0.470,\qquad
\varepsilon w_2\approx0.695\times0.255=0.177,\qquad
\varepsilon w_3\approx0.695\times0.069=0.048.
\]
The diagonal probabilities add the exploitation probability:
\[
p_{11}=1-\varepsilon+\varepsilon w_1\approx0.305+0.470=0.775,\hspace{.5em}
p_{22}=1-\varepsilon+\varepsilon w_2\approx0.305+0.177=0.482,\hspace{.5em}
p_{33}=1-\varepsilon+\varepsilon w_3\approx0.305+0.048=0.353.
\]

Therefore, the heterogeneous stochastic matrix is
\[
\mathbf P^{het}
\approx
\begin{pmatrix}
0.775 & 0.470 & 0.470 \\
0.177 & 0.482 & 0.177 \\
0.048 & 0.048 & 0.353
\end{pmatrix}.
\]
Rows correspond to displayed arms and columns correspond to greedy arms. The arm-specific privacy guarantees can be verified using Equation~\eqref{eq:privacy-ratio-final}. For arm 1, $\frac{p_{11}}{p_{12}}=\frac{0.775}{0.470}\approx1.65=e^{0.5}$; for arm 2, $\frac{p_{22}}{p_{21}}=\frac{0.482}{0.177}\approx2.72=e^1$; and for arm 3, $\frac{p_{33}}{p_{31}}=\frac{0.353}{0.048}\approx7.39=e^2$.
Therefore,
\[
\log\left(\frac{p_{11}}{p_{12}}\right)\approx 0.5,
\qquad
\log\left(\frac{p_{22}}{p_{21}}\right)\approx 1,
\qquad
\log\left(\frac{p_{33}}{p_{31}}\right)\approx 2.
\]

In this example, arm 1 is treated as the most privacy-sensitive arm because it has the smallest privacy parameter, $\xi_1=0.5$. It therefore receives the largest exploration weight, $w_1\approx 0.676$. This makes arm 1 appear often even when it is not the greedy arm, so observing arm 1 is less informative.

Arm 3 is treated as the least privacy-sensitive arm because it has the largest privacy parameter, $\xi_3=2$. It receives the smallest exploration weight, $w_3\approx 0.069$. As a result, observing arm 3 is more informative about whether arm 3 was the greedy arm.

This heterogeneous policy nests the standard symmetric mechanism. If $\xi_j=\xi$ for all arms $j$, then Equation~\eqref{eq:weights-heterogeneous}
implies
\[
w_j=\frac{1}{k}
\qquad \text{for all } j,
\]
and Equation~\eqref{eq:epsilon-heterogeneous} becomes
\[
\varepsilon
=
\frac{k}{k+e^\xi-1}.
\]
Thus, the weighted $\varepsilon$-greedy policy collapses to the standard $\varepsilon$-greedy policy with uniform exploration.

\subsection{Regret bound under weighted exploration}
The regret analysis changes when privacy guarantees vary by arm. In the symmetric mechanism, exploration is uniform, so each arm is explored with probability $\varepsilon_t/k$. This is why the confidence radius depends on $t\varepsilon_t/k$ (see Equation~\ref{unionboundd}). With arm-specific privacy guarantees, exploration is weighted. Following Equation~\eqref{eq:weighted-epsilon-greedy}, arm $j$ is explored with probability
\begin{equation}
\label{eq:qjt-def}
q_{j,t}
=
\varepsilon_t w_{j,t}.
\end{equation}

The worst-case confidence radius is therefore governed by the least-explored arm,
\begin{equation}
\label{eq:qmin-def}
q_{\min,t}
=
\min_j q_{j,t}.
\end{equation}

Using the same clean-event argument as in the symmetric regret analysis, the per-round regret bound becomes
\begin{equation}
\label{eq:het-regret-general}
E[\tilde R^{het}(t)]
\leq
\varepsilon_t^{het}
+
2\sqrt{
\frac{2\log t}{tq_{\min,t}}
}.
\end{equation}
The first term in Equation~\eqref{eq:het-regret-general} bounds the regret from exploration, while the second term bounds the regret from exploiting an incorrect greedy arm due to estimation error. Given the vector of privacy parameters $(\xi_{1,t},\dots,\xi_{k,t})$, the total exploration probability is
\begin{equation}
\label{eq:epsilon-het}
\varepsilon_t^{het}
=
\frac{
\sum_{j=1}^k \frac{1}{e^{\xi_{j,t}}-1}
}{
1+\sum_{j=1}^k \frac{1}{e^{\xi_{j,t}}-1}
}.
\end{equation}

The corresponding exploration weights are
\begin{equation}
\label{eq:weights-het-time}
w_{j,t}
=
\frac{
\frac{1}{e^{\xi_{j,t}}-1}
}{
\sum_{m=1}^k \frac{1}{e^{\xi_{m,t}}-1}
}.
\end{equation}
Therefore, the probability that arm $j$ is explored is
\begin{equation}
\label{eq:qjt-privacy}
q_{j,t}
=
\varepsilon_t^{het}w_{j,t}
=
\frac{
\frac{1}{e^{\xi_{j,t}}-1}
}{
1+\sum_{m=1}^k \frac{1}{e^{\xi_{m,t}}-1}
}.
\end{equation}

\subsubsection{Constant and dynamic arm-level privacy schedules}
The regret bound in Equation~\eqref{eq:het-regret-general} applies to both constant and dynamic heterogeneous privacy strategies. Under a constant heterogeneous strategy, each arm has a fixed privacy budget $\gamma_j$, so
\begin{equation}
\label{eq:constant-het-xi}
\xi_{j,t}
\equiv
\gamma_j.
\end{equation}

Under a dynamic heterogeneous strategy, each arm has a budget $\gamma_j$, but the uncapped dynamic privacy level changes over time. Let
\begin{equation}
\label{eq:uncapped-dynamic-xi}
\xi_t^{dyn}
=
\log\left(
1-k+\frac{k}{\left(\frac{k\log t}{t}\right)^{1/3}}
\right)
\end{equation}
denote the privacy level implied by the uncapped dynamic strategy. The arm-specific dynamic privacy level is
\begin{equation}
\label{eq:dynamic-het-xi}
\xi_{j,t}
=
\min\left\{
\gamma_j,\,
\xi_t^{dyn}
\right\}.
\end{equation}

Thus, before any arm-specific budget binds, all arms receive the same privacy level and the policy is symmetric. Once some budgets bind and others do not, privacy guarantees differ across arms and the policy becomes heterogeneous.

\subsubsection{Symmetric benchmark with the same global privacy guarantee}

We compare the heterogeneous mechanism to a symmetric benchmark with the same
global privacy guarantee at time $t$. Let
\begin{equation}
\label{eq:xi-max-def}
\xi_{\max,t}
=
\max_j \xi_{j,t}.
\end{equation}
The symmetric benchmark applies privacy level $\xi_{\max,t}$ to every arm. Thus, both mechanisms satisfy the same global privacy guarantee at time $t$. Because $\xi_{\max,t}$ is the largest privacy parameter, the least-explored arm under the heterogeneous mechanism is the arm with privacy parameter $\xi_{\max,t}$. Therefore, from Equation~\eqref{eq:qjt-privacy},
\begin{equation}
\label{eq:qmin-privacy}
q_{\min,t}
=
\frac{
\frac{1}{e^{\xi_{\max,t}}-1}
}{
1+\sum_{m=1}^k \frac{1}{e^{\xi_{m,t}}-1}
}.
\end{equation}

Substituting Equations~\eqref{eq:epsilon-het} and \eqref{eq:qmin-privacy} into Equation~\eqref{eq:het-regret-general} gives
\begin{equation}
\label{eq:het-regret-expanded}
E[\tilde R^{het}(t)]
\leq
\frac{
\sum_{j=1}^k \frac{1}{e^{\xi_{j,t}}-1}
}{
1+\sum_{j=1}^k \frac{1}{e^{\xi_{j,t}}-1}
}
+
2\sqrt{
\frac{
2\left(1+\sum_{j=1}^k \frac{1}{e^{\xi_{j,t}}-1}\right)\log t
}{
t\frac{1}{e^{\xi_{\max,t}}-1}
}
}.
\end{equation}

The symmetric benchmark corresponds to setting
\begin{equation}
\label{eq:sym-privacy}
\xi_{j,t}
=
\xi_{\max,t}
\qquad \text{for all } j.
\end{equation}
Under Equation~\eqref{eq:sym-privacy}, the exploration probability is
\begin{equation}
\label{eq:epsilon-sym}
\varepsilon_t^{sym}
=
\frac{
\frac{k}{e^{\xi_{\max,t}}-1}
}{
1+\frac{k}{e^{\xi_{\max,t}}-1}
}
=
\frac{k}{k+e^{\xi_{\max,t}}-1}.
\end{equation}
which corresponds to Equation \eqref{eq:differential-greedy}. Each arm is explored with probability
\begin{equation}
\label{eq:q-sym}
q_t^{sym}
=
\frac{
\frac{1}{e^{\xi_{\max,t}}-1}
}{
1+\frac{k}{e^{\xi_{\max,t}}-1}
}
=
\frac{1}{k+e^{\xi_{\max,t}}-1}.
\end{equation}

The symmetric worst-case regret bound is therefore
\begin{equation}
\label{eq:sym-regret-bound}
E[\tilde R^{sym}(t)]
\leq
\frac{
\frac{k}{e^{\xi_{\max,t}}-1}
}{
1+\frac{k}{e^{\xi_{\max,t}}-1}
}
+
2\sqrt{
\frac{
2\left(1+\frac{k}{e^{\xi_{\max,t}}-1}\right)\log t
}{
t\frac{1}{e^{\xi_{\max,t}}-1}
}
}.
\end{equation}

\subsubsection{Why arm-level heterogeneity increases the regret bound}
We now compare the heterogeneous bound in Equation~\eqref{eq:het-regret-expanded} with the symmetric bound in Equation~\eqref{eq:sym-regret-bound}. By definition, $\xi_{j,t}\leq \xi_{\max,t}$ for all arms. If privacy guarantees are heterogeneous at time $t$, then this inequality is strict for at least one arm. Since the function $x\mapsto 1/(e^x-1)$ is strictly decreasing, it follows that
\begin{equation}
\label{eq:sum-privacy-ineq}
\sum_{j=1}^k
\frac{1}{e^{\xi_{j,t}}-1}
>
\frac{k}{e^{\xi_{\max,t}}-1}.
\end{equation}

By Equations~\eqref{eq:epsilon-het}, \eqref{eq:epsilon-sym}, and \eqref{eq:sum-privacy-ineq}, the heterogeneous exploration probability is larger than the symmetric exploration probability:
\begin{equation}
\label{eq:epsilon-het-greater}
\varepsilon_t^{het}
>
\varepsilon_t^{sym}.
\end{equation}

Equation~\eqref{eq:sum-privacy-ineq} also implies that the least-explored arm under heterogeneous privacy is less explored than each arm under the symmetric benchmark. Using Equations~\eqref{eq:qmin-privacy} and \eqref{eq:q-sym},
\begin{equation}
\label{eq:qmin-less}
q_{\min,t}
=
\frac{
\frac{1}{e^{\xi_{\max,t}}-1}
}{
1+\sum_{m=1}^k \frac{1}{e^{\xi_{m,t}}-1}
}
<
\frac{
\frac{1}{e^{\xi_{\max,t}}-1}
}{
1+\frac{k}{e^{\xi_{\max,t}}-1}
}
=
q_t^{sym}.
\end{equation}

Because the confidence term in Equation~\eqref{eq:het-regret-general} is decreasing in the exploration probability of the least-explored arm, Equation~\eqref{eq:qmin-less} implies
\begin{equation}
\label{eq:confidence-term-greater}
2\sqrt{
\frac{2\log t}{tq_{\min,t}}
}
>
2\sqrt{
\frac{2\log t}{tq_t^{sym}}
}.
\end{equation}

Equations~\eqref{eq:epsilon-het-greater} and
\eqref{eq:confidence-term-greater} show that both components of the worst-case regret bound are larger under heterogeneous privacy. Equation~\eqref{eq:epsilon-het-greater} compares the exploration component: heterogeneous privacy increases the total privacy-induced exploration probability. Equation~\eqref{eq:confidence-term-greater} compares the exploitation component: because heterogeneous privacy makes exploration uneven across arms, the least-explored arm receives less exploration than under the symmetric benchmark, which increases the confidence radius and therefore the regret from exploiting an incorrectly estimated greedy arm.

Thus, relative to a symmetric mechanism with the same global privacy guarantee $\xi_{\max,t}$, arm-level privacy heterogeneity increases the worst-case per-round regret bound whenever privacy guarantees differ across arms at time $t$.

\subsubsection{Quantifying the regret cost of heterogeneity.}\label{app:costheterogeneity}
Holding $\xi_{\max,t}$ fixed, we can quantify the increase in the worst-case regret bound from arm-level privacy heterogeneity. Define
\begin{equation}
\label{eq:heterogeneity-gap}
\Delta_t^{het}
=
\sum_{j=1}^k
\frac{1}{e^{\xi_{j,t}}-1}
-
\frac{k}{e^{\xi_{\max,t}}-1}.
\end{equation}
The term $\Delta_t^{het}$ measures the distance from the symmetric benchmark on the exploration scale. It equals zero when all arms have privacy parameter $\xi_{\max,t}$ and is positive whenever privacy guarantees differ across arms.

Let $B_t^{het}$ and $B_t^{sym}$ denote the right-hand sides of Equations~\eqref{eq:het-regret-expanded} and \eqref{eq:sym-regret-bound}, respectively. Using Equation~\eqref{eq:heterogeneity-gap}, the heterogeneous bound can be written as
\begin{equation}
\label{eq:het-bound-gap}
B_t^{het}
=
\frac{
\frac{k}{e^{\xi_{\max,t}}-1}
+
\Delta_t^{het}
}{
1+
\frac{k}{e^{\xi_{\max,t}}-1}
+
\Delta_t^{het}
}
+
2\sqrt{
\frac{
2\left(
1+\frac{k}{e^{\xi_{\max,t}}-1}
+\Delta_t^{het}
\right)
(e^{\xi_{\max,t}}-1)\log t
}{
t
}
}.
\end{equation}

The symmetric benchmark corresponds to $\Delta_t^{het}=0$. Therefore, the increase in the worst-case per-round regret bound due to heterogeneity is
\begin{equation}
\label{eq:regret-increase-heterogeneity}
B_t^{het}-B_t^{sym}
=
\left[
\frac{
\frac{k}{e^{\xi_{\max,t}}-1}
+
\Delta_t^{het}
}{
1+
\frac{k}{e^{\xi_{\max,t}}-1}
+
\Delta_t^{het}
}
-
\frac{
\frac{k}{e^{\xi_{\max,t}}-1}
}{
1+
\frac{k}{e^{\xi_{\max,t}}-1}
}
\right]
+
2\sqrt{
\frac{
2(e^{\xi_{\max,t}}-1)\log t
}{
t
}
}
\left[
\sqrt{
1+\frac{k}{e^{\xi_{\max,t}}-1}
+\Delta_t^{het}
}
-
\sqrt{
1+\frac{k}{e^{\xi_{\max,t}}-1}
}
\right].
\end{equation}

Equivalently, using
\[
\frac{x+\Delta}{1+x+\Delta}-\frac{x}{1+x}
=
\frac{\Delta}{(1+x+\Delta)(1+x)}
\]
and
\[
\sqrt{x+\Delta}-\sqrt{x}
=
\frac{\Delta}{\sqrt{x+\Delta}+\sqrt{x}},
\]
Equation~\eqref{eq:regret-increase-heterogeneity} can be written as
\begin{equation}
\label{eq:regret-increase-heterogeneity-factored}
B_t^{het}-B_t^{sym}
=
\Delta_t^{het}
\left[
\frac{1}
{
\left(
1+\frac{k}{e^{\xi_{\max,t}}-1}+\Delta_t^{het}
\right)
\left(
1+\frac{k}{e^{\xi_{\max,t}}-1}
\right)
}
+
\frac{
2\sqrt{
\frac{
2(e^{\xi_{\max,t}}-1)\log t
}{
t
}
}
}
{
\sqrt{
1+\frac{k}{e^{\xi_{\max,t}}-1}+\Delta_t^{het}
}
+
\sqrt{
1+\frac{k}{e^{\xi_{\max,t}}-1}
}
}
\right].
\end{equation}

Equation~\eqref{eq:regret-increase-heterogeneity-factored} isolates the additional regret due to arm-level privacy heterogeneity. The increase in the bound is proportional to $\Delta_t^{het}$, the additional exploration scale required by heterogeneous privacy guarantees. The bracketed term is strictly positive. Hence, whenever $\Delta_t^{het}>0$, the heterogeneous privacy policy has a larger worst-case regret bound than the symmetric benchmark with the same global privacy guarantee.

Moreover, because $\Delta_t^{het}>0$, the term
\[
1+\frac{k}{e^{\xi_{\max,t}}-1}+\Delta_t^{het}
\]
is larger than its value under the symmetric benchmark, where
$\Delta_t^{het}=0$. Hence, both denominators in the bracketed term of
Equation~\eqref{eq:regret-increase-heterogeneity-factored} are larger than
their values at $\Delta_t^{het}=0$. Since the corresponding numerators are
positive and do not depend on $\Delta_t^{het}$, the bracketed term is largest
at $\Delta_t^{het}=0$. Evaluating the bracket at this value therefore gives
an upper bound on the additional regret from heterogeneity. Therefore,
\begin{equation}
\label{eq:regret-increase-heterogeneity-upper-bound}
B_t^{het}-B_t^{sym}
\leq
\Delta_t^{het}
\left[
\frac{1}
{
\left(
1+\frac{k}{e^{\xi_{\max,t}}-1}
\right)^2
}
+
\frac{
\sqrt{
\frac{
2(e^{\xi_{\max,t}}-1)\log t
}{
t
}
}
}
{
\sqrt{
1+\frac{k}{e^{\xi_{\max,t}}-1}
}
}
\right].
\end{equation}

Thus, holding $\xi_{\max,t}$ fixed, the additional regret from arm-level privacy heterogeneity is bounded above by a term that is linear in $\Delta_t^{het}$. The coefficient multiplying $\Delta_t^{het}$ is the marginal cost of heterogeneity evaluated at the symmetric benchmark. Since this coefficient is positive, any $\Delta_t^{het}>0$ increases the worst-case regret bound relative to the symmetric benchmark.
\appendix
\end{APPENDICES}


\bibliographystyle{informs2014} 
\bibliography{references} 


\end{document}